\documentstyle[float,epsfig,12pt]{article}                                                                                     
                                                                                      
\newlength{\dinwidth}                                                                                      
\newlength{\dinmargin}                                                                                      
\def\lapproxeq{\lower .7ex\hbox{$\;\stackrel{\textstyle                                                                                      
<}{\sim}\;$}}                                                                                      
\def\gapproxeq{\lower .7ex\hbox{$\;\stackrel{\textstyle                                                                                      
>}{\sim}\;$}}                                                                                      
\def\btop{\mathchar"1339}                                                                                      
\def\bmid{\mathchar"133D}                                                                                      
\def\bbot{\mathchar"133B}                                                                                      
                                                                                     
\def\msb{\overline{\rm MS}}                                                                                      
\def\GeV{{\rm GeV}}                                                                                      
\def\TeV{{\rm TeV}}                                                                                      
                                                                                      
\def\gup{{g\uparrow}} 
\def\gdown{{g\downarrow}} 
\def\asup{{\alpha_S\uparrow\uparrow}}                                                   
\def\asdown{{\alpha_S\downarrow\downarrow}}                                                   
\def\beq{\begin{equation}}                                                   
\def\eeq{\end{equation}}                                                   
\def\be{\begin{equation}}                                                                                      
\def\ee{\end{equation}}                                                                                      
\def\bea{\begin{eqnarray}}                                                                                      
\def\eea{\end{eqnarray}}                                                                                      
                                                                                      
\begin{document}                                                                                      
\titlepage                                                                                      
\begin{flushright}                                                                                      
DTP/98/10  \\                                                                                      
RAL-TR-98-029 \\ 
hep-ph/9803445\\ 
March 1998                                                                                      
\end{flushright}                                                                                      
                                                                                      
\begin{center}
\vspace*{2cm}                                                                                      
{\Large \bf Parton distributions :  a new global analysis} \\                                                                                      
\vspace*{1cm}                                                                                      
A.\ D.\ Martin$^a$, R.\ G.\ Roberts$^b$, W.\ J.\ Stirling$^{a,c}$ and R.\ S.\                                             
Thorne$^d$ \\                                                                                      
                                                                                      
\vspace*{0.5cm}                                                                                      
$^a \; $ {\it Department of Physics, University of Durham,                                                                                      
Durham, DH1 3LE }\\                                                                                      
                                                                                      
$^b \; $ {\it Rutherford Appleton Laboratory, Chilton,                                                                                      
Didcot, Oxon, OX11 0QX}\\                                                                                      
                                            
$^c \; $ {\it Department of Mathematical Sciences, University of Durham,                                             
Durham, DH1 3LE} \\                                            
                                                                                      
$^d \; $ {\it Jesus College, University of Oxford, Oxford, OX1 3DW}                                                                                      
\end{center}                                                                                      
                                                                                      
\vspace*{1.5cm}                                                                                      
\begin{abstract}                                                                                      
We present a new analysis of parton distributions of the proton.  This incorporates a                                        
wide range of new data, an improved treatment of heavy flavours and a
re-examination of prompt                                       
photon production.  The new set (MRST) shows systematic differences from previous                                        
sets of partons which can be identified with particular features of the new data and                                        
with improvements in the analysis.  We also investigate the sensitivities of the results                                        
to (i) the uncertainty in the determination of the gluon at large $x$, (ii) the value of                                        
$\alpha_S (M_Z^2)$ and (iii) the minimum $Q^2$ cut on the data that are included in     
the global fit.                                                                                  
\end{abstract}                                                                                      
                                                                                      
\newpage                                                                                      
                                                                                      
\noindent{\large \bf 1. Introduction}                                                                                      
                                                                                      
The last few years have seen a spectacular improvement in the precision                                                                                       
and in the kinematic range of the experimental measurements of deep inelastic                                                                                      
and related hard scattering processes.  As a consequence the parton distributions                                                                                      
of the proton are much better known, with tight constraints on the gluon and the                                                                                      
quark sea for Bjorken $x$ as low as $10^{-4}$.  However, several significant                                                                                      
sets of new data have become available recently which, when incorporated in a                                                                                       
global analysis, will increase considerably our knowledge of the parton                                                                                       
distributions.                                                                                      
                                                                                      
First let us summarize the new experimental measurements that have become                                                                                      
available, and their implications.  The new information includes the following.                                                                                      
\begin{itemize}                                                                                      
\item[(i)]  We have new, more precise, measurements of the structure function                                                                                      
$F_2^{ep}$ for deep inelastic electron-proton scattering by the H1 and                                                                                       
ZEUS collaborations at HERA \cite{H1,ZEUS}.  The data now extend over a                                                                                      
much wider kinematic range.  Loosely speaking $F_2$ and $\partial F_2/\partial                                                                                      
\ln Q^2$ serve to constrain the sea quark and gluon\footnote{At the lower values                                                                                      
of $Q^2$ there is a significant correlation between $g (x, Q^2)$ and the                                                                                       
value of the strong coupling $\alpha_S (Q^2)$.} distributions in the region                                                                                      
$10^{-4} \lapproxeq x \lapproxeq 10^{-2}$.                                                                                      
                                                                                      
\item[(ii)] A re-analysis of the CCFR neutrino data has led to a new set of                                                                                      
$F_2^{\nu N}$ and $x F_3^{\nu N}$ structure function measurements \cite{CCFR2}.                                                                                        
Besides affecting                                                                                      
the quark densities, the new structure functions give a value of the strong                                                                                      
coupling $\alpha_S$ much more in line with the world average determination                                                                                       
than the original CCFR data \cite{CCFR1}.                                                                                      
                                                                                      
\item[(iii)]  There now exist measurements of the charm component of $F_2^{ep}$                                                                                      
in electron-proton deep inelastic scattering at HERA \cite{H1C,ZEUSC}.  These new                                                                                      
data sample $x \sim 10^{-3}$ and complement the existing EMC charm data with                                                                                      
$x \sim 10^{-1}$ \cite{EMC}.  Such data constrain both the charm sea and gluon                                                                                      
distributions via the subprocesses $\gamma^* c \rightarrow c$ and $\gamma^* g                                                                                       
\rightarrow c \bar{c}$ respectively.                                                                                      
                                                                                      
\item[(iv)]  Very precise measurements of prompt photon production, $pp                                                                                       
\rightarrow \gamma X$, have become available from the E706 collaboration                                                                                       
\cite{E706}.  These data motivate us to look again at our treatment of this                                                                                      
reaction, and in particular of the WA70 prompt photon measurements \cite{WA70}.                                                                                      
We emphasize that such prompt photon data are the main constraints on the                                                                                       
gluon\footnote{Other reactions can in the future offer important constraints on                                                                                      
the gluon, see (ix) below.} outside the HERA small $x$ $(x \sim 10^{-3})$                                                                                       
domain, apart of course from the global momentum sum rule constraint.                                                                                      
                                                                                      
\item[(v)] The E866 collaboration \cite{E866} have measured the asymmetry in                                                                                       
Drell-Yan production in $pp$ and $pn$ collisions over an extended $x$ range,                                                                                       
$0.03 \lapproxeq x \lapproxeq 0.35$.  The asymmetry data provide direct information                                             
on the $x$ dependence of the difference, $\bar{u} - \bar{d}$, of the                                                                                       
sea quark densities.                                                                                      
Previously there existed only the single measurement of NA51 \cite{NA51} at                                                                                      
$x = 0.18$, which revealed that $\bar{d} \simeq 2\bar{u}$ at this $x$ value.  Now                                                                                      
much more information on $\bar{u} - \bar{d}$ is available\footnote{See also                                       
Section~7.2 where information on $\bar{u} - \bar{d}$ from semi-inclusive deep                                       
inelastic data is discussed.}.                                                                                      
                                                                                      
\item[(vi)]  The CDF collaboration \cite{CDF} have been able to considerably                                                                                      
improve the precision of their measurements of the asymmetry of the rapidity                                                                                      
distributions of the charged lepton from $W^\pm \rightarrow l^\pm \nu$ decays                                                                                      
at the Tevatron $p \bar{p}$ collider.  The new data extend to larger                                                                                       
values of lepton rapidity.  These data offer a tight constraint                                                                                      
on the $u/d$ ratio of parton densities.                                                                                      
                                                                                      
\item[(vii)]  NMC have now completed their structure function analysis, and                                                                                       
supplied further data on $F_2^{\mu p}$, $F_2^{\mu d}$ and the ratio $F_2^{\mu                                           
d}/F_2^{\mu p}$ \cite{NMC}.  The dedicated measurement of the ratio provides                                                                                       
a valuable constraint on the $u/d$ ratio.                                                                                      
                                                                                      
\item[(viii)]The data on Drell-Yan production obtained by the E772 collaboration                                                                                      
\cite{E772} cover a wider range of $x_F$ than the E605 data \cite{E605} which we                                             
have used to constrain the sea.  For $x_F \sim 0$ both experiments provide a useful                                                                                      
measure of the quark sea at larger values of $x$, typically $x \lapproxeq 0.3$.  For                                                                                      
larger $x_F$ the E772 data probe, in principle, the valence quarks at $x \sim 0.5$ and                                       
the sea quarks at $x \sim 0.025$.                                                                                      
                                                                                      
\item[(ix)] There are several other data sets which have the potential to provide                                                                                      
important information on partons in the future.  These include jet, $W +$ jet                                                                                      
and heavy quark $(b, t)$ production at Fermilab, and diffractive $J/\psi$ and                                                                                      
dijet production at HERA, as well as jet production in deep inelastic scattering.                                                                                      
\end{itemize}  
 
The paper is organized as follows.  In Section~2 we describe our procedure and the  
input parametrization of the partons.  The optimum global set of partons, which we  
denote simply MRST, is presented together with two other sets which represent the  
possible range of behaviour of the gluon\footnote{The FORTRAN code for all the parton sets  
described in this paper can be obtained from {\tt http://durpdg.dur.ac.uk/HEPDATA/PDF},  
or by contacting {\tt W.J.Stirling@durham.ac.uk}.}.  The quality of the description  
of the deep inelastic scattering data is shown in Section~3.  The sensitivity to the cuts  
imposed on the data that are fitted and to the value of $\alpha_S$ are also discussed.   
In Section~4 we study the impact of the prompt photon data on the determination of  
the gluon.  We pay particular attention to the transverse momentum $(k_T)$ smearing  
of the incoming partons.  In Section~5 we present a self-contained discussion of our  
new treatment of the heavy flavour $(c, b)$ contributions to the structure functions.   
The description of the data for the Drell-Yan process is given in Section~6 and  
 in Section~7 we discuss the $(u, d, s)$ flavour decomposition of the sea.   
Section~8 is devoted to the constraint imposed by the new $W$ asymmetry data.   
Some implications of the impact of our new partons for processes observed at the  
Fermilab Tevatron collider are presented in Section~9.  Finally in Section~10 we  
summarize the key features of our analysis. \\ 
                                                                                      
\noindent{\large \bf 2. Global parton analysis}                                                                                      
                                                                                      
The parton distributions $f_i$ are determined from a global fit to a wide range                                                                                      
of deep inelastic and related hard scattering data. The basic procedure is to                                                                                      
parametrize the $f_i(x,Q^2)$ at a low value of $Q^2=Q^2_0$ such that the                                                                                       
$f_i(x,Q^2)$ can be calculated at higher $Q^2$ by using next-to-leading-order                                                                                       
(NLO) DGLAP (Dokshitzer-Gribov-Lipatov-Altarelli-Parisi) evolution equations.                                                                                       
Data are fitted for all $Q^2>Q^2_1$, where $Q^2_1>Q^2_0$ is a value of $Q^2$                                                                                       
where perturbative QCD is believed to be the dominant contribution. We shall                                                                                      
study the                                                                                       
sensitivity of the results to variation of the choice of $Q^2_1$.                                                                                      
                                                                                      
We parametrize the starting parton distributions at $Q^2 = Q^2_0                                                                                       
= 1$~GeV$^2$ where the number of active flavours is $n_f=3$. We work in the                                                                                       
$\rm\overline{MS}$ renormalization scheme and use the starting parametric forms                                                                                      
\begin{eqnarray}                                                                                      
xu_v &=& A_{u}x^{\eta_1} (1-x)^{\eta_2} (1+\epsilon_{u}\sqrt{x}+\gamma_{u}x)                                                                                      
\label{eq:a1} \\                                                                                      
xd_v &=& A_{d}x^{\eta_3} (1-x)^{\eta_4} (1+\epsilon_{d}\sqrt{x}+\gamma_{d}x)                                                                                      
\label{eq:a2} \\                                                                                      
xS &=& A_{S}x^{-\lambda_S} (1-x)^{\eta_S}                                                                                      
(1+\epsilon_{S}\sqrt{x}+\gamma_{S}x) \label{eq:a3} \\                                                                                      
xg &=& A_{g}x^{-\lambda g} (1-x)^{\eta g}                                                                                      
(1+\epsilon_{g}\sqrt{x}+\gamma_{g}x).   \label{eq:a4}                                                                                   
\end{eqnarray}                                                                                      
\noindent The flavour structure of the light quark sea at $Q_0^2$ is taken to be                                                                                      
\begin{eqnarray}                                                                                      
2\bar{u} &=& 0.4 S - \Delta \label{eq:a5} \\                                                                                      
2\bar{d} &=& 0.4 S + \Delta \label{eq:a6} \\                                                                                      
2\bar{s} &=& 0.2 S   \label{eq:a7}                                                                                   
\end{eqnarray}                                                                                      
\noindent where the $2\bar{s}/(\bar{u}+\bar{d})$ ratio of 0.5 is chosen to                                                                                      
obtain a $\bar{s}$ density consistent with the CCFR data on dimuon production                                                                                       
\cite{CCFRMM}, see Fig.~\ref{fig:strange} of Section~7. The parametric                                                                                       
form of $\Delta$, which specifies the difference                                                                                       
between $\bar{u}$ and $\bar{d}$, is taken to be                                                                                      
\begin{equation}                                                                                      
x\Delta \equiv x(\bar{d} - \bar{u}) = A_{\Delta}x^{\eta_\Delta}(1-x)^{\eta_{S}+2}                                                                                      
(1+\gamma_{\Delta}x+\delta_{\Delta}x^{2}).   \label{eq:a8}                                                                                   
\end{equation}                                                                                      
\noindent The data require the integral $\int^{1}_{0} dx(\bar{d}-\bar{u})$,                                                                                       
which occurs in the Gottfried sum rule, to be positive. For the parton sets                                                                                        
obtained in this analysis the integral is approximately equal to 0.1 giving                                                                                       
a Gottfried sum                                                                                      
\begin{equation}                                                                                      
I_{\rm GS} \ \equiv \ \int_{0}^{1} \frac{dx}{x} \left[ F_2^p-F_2^n \right] \                                                                                       
\approx \ 0.27   \label{eq:a9}                                                                                   
\end{equation}                                                                                      
\noindent in the region $Q^2 \approx 5$~GeV$^2$. The relevant data are the                                                                                       
measurements of the asymmetry in Drell-Yan production in $pp$ and $pn$                                                                                       
collisions. The pioneering experiment of the NA51 collaboration \cite{NA51}                                                                                      
measured the asymmetry at one value of $x$, $x=0.18$. Very recently the E866                                                                                      
collaboration \cite{E866} have made measurements over a range of $x$. The new                                                                                      
data indicate that, while indeed $x\Delta>0$ for $x\lapproxeq 0.2$, for larger                                                                                      
$x$ values $\Delta \equiv \bar{d} - \bar{u}$ becomes small and may even become                                             
negative. The structure of $x\Delta$ is accommodated in our                                                                                       
parametrizations by negative values of $\delta_\Delta$. To ensure that the                                                                                       
individual densities $\bar{u}$ and $\bar{d}$ are both positive at all $x$, we                                                                                      
suppress the difference $\Delta$ at very large $x$ by an extra factor of 2 in                                                                                      
the exponent of $(1-x)$, see (8).                                                                                      
                                                                                      
For the first time our treatment of the heavy flavour densities, charm and                                                                                       
bottom, is on a firm theoretical footing. These densities are determined by the                                                                                      
other parton distributions and no extra parameters are introduced apart from the                                                                                      
heavy quark masses. At very low $Q^2$ the structure functions $F^H_2(x,Q^2)$,                                                                                      
with $H=c,b,$ are described by boson-gluon fusion and the heavy quark densities                                                                                      
turn on at $Q^2\simeq m^2_H$. The procedure for ensuring a smooth continuation                                                                                       
in the behaviour of $F^H_2$ in the threshold region is described in Section 5.                                                                                      
                                                                                      
A new feature of our analysis is the particular attention to the uncertainties                                                                                       
in the gluon                                                                                      
distribution at large $x$. The main constraints in this region are data on                                                                                      
prompt photon production in $pp$ or $pA$ collisions from the WA70 \cite{WA70}                                                                                      
and the E706 \cite{E706} experiments. The latter data confirm the 
implication from other high energy prompt photon experiments \cite{PPEXPT} that a
significant initial state partonic transverse momenta is needed to obtain                                                                                      
agreement with the NLO QCD prediction \cite{PTTUNG}. This naturally raises the                                                                                      
question whether such a transverse momentum component should be included when                                                                                       
determining the behaviour of the gluon at large $x (x\simeq 0.4)$ from the                                                                                      
lower energy ($\sqrt{s} = 23$~GeV) WA70 measurements --- even though the WA70                                             
data can be adequately described                                                                                       
without such a component\footnote{In previous MRS analyses of the WA70                                                                                       
data we did not include initial state transverse momentum.}. We find\footnote{The                                                                                      
calculation is described in Section 4 and compared with existing descriptions                                                                                      
of the E706 data.} that the E706 data, which correspond to $\sqrt{s} =                                                                                       
31.5$ and 38.8~GeV, require                                                                                       
the average value of the transverse momentum of the initial partonic system                                                                                       
$\left<k_T\right>\sim 1$~GeV, and we expect this to be less for experiments                                                                                       
at lower energies.                                                                                      
We therefore begin by taking a canonical value of $\left<k_T\right> = 0.4$~GeV                                                                                       
for the                                                                                       
analysis of the WA70 data at $\sqrt{s} = 23$~GeV. We then explore a range of                                                                                       
gluon distributions which result from global analyses in which $\left<k_T\right>$                                                                                        
goes from one extreme of $\left<k_T\right> =0$ to the other $\left<k_T\right> =                                                                                      
0.64$~GeV, which is the                                                                                       
maximum value that we find compatible with a reasonable description of the WA70                                                                                      
data, see Section 4. We call the gluon distributions which correspond to                                             
$\left<k_T\right>=$                                                                                      
0, 0.4 and 0.64~GeV the higher, central and lower (large $x$) gluons respectively                                                                                      
--- since a smaller gluon density is compensated by a larger $\left<k_T\right>$.                                                                                      
We denote the corresponding parton sets by MRST($\gup$), MRST and                                                                                       
MRST($\gdown$).
Of course the inclusion of intrinsic $\left<k_T\right>$ in this way can only be 
a crude approximation to the actual underlying physics, which presumably 
incorporates multiple emission of soft gluons as well as genuine non-perturbative
higher-twist effects. However, unlike for the Drell-Yan process, there is as yet no
complete theoretical treatment of this process, although a quantitative
estimate of the effects of soft gluon emission has recently been made in
Ref.~\cite{SMEAR}. However,
the three choices of the large $x$ gluon behaviour do give, we believe, a realistic                                                                                       
indication of the uncertainty of the gluon distribution due to these effects.                                                                                      
Table I lists the                                                                                       
values of the parameters of (1)-(4),(8) for the three parton sets.                                                                                      
                                                                                      
The optimum global MRST description has the QCD parameter                                                                                       
$\Lambda_{\rm\overline{MS}} (n_f=4) = 300$~MeV, which corresponds to                                                                                       
$\alpha_S(M_Z^2) = 0.1175$, in excellent agreement with the world average                                                                                       
value $\alpha_S(M_Z^2)=0.118$ \cite{ALPHA}. The same $\alpha_S$ value is also                                                                                       
used for the MRST($\gup$ and $\gdown$)                                                                                       
partons\footnote{The optimum value of $\alpha_S(M_Z^2)$                                                                                      
for the ($\gup$ and $\gdown$) parton sets is close to that of the                                                                                       
central gluon fit.}.                                                                                      
In the literature errors between $\pm 0.005$ and $\pm 0.003$ are quoted on the                                                                                       
world average value of $\alpha_S(M_Z^2)$. In Section~3.3 we will present                                                                                       
parton sets which cover the range given by the more conservative error of                                                                                      
$\pm 0.005$, in each case taking $\left<k_T\right> = 0.4$~GeV when analysing the                                                                                      
WA70 data.                                                                                      
                                                                                      
~From Table I we see that the values of $\lambda_g$ are negative, which                                                                                       
imply a `valence' type of behaviour for the gluon at low $x$ for                                                                                       
$Q^2=1$~GeV$^2$. Care should be taken not to attach physical significance to                                                                                       
this behaviour, as it arises from an \lq extrapolation\rq~outside the domain                                                                                       
of the                                                                                       
fitted data. Indeed as $Q^2$ increases the behaviour rapidly changes so that by                                                                                      
$Q^2=2$~GeV$^2$ the gluon distributions are approximately \lq flat\rq~in $x$.                                                                                      
Fig.~\ref{fig:gluon} shows the three gluon solutions  as $Q^2$  varies from 2 to                                                                                       
100~GeV$^2$ and we can see how evolution up in $Q^2$ soon blurs the                                                                                       
distinction between the initial starting distributions. Notice also that at low                                                                                      
$Q^2$ the distinct behaviour at $x \sim 0.4$ is compensated by the opposite                                                                                       
behaviour at $x \sim 0.05$ so that in each case the momentum fraction carried                                                                                       
by the                                                                                       
gluon remains at roughly $35\%$ at $Q^2_0=1$~GeV$^2$. This structure is more                                                                                       
evident in Fig.~\ref{fig:gluonratio} which shows the ratio of the gluons at                                                                                      
$Q^2=10$ and $10^4$~GeV$^2$. We see that all our three gluon distributions                                                                                      
converge for $x \lapproxeq 0.01$ due to the requirement of fitting the HERA                                                                                       
data. The CTEQ4M \cite{CTEQ4M} gluon distribution  is also shown\footnote{Very                                             
recently CTEQ \cite{CTEQG} have attempted to estimate the uncertainty on the                                             
gluon, without using any constraints from prompt photon data.} and the comparison                                             
will be discussed later.                                                                                      
                                                                                      
Table II lists the fraction of momentum carried by the individual partons as a                                                                                       
function of $Q^2$ for the central solution, MRST. For $Q^2=200$~GeV$^2$, for                                                                                      
example, $46\%, 31\%$ and $23\%$ is carried by the gluon, valence and sea quarks                                                                                      
respectively. In fact the flavour decomposition of the sea momentum fraction is                                                                                      
\begin{equation}                                                                                      
u,\ d,\ s,\ c,\ b \ =\ 7,\ 8,\ 5,\ 3,\ 1\%,   \label{eq:a10}                                                                                   
\end{equation}                                                                                      
\noindent demonstrating the growth of the strange, charm and bottom                                                                                       
distributions with increasing $Q^2$.                                                                                      
                                                                                      
The wide range of processes used in the global analyses is listed in Table III.                                                                                      
We include deep inelastic scattering (DIS) data from                                                                                       
H1\cite{H1}, ZEUS\cite{ZEUS}, BCDMS\cite{BCDMS}, NMC\cite{NMC},                                                                                      
E665\cite{E665},                                                                                      
SLAC\cite{SLAC} and CCFR\cite{CCFR2}. In general we only fit to DIS data with                                                                                      
$Q^2 > 2$~GeV$^2$ and $W^2 > 10$~GeV$^2$, but in order to include very small                                                                                      
$x$ measurements of $F^{ep}_2$ we admit the HERA data for $Q^2$ down to                                                                                       
1.5~GeV$^2$. Compared to our 1996 global analysis \cite{MRSR}, the HERA                                                                                      
data are updated to include the H1 shifted vertex as well as nominal vertex data, the                                             
NMC data now include all five beam energies and, finally, we use the reanalysed                                             
CCFR neutrino data. Table IV shows the $\chi^2$ values for all these DIS data for the                                             
three \lq gluon\rq~fits described above. The important constraints from                                                                                       
non-DIS data are discussed in detail in later sections.                                                                                      
Figures~\ref{fig:MRSTpartsa}  and \ref{fig:MRSTpartsb}
show the MRST parton distributions                                          
as a function of $x$ for $Q^2 = 20$~GeV$^2$ and $Q^2 = 10^4$~GeV$^2$
respectively, while Fig.~\ref{fig:MRSTvsR2} compares them with                                             
those of MRS(R2), our favoured set of the previous analysis \cite{MRSR}. We                                          
discuss the comparison of the new partons with previous sets in Section~10. \\                                            
                                                                                      
\noindent {\large \bf 3.  Description of DIS data}                                                                                      
                                                                                      
The description of the DIS structure function data by the MRST partons is shown in                                                                                      
Figs.~\ref{fig:DISsmallx}--\ref{fig:CCFRb}.  Overall the quality of the fit is                                         
satisfactory as                                                                                
reflected by the                                                                                      
$\chi^2$ values listed in Table IV.  However, it is informative to note special features                                                                                      
of the fit and, in particular, to highlight those areas in which the description of the DIS                                                                                      
data is systematically poorer than average.                                                                                      
                                                                                      
A comparison with the small $x$ data that are used in the fit is shown in                                             
Fig.~\ref{fig:DISsmallx}.                                                                                       
For the purposes of illustration, data at adjacent $x$ values are grouped together at a                                                                                      
mean value, together with the MRST fit.  Very recently the H1 collaboration                                             
\cite{F2H196} have made available unpublished preliminary measurements of $F_2$  from                                           
the 1995 and 1996 runs.  These data are not included in the fit but a comparison with                                   
MRST is                                             
shown in Fig.~\ref{fig:F2H196}.  It is clear that the data in Fig.~\ref{fig:DISsmallx}                                          
are  sufficient to                                                                                      
put strong constraints on the small $x$ behaviour of both the sea quark and gluon                                                                                      
distributions.  Loosely speaking $F_2$ and $\partial F_2/\partial \ln Q^2$ determine                                                                                      
the $x^{- \lambda}$ exponents, $\lambda_S$ and $\lambda_g$, of the 
sea and gluon                                                                                      
distributions respectively, as well as constraining the overall normalization.  The                                                                                      
values of $\lambda_S$ and, particularly, $\lambda_g$ depend sensitively on the value                                                                                      
of $Q^2 = Q_0^2$ chosen to parametrize the input distributions.                                            
                                                                                      
We may look at how the exponents $\lambda_S$ and $\lambda_g$ vary with $Q^2$                                                                                      
by fitting the sea quark and gluon distributions of the MRST partons to the forms                                                                                      
\begin{equation}  
x f_i (x, Q^2) \; = \; A (Q^2) x^{- \lambda_i (Q^2)} \label{eq:a11}                                                                                   
\end{equation}                                                                                      
as $x \rightarrow 0$.  The results are shown in Fig.~\ref{fig:lambda}.  As $Q^2$                                                                                      
increases from the input scale, $Q_0^2 = 1$~GeV$^2$, we see that the valence-like                                                                                      
character of the gluon rapidly disappears due to evolution being driven by the much                                             
steeper sea, and that by $Q^2 \simeq 2$~GeV$^2$ the                                                                                      
gluon is \lq flat\rq~in $x$, that is $\lambda_g = 0$.  By $Q^2 \simeq 6$~GeV$^2$ we                                                                                      
see that $\lambda_g = \lambda_S$, which incidentally is close to the                                             
assumption\footnote{Approaches in which $\lambda_S$ is tied to $\lambda_g$ at                                             
lower scales (say $Q_0^2 < 2$~GeV$^2$), as in the dynamical GRV model                                             
\cite{GRV}, will clearly have difficulty in fitting the new HERA data.} made in                                                                                      
the early NLO global analyses in which the input scale was chosen to be $Q_0^2 =                                                                                      
4$~GeV$^2$.  For higher values of $Q^2$ the gluon exponent \lq leads\rq~that of the                                                                                      
sea, $\lambda_g > \lambda_S$, since the gluon drives the sea quark distribution via                                                                                      
the $g \rightarrow q\overline{q}$ transition.                                                                                      
                                                                                      
Fig.~\ref{fig:slope} is an alternative way of looking at the quality of the                                                                                      
description of $\partial F_2/\partial \ln Q^2$ at low $x$.  The continuous curves on     
the plots show the values of                                                                                      
the slope $\partial F_2/\partial \ln Q^2$ versus $x$ for both the H1 and ZEUS data,                                                                                      
compared with the slope found in the MRST fit (evaluated at the particular values of                                                                                      
$Q^2$ appropriate to the experimental data).  Though the overall description is                                                                                      
satisfactory, it is possible that for $x \lapproxeq 10^{-3}$ there may be a systematic                                                                                      
difference between the data and the fit which reflects the onset of $\ln 1/x$                                                                                      
contributions which are outside the scope of our NLO DGLAP analysis.  This                                             
systematic trend is even more evident in the preliminary H1 data from the 1995/96 runs  
\cite{F2H196}.  As the                                                                                      
precision of the HERA measurements of $F_2$ improves, it will be interesting to see                                                                                      
whether or not the statistical significance of the discrepancy increases.                                                                                      
                                                                                      
The description of the NMC DIS data \cite{NMC} for $F_2^p$ and $F_2^d$ is                                                                                      
shown in Fig.~\ref{fig:NMCa} and Fig.~\ref{fig:NMCb}.  It is apparent                                                                                      
that the data have systematically a larger slope, $\partial                                                                                      
F_2/\partial \ln Q^2$, than the fit.  This is well illustrated by the continuous curve in     
Fig.~\ref{fig:NMCslope}                                             
which shows $\partial F_2/\partial \ln Q^2$ for the NMC $F_2^p$ data.  The                                             
discrepancy indicates that the NMC data would prefer a larger gluon in this $x$ region                                             
and/or a larger $\alpha_S$ value than that of MRST.  In turn the larger gluon would                                                                                      
imply (by momentum conservation) a smaller gluon in the very small $x$ domain,                                             
contrary to the HERA data.  The fit is a compromise between these data sets, but it                                                                                      
demonstrates the tight constraints now imposed by the increased precision of the                                                                                      
$F_2$ data.  A particular virtue of the NMC data is the accurate measurement of                                                                                      
$F_2^n/F_2^p$.  We postpone a study of their implications                                                                                      
until we discuss the description of the asymmetry in Drell-Yan production in $pp$ and                                                                                      
$pn$ collisions and the rapidity asymmetry in the processes $p\bar{p}                                                                                      
\rightarrow W^\pm X$.                                                                                      
                                                                                      
The BCDMS data \cite{BCDMS} cover the range $0.1 \lapproxeq x \lapproxeq 0.75$     
and are the most precise data at large $x$, see Fig.~\ref{fig:BCDMS}.  From the  
figure it is apparent that the data would prefer a smaller value of    
$\alpha_S$                                             
than that found in the global fit, namely $\alpha_S (M_Z^2) = 0.1175$.  Indeed if the                                   
BCDMS data                                             
are analysed on their own (apart from including SLAC data \cite{SLAC} to constrain                                             
the higher twist contribution) then $\alpha_S (M_Z^2)$ is found to be $0.113 \pm                                             
0.005$ \cite{ALP}.  Our optimum value of $\alpha_S$ is therefore within a standard                                             
deviation of the BCDMS determination.                                            
                                                                                      
The re-analysed CCFR neutrino measurements of $F_2^{\nu N}$ and $xF_3^{\nu                                                                                      
N}$ are compared with the MRST values in Figs.~\ref{fig:CCFRa} and                                         
\ref{fig:CCFRb}.                                              
The long-standing discrepancy between the CCFR $F_2^{\nu N}$                                                                                      
and the NMC $F_2^{\mu d}$ measurements for $x \lapproxeq 0.1$ remains:  in this                                                                                      
$x$ region the $F_2^{\nu N}$ measurements are in excess of the $F_2^{\mu d}$ data                                                                                      
by a significantly larger amount than that implied by the strange quark                                                                                      
distribution\footnote{Recall, at LO, that $xs (x) = \frac{5}{6} F_2^{\nu N} (x) - 3                                                                                      
F_2^{\mu d} (x)$.} determined from dimuon data, see Fig.~\ref{fig:strange} of  
Section 7.1. However, it is important to note that the neutrino data must be corrected     
for heavy                                                                                      
target effects.  In Figs.~\ref{fig:CCFRa} and \ref{fig:CCFRb} we have subjected the                                                                                      
MRST curves to a heavy target correction.  The parametric form that we use for the                                                                                      
heavy target correction factor $R_{\rm HT}$ is deduced from a $Q^2$-independent fit
to the EMC effect for the scattering of muons on a heavy nuclear target ($A=56$).
Explicitly,
\begin{equation}  
R_{\rm HT} = \left\{ \begin{array}{ll}
1.238 + 0.203 \log_{10}x & \qquad\mbox{for $x < 0.0903$} \\
1.026                    & \qquad\mbox{for $0.0903 < x < 0.234$} \\
0.783 - 0.385 \log_{10}x & \qquad\mbox{for $0.234 < x .$}
\end{array}\right. 
\label{eq:heavycorr}                                                                                   
\end{equation}                                                                                      
Note that the correction factor that we obtain in this way is more severe at low                                                                                      
$x$ than that implied by shadowing.  This is one reason why the (dashed) curves are                                                                                      
considerably below the $x < 0.1$ neutrino data.  It is not clear whether the correction                                                                                      
factor should be the same for neutral current and charged current DIS data, or be the                                                                                      
same for $F_2$ and $xF_3$ neutrino data.  For these reasons we do not include the                                                                                      
CCFR heavy target data for $x < 0.1$ in the fit.  The remaining neutrino data are well                                                                                      
described.  However, it is possible to see that the fit slightly underestimates the slopes                                                                                      
as a function of $\ln Q^2$, which reflects the value $\alpha_S (M_Z^2) = 0.119$                                                                                      
obtained from fitting to the neutrino data alone \cite{CCFR2}.                                                                                   
                                  
\medskip                                  
\noindent{\bf 3.1~$F_L$ and implications for partons}                     
    
In the DIS experiments it is not the structure function     
$F_2(x,Q^2)$ which is measured directly but the differential cross section.    
Defining the rescaled differential cross section    
\begin{equation}     
\tilde \sigma(x,Q^2)=\frac{Q^4x}{2\pi\alpha^2}\frac{1}{[1 + (1 - y)^2]}    
\frac{d^2\sigma}{dxdQ^2}, \label{eq:fl1}    
\end{equation}     
where $y=Q^2/xs$, we have    
\begin{equation}    
\tilde \sigma(x,Q^2)= F_2(x,Q^2)-{y^2\over [1 + (1-y)^2]}    
F_L(x,Q^2). \label{eq:fl2}    
\end{equation}    
Since both $y$ and $F_L$ are usually small the latter term in this expression     
is usually negligible, and the measurement of $F_2$ is effectively direct.    
       
However, the analysis of data on the longitudinal structure function    
$F_L(x,Q^2)$ is in principle an important probe of the parton     
distributions.    
This is particularly the case for the gluon at small $x$ since     
in this region, to a good    
approximation, we have the relationship 
\begin{equation} 
xg(x,Q^2)=5.9(3\alpha_S/4\pi)  F_L(0.4x, Q^2) 
\label{eq:fl3} 
\end{equation} 
for three massless flavours \cite{flrel} at leading order    
in $\alpha_S$    
(though the next-to-leading correction leads to an increased gluon relative    
to a fixed $F_L$).    
Nevertheless, until recently there     
have been little data on $F_L$, and these have been at high $x$    
(where there are likely to be important higher twist effects) and have     
very large errors. The situation is now beginning to change.      
The HERA experiments measure the differential cross section at $y>0.5$, and     
are thus sensitive to the component due to  $F_L$. Also, the NMCollaboration have  
direct measurements of $F_L$ for $0.1\gapproxeq x    
\gapproxeq 0.01$, obtained     
by data runs for different beam energies, and hence different values of $y$.     
    
The only consistent way in which to analyse the HERA data at large $y$ is to    
calculate both the NLO expressions for $F_2$ and $F_L$     
(i.e. using the ${\cal O}(\alpha_S^2)$ coefficient functions for $F_L$    
\cite{nlofl,NLOCF},     
where those for heavy quarks use the prescription of \cite{RT})     
and to compare (\ref{eq:fl2}) with the measured $\tilde \sigma$. Of course, this is     
equivalent to the correction of the extracted values of $F_2(x,Q^2)$ to take    
account of the predicted values of $F_L(x,Q^2)$, and it is this latter     
procedure that we employ when fitting to the HERA data. (This results in     
corrections of at most $2-3\%$ to the values of $F_2(x,Q^2 )$ quoted in     
\cite{H1,ZEUS}.) Since in the published data the value of $y$ is nearly    
always $<0.6$, there is relatively little sensitivity to the value of     
$F_L$. However, the H1 collaboration have also published a number of    
measurements of $\tilde \sigma$ for $y=0.7$ \cite{h1fl}, and in the     
preliminary 1995/96 data have reproduced these measurements, and also produced     
measurements at $y=0.82$ \cite{F2H196}.     
In Fig.~\ref{fig:sigFL} we show a comparison of     
the prediction for $\tilde \sigma(x,Q^2)$ obtained from the MRST partons     
with these preliminary H1 data. It is clear that the $y=0.82$     
points (the first data point in each plot) lie below the theoretical curves in general  
(which implies that the predicted $F_L$  is too small). However, it is also     
clear that the points for $0.01\gapproxeq x \gapproxeq 0.001$     
tend to lie above the curve. Indeed,    
the curves in Fig.~\ref{fig:sigFL} resemble the curves in     
Fig.~\ref{fig:slope}: there is a tendency for $\partial F_2(x,Q^2)/    
\partial \ln Q^2$ to be too small at $x \sim 0.005$ and hence      
$F_2(x,Q^2)$ tends to be too small at high $Q^2$,     
and at $x\sim 0.0005$ there is           
a tendency for $\partial F_2(x,Q^2)/\partial \ln Q^2$ to be too large and hence     
$F_2(x,Q^2)$ tends to be too large at high $Q^2$. Thus, also     
bearing in mind the large errors on the points at $y=0.82$, we feel that it is     
premature to claim any inconsistency in the NLO prediction for     
$F_L(x,Q^2)$.\footnote{We note that in the NLO fit performed by H1 there is    
no direct constraint on the gluon at large $x$. From the momentum sum rule    
this leads to more flexibility for the gluon at small $x$, and hence the     
details of their best fit are rather different from ours in this region.}    
Alternative theoretical treatments to ours, in particular the inclusion    
of leading $\ln(1/x)$ terms, lead to different predictions for $F_L$ for    
similar fits to $F_2$, and measurements at high $y$ are therefore an     
important test of such approaches.\footnote{For a discussion of such tests,    
and in particular a demonstration that the high $y$ data gives strong evidence    
against the validity of a LO-in-$\alpha_S$ fit, see \cite{RTFL}.}      
However, direct measurements of $F_L$     
would provide an even better test. We exhibit the predictions for $F_L$ in the     
HERA kinematic range    
obtained using the MRST partons in Fig.~\ref{fig:predFL}.     
    
In Fig.~\ref{fig:NMCFL} we compare our predictions for $F_L(x,Q^2)$, made    
using the partons resulting from each of the three parton sets, MRST,    
MRST$(\gup)$ and MRST$(\gdown)$, with the direct measurements     
made by NMC\cite{NMC}. There is little variation     
in the predictions, and each provides a perfectly satisfactory     
description of the data. Hence, the NLO calculation of structure functions     
seems to be compatible with both direct and indirect data on $F_L$.    
                                                                                   
\medskip                                                                                      
\noindent {\bf 3.2~Sensitivity to cuts on the data fitted}                                                                                   
                                            
Given that there are potentially important higher twist and $\ln (1/x)$ contributions at                                             
small $Q^2$ and/or small $x$ it is instructive to explore the sensitivity of our fits to                                             
the minimum $Q^2$ and/or $x$ cuts.  Recall that our minimum $Q^2$ cut is at                                             
$Q_1^2 = 2$~GeV$^2$, except for the HERA data where the cut is at 1.5~GeV$^2$.                                              
We have imposed no minimum cut on $x$.  We have made repeated global analyses                                             
with different minimum $Q^2$ cuts up to a value $Q_1^2 = 10$~GeV$^2$.  Note that                                             
increasing $Q_1^2$ has the effect of removing much of the lowest $x$ data from the                                   
fit.  The dashed curve in Fig.~\ref{fig:slope} shows the effect on $\partial F_2/\partial                                   
\ln Q^2$ when the cut is increased to $Q_1^2 = 10$~GeV$^2$.  Not surprisingly the                                   
description of the slope determined  from                                            
the HERA data with $Q^2 \gapproxeq 10$~GeV$^2$ is much improved.  We                                             
should note that with $Q_1^2 = 10$~GeV$^2$ a significant fraction of the NMC data                                             
is also excluded, see Fig.~\ref{fig:NMCa}.  This removes a strong constraint on the                                   
behaviour of the gluon at intermediate $x$ thus allowing the gluon to increase at small                                   
$x$, which improves the fit to the HERA data.  As we can see from                                   
Fig.~\ref{fig:gluoncut} the gluon                                          
obtained from                                             
the $Q_1^2 = 10$~GeV$^2$ fit is larger at small $x$ than the standard MRST                                             
gluon.  As expected the difference between the gluons decreases rapidly with     
increasing $Q^2$.  The effect of an intermediate choice of $Q_1^2$ can be     
anticipated by interpolating the $Q_1^2 = 2$ and $Q_1^2 = 10$~GeV$^2$ results.      
    
Fig.~\ref{fig:partoncut} shows the sensitivity of the parton 
distributions at $Q^2 =     
10$~GeV$^2$ to the minimum $Q^2$ cut on the structure function data that are     
included in the fit.  Recall that only data with $Q^2 > Q_1^2$ are included.      
Fig~\ref{fig:partoncut} compares the results of analyses with cuts at $Q_1^2 = 5$ and     
$Q_1^2 = 10$~GeV$^2$ to our default MRST set obtained by taking the cut at     
$Q_1^2 = 2$~GeV$^2$.  The plot is interesting because the variation of the values of     
the partons with the $Q_1^2$ cut reflects the interplay of the constraints imposed by     
the various data sets.  For instance the $u$ and $d$ quarks for $0.01 \lapproxeq x     
\lapproxeq 0.5$ are stable to a choice of $Q_1^2$ in the range 2--10~GeV$^2$, since     
precise data remain in this domain even for the highest $Q_1^2$ cut.  For the higher     
$Q_1^2$ values the small enhancement (up to at most 2\%) in the $u$ distribution in     
the region $0.01 \lapproxeq x \lapproxeq 0.1$ can be understood by looking at the     
description of the highest $Q^2$ NMC $F_2^p$ data points in this $x$ range, see     
Fig.~\ref{fig:NMCa}.  The variation of the gluon with $Q_1^2$ has been discussed     
above, and is responsible for the similar variation of the charm distribution.  The     
decrease of the light quarks at small $x$ ($x <$~few~$\times 10^{-3}$) with     
increasing $Q_1^2$ is partly due to compensation for the increase of the charm  
contribution to $F_2^p$ and partly due to the decrease of $F_2^p$ itself induced by  
the larger slope $\partial F_2/\partial \ln Q^2$ required by the higher $Q^2$ data.    
                                            
We have also repeated the global analysis with various $x$ cuts on the data up to                                             
$x_{\rm min} = 0.01$.  This removes more than half of the HERA data, but leaves the                                             
NMC and other fixed target data virtually untouched.  The removal of the constraint                                   
on the gluon at very                                             
small $x$ allows a larger gluon in the region $x \sim 0.1$, leading to an improvement                                             
in the description of the NMC data illustrated in Fig.~\ref{fig:NMCslope}.                                    
Nevertheless, as can be seen, the improvement is not as great as might be expected.                                             
                                                                                   
\medskip                                                                                      
\noindent {\bf 3.3~Sensitivity to $\alpha_S$}                                                                                      
                                                                                      
We have mentioned that $\alpha_S(M_Z^2)=0.1175$ yields the optimum $\chi^2$ of                                                                                      
the global fit to the combined data sets. Next we explore the sensitivity to the                                                                                      
variation of the value of $\alpha_S$.  To do this we perform global analyses with                                                                                      
fixed values of $\alpha_S$ in the range $\pm 0.005$ of our optimum value.  In each                                                                                      
case we use $\langle k_T \rangle = 0.4$~GeV when analysing the WA70 data, which                                                                                      
corresponds to the central gluon distribution of the three \lq gluon\rq~fits described in                                                                                      
the previous section.  The contributions to the total $\chi^2$ coming from the various                                                                                      
data sets are plotted as a function of $\alpha_S$ in Fig.~\ref{fig:alpha}.  We                                                                                      
emphasize that this is not the optimum $\chi^2$ for a particular data set fitted on its                                                                                      
own, but is the contribution to $\chi^2$ for the global fit which, of necessity, has to                                                                                      
make compromises between the descriptions of the various data sets.  As expected from                                                                                      
our previous discussion, we see the opposite trend for the $\chi^2$ of the                                                                                      
BCDMS data (which favour a smaller $\alpha_S$) and the CCFR data (which favour                                                                                      
a larger $\alpha_S$).  Similarly the NMC data favour a larger $\alpha_S$ to                                                                                      
compensate for the global fit yielding a smaller gluon than that which would be                                                                                      
obtained by fitting to the data set on its own.  It is noticeable that the recent, more                                                                                      
precise, HERA data give a contribution to $\chi^2$ which is less sensitive to variation                                                                                      
in $\alpha_S$ than the earlier HERA measurements (see ref.~\cite{oldalpha}).  In                                                                                      
summary we see that the overall minimum value $\alpha_S (M_Z^2) = 0.1175$ in the                                                                                      
{\it global fit} is a pinch between the BCDMS and HERA data favouring smaller                                                                                      
$\alpha_S$ values and the NMC, SLAC and CCFR data favouring larger values.                                                                                      
                                                                                      
An independent sensitive measure of the value of $\alpha_S$ is the single jet inclusive                                                                                      
$E_T$ distribution measured in the Fermilab $p\bar{p}$ 
experiments \cite{CDFJ,D0J}.  
We shall                                                                                      
see in Section~9 that these data favour values of $\alpha_S (M_Z^2)$ in the region                                                                                      
0.115--0.118.  There thus seems to be general agreement that the value of $\alpha_S                                                                                      
(M_Z^2)$ is in the region of our optimal value $0.1175$ with a spread of about $\pm                                                                                      
0.0025$.  It is useful to have a range of parton sets available for different values of                                                                                      
$\alpha_S$, so we present four additional sets which cover a conservative range of                                                                                      
$\pm 0.005$ about our optimal value.  We denote these by
 MRST($\asdown$), MRST(${\alpha_S \downarrow}$), MRST(${\alpha_S\uparrow}$)
and MRST($\asup$) corresponding to $\alpha_S (M_Z^2) =                                             
0.1125$, 0.1150, 0.1200 and 0.1225 respectively.    
    
Given the recent interest in the very high $Q^2$ region at HERA it is important to     
quantify the uncertainty in the extrapolation of $F_2^p$ to high $Q^2$ at a large     
value of $x$.  In this $x$ region the main effect comes from the uncertainty in the     
value of $\alpha_S$ in the DGLAP evolution of $F_2^p$.  The effect is illustrated at     
$x = 0.45$ in Fig.~\ref{fig:x45plot}, which shows the spread in the extrapolated values     
of $F_2^p$ arising from $\alpha_S$ varying across the interval $0.1175 \pm 0.005$. \\                                                                                      
                                                                                      
\noindent {\large \bf 4.  Prompt photon production and the gluon at large $x$}                                                                                      
                                                                                      
In previous MRS parton analyses the WA70 data for $pp \rightarrow \gamma X$                                                                                      
\cite{WA70} have been a key constraint on the gluon distribution for $x \sim                                             
0.3$--0.5.  In these analyses we have not included any initial state partonic transverse                                                                                      
momenta, that is we have taken $\langle k_T \rangle = 0$, in fitting to the prompt                                                                                      
photon data.  If we continue to fit the WA70 photon $p_T$ spectrum in this way then                                                                                      
the global analysis yields the set of partons that we have called MRST($\gup$)                                                                                      
in Table I.  We use the $\overline{\rm MS}$ renormalization and factorization                                             
prescriptions, with a common scale $Q = p_T/2$.  We perform a full NLO calculation                                            
including the effects of fragmentation \cite{GV}\footnote{We thank Werner                                            
Vogelsang for performing the relevant calculations.}.  A good description of                                                                                      
the WA70 data is achieved.                                                                                      
                                                                                      
However, there is now compelling evidence for the need to include non-zero                                                                                      
transverse momentum $k_T$ of the incoming partons (arising from parton multigluon                                                                                      
emission\footnote{An estimate of the amount of smearing expected from the                                            
perturbative component based on the resummation of leading logs has been made in  
Ref.~\cite{SMEAR}.} and from non-perturbative \lq intrinsic\rq~partonic transverse                                            
momentum).  The  reason is apparent if we consider all the data for prompt photon                                                                                      
production in high energy $pp$ and $p\bar{p}$ collisions simultaneously.                                                                                       
Collectively these data span the entire interval $0.1 \lapproxeq x \lapproxeq 0.5$.  A                                                                                      
major experimental challenge in these experiments is to cleanly extract the prompt                                                                                      
photon signal from the copious background of $\pi^0$ and $\eta$ decays.  There is a                                                                                      
pattern of deviation between theory and experiment in the shape of the photon $p_T$                                                                                      
spectrum.  The data fall off more steeply with increasing $p_T$ than the NLO QCD                                                                                      
predictions.  Neither changes of scale nor the introduction of fragmentation effects can                                                                                      
resolve the discrepancy\footnote{Vogelsang and Vogt \cite{VV} have demonstrated                                                                                      
that these effects can improve the description of a single prompt photon experiment.                                                                                       
However, experiments at different $\sqrt{s}$ reproduce a similar pattern, but in                                                                                      
different $x$ intervals.} since the various experiments probe different ranges of $x                                                                                      
\simeq x_T \equiv 2 p_T/\sqrt{s}$.  On the other hand it has been shown                                                                                      
\cite{CTEQPP} that the discrepancy can be removed by a broadening of the initial                                                                                      
state parton $k_T$ which increases with the energy $\sqrt{s}$.                                                                                      
As already mentioned in Section~2, there is no complete theoretical treatment
currently available which would allow a parameter free description 
of the broadening of the $p_T$ distribution. There is evidence 
from $\pi^0\pi^0$ and $\gamma\gamma$ production that an approximately
Gaussian smearing form reproduces the observed broadening
(see, for example, Ref.~\cite{E706}) and that the width of the Gaussian increases
with energy.
                                                                                      
The parton $k_T$ effect is found to be least in fitting the data due to WA70 --- the                                                                                      
lowest energy prompt photon experiment.  Moreover these data do not exhibit the                                                                                      
$p_T$ shape discrepancy with QCD that is seen in the other experiments                                   
\cite{PPEXPT,E706}.  For these                                                                                      
reasons we have set $\langle k_T \rangle = 0$ in our previous analyses.  However, if     
we                                                                                      
include this type of analysis in our new global fit and use the resulting partons,                                                                                      
MRST($\gup$), to predict the high precision E706 prompt photon $p_T$ spectra                                                                                      
the description is disastrous.  To reconcile our prediction with the E706 data we may                                                                                      
fold in a Gaussian $k_T$ spectrum\footnote{This is in qualitative agreement with the                                                                                      
findings of the E706 collaboration \cite{JH}, although the detailed prescriptions for                                                                                      
the $k_T$ smearing are different.  We smear the perturbative QCD distribution by                                                                                      
first making an analytic continuation for $p_T < p_0 = 3$~GeV
of $(d\sigma/dp_T^2)_{\rm QCD}$ of the form $\exp(\sum_{i=0}^{4} a_i p_T^i)$ 
to regulate the infra-red singularity at $p_T = 0$, 
and then we convolute with a Gaussian form 
$(1/\pi \sigma) \exp (-k_T^2/\sigma)$ 
where $\sigma=(4/\pi) \langle k_T \rangle^2$. The results are not sensitive to
the particular choice of $p_0$.} 
with $\langle k_T \rangle \simeq 1$~GeV, or to be                                                                                      
precise $\langle k_T \rangle = 0.87 (0.97)$~GeV for data taken at laboratory     
momentum $p_{\rm                                            
lab} = 530 (800)$~GeV.  As a consequence it can be argued that                                                                                      
our $\langle k_T \rangle = 0$ analysis of the WA70 data is inconsistent.  We should                                                                                      
include $\langle k_T \rangle \neq 0$ in the description of these lower energy data but                                                                                      
with a smaller value\footnote{The empirical evidence that $\langle k_T \rangle$                                                                                      
increases with $\sqrt{s}$ is supported by a similar effect in Drell-Yan production                                                                                      
\cite{DYKT}.} of $\langle k_T \rangle$ than that needed to describe the E706 data.                                                                                      
                                                                                      
We therefore repeat the global analysis but fit to the WA70 data using a Gaussian                                                                                      
partonic $k_T$ spectrum with $\langle k_T \rangle = 0.4$~GeV, corresponding to                                                                                      
280~MeV per incoming parton.  The description of the WA70 data is shown by the                                            
continuous curve in Fig.~\ref{fig:WA70}.  For comparison the dashed curve shows                                          
the                                            
unsmeared prediction, which of necessity undershoots the data.  The resulting set of                                            
partons are labelled simply                                                                                      
MRST in Table I, and give an equally good fit to those of MRST($\gup$) with                                                                                      
$\langle k_T \rangle = 0$.  How well are the E706 photon data described by the                                                                                      
MRST partons, and in particular by the new gluon which is 
smaller at $x \sim 0.4$?       From the continuous curves in 
Figs.~\ref{fig:E706a} and \ref{fig:E706b} we see an                                            
excellent description of the $p_{\rm lab} = 530$~(800)~GeV E706 data                                                                                      
is obtained if we take $\langle k_T \rangle = 0.92$~(1.01)~GeV.                                           
                                                                                      
We can regard the MRST($\gup$) partons with $\langle k_T \rangle = 0$ as one                                                                                      
extremum.  Conversely how large can we take $\langle k_T \rangle$ to be to describe                                                                                      
the WA70 data and still retain a satisfactory fit?  We find that we cannot choose                                                                                      
$\langle k_T \rangle$ to be arbitrarily large because not only does the gluon become                                                                                      
smaller, but it becomes steeper and eventually the shape of the $p_T$ spectrum is not                                                                                      
reproduced.  The maximum value of $\langle k_T \rangle$ for which a reasonable fit                                                                                      
to the WA70 data can still be found is 0.64~GeV. The parton                                                                                      
set corresponding to this upper extremum for $\langle k_T \rangle$ is labelled                                                                                      
MRST($\gdown$), to indicate that it has the smallest gluon at $x \sim 0.4$.  The                                                                                      
E706 data are well described by the MRST($\gdown$) partons provided we take                                                                                      
$\langle k_T \rangle = 0.97$~(1.04)~GeV at $p_{\rm lab} = 530$~(800)~GeV.                                                                                      
                                                                                      
So far we have considered the variation of the partons, and in particular of the gluon,                                                                                      
due to the uncertainties in the $\langle k_T \rangle$ smearing.  Our preferred set of                                                                                      
partons with $\langle k_T \rangle = 0.4$~GeV is MRST.  The \lq extremum\rq~parton                                                                                      
sets with $\langle k_T \rangle = 0$ and 0.64~GeV are labelled MRST($\gup$)                                                                                      
and MRST($\gdown$) in Table I, with gluons which are respectively larger and                                                                                      
smaller than the MRST gluon at $x \sim 0.4$.                                           
                                           
There is also a non-negligible dependence on the choice of scale.  For instance the                                            
effect of changing the scale from $Q = p_T/2$ to $Q = p_T$ is shown in                                            
Fig.~\ref{fig:E706a}.  We see that the unsmeared cross section is decreased by some                                            
30\%, which can be compensated by a relatively modest increase in the size of the                                   
gluon distribution and/or in                                            
$\langle k_T \rangle$.  That is the effect of the change of scale is considerably less                                            
than the uncertainties associated with smearing.  In principle the former can be                                            
reduced by a knowledge of the NNLO perturbative contributions, while a reduction in                                            
the latter will require a more detailed theoretical understanding of the origin of the                                   
partonic transverse momentum. \\

\noindent{\large \bf 5.  Treatment of Heavy Flavours}                                                                        
                                                                        
Until recently the treatment of heavy quark distributions in MRS and CTEQ global                               
analyses has been rather naive.  In previous analyses the charm                                                                         
and bottom quarks were regarded as infinitely massive below a threshold                               
$Q^2=m_H^2$, and then                                                                         
being treated as massless, and thus evolving according to the normal                                                                        
massless evolution equations above this threshold. Up to NLO in $\alpha_S$                                                                        
this prescription guarantees that the correct results will be obtained                                                                         
asymptotically, but is clearly rather unsatisfactory near threshold                                                                         
where there should be a smooth threshold at $W^2=Q^2(1/x -1)=4m_H^2$,                                                                        
where $W^2$ is the invariant mass of the hadronic system, rather than an                                                                         
abrupt threshold in $Q^2=m_H^2$. Nevertheless, choosing the slightly high                                                                        
value of $m_c^2=2.7$~GeV$^2$, a reasonable match to the EMC data                                                                        
\cite{EMC} on the charm                                                                        
structure function for $Q^2>4$~GeV$^2$ was obtained\footnote{In the                                                                         
MRS global analysis of \cite{MRSR} the charm evolved from the low value of     
$Q^2=1~\GeV^2$                                                                        
but was suppressed by a phenomenological damping factor.}, and the low contribution                                                                         
due to charm for the total structure function rendered a more complete                                                                         
treatment of heavy quark contributions unnecessary in this $x$ and $Q^2$ range.                              
                                                                        
An alternative procedure to that outlined above is where all charm is                                                                        
regarded as being produced from the hard scatter between the electroweak                                                                        
boson and a light parton, i.e. the number of active flavours, $n_f$,                                                                         
is 3 and the charm cross section is generated (mainly) by photon-gluon fusion (PGF).                                                                         
This corresponds to the                                                                         
so-called fixed flavour number scheme                                                                         
(FFNS) and it incorporates the correct threshold behaviour automatically.                                                                        
For example, at order ${\cal O}(\alpha_S)$ the charm structure function is given by                                                                        
\begin{equation}                                                                        
F_2^c(x,Q^2,m_c^2) \; = \;\frac{\alpha_S(\mu^2)}{2\pi}                                                                         
\;C_g^{(1)\;{\rm FF}}(Q^2/m_c^2) \:\otimes\:g_{n_f=3}(\mu^2),                                                                        
\label{eq:lof2cffns}                                                                        
\end{equation}                                                                        
where the coefficient function (CF), which is convoluted with LO evolved gluon                               
density                                                                         
$g_{n_f=3}$, is                                                                        
\bea                                                                        
 C^{(1)\;{\rm FF}}_g (z,Q^2/m_c^2) & = &                                                                        
\biggl [ \;(P^0_{qg}(z)                                                                        
+ 4\frac{m_c^2}{Q^2} z (1-3z) -8\biggl(\frac{m_c^2}{Q^2}\biggr)^2 z^2)\;     
\ln  \biggl ( \frac{1+v}{1-v}                                                                        
\biggr ) \nonumber \\                                                                        
& + &  (8z(1-z)-1-4\frac{m_c^2}{Q^2} z(1-z))v \biggr ] \;    
\theta(\hat W^2-4m_c^2)                                                                         
\label{eq:lochgffns}                                                                        
\eea                                                                        
where $\hat W^2=Q^2(1/z-1)$                                                                         
is the gluon quark centre of mass energy,                                                                        
$v$ is the velocity of the charm quark                                                                         
or antiquark in the                                                                         
photon--gluon centre--of--mass frame, defined by $v^2=1-4m_c^2/\hat W^2$,                                                                        
and $P^0_{qg}(z) = z^2 + (1-z)^2$, the LO quark-gluon splitting function.                                                                         
 These $v$--dependent terms ensure                                                                         
that the coefficient function tends to zero                                                                        
smoothly as $\hat W^2=4m_c^2$ is approached from below,                                                                        
and hence the structure function has a smooth threshold in                                                                         
$W^2$.                                                                        
This method does not sum potentially large logarithms in $Q^2/m_c^2$,                                                                         
and thus is unsuitable for $Q^2 \gg m_c^2$, but                                                                        
provides an acceptable description provided $Q^2$ is not                                                                        
large and one is not interested in the concept of a charm quark density.                                                                        
It is the method used to produce the most recent GRV structure functions                                                                        
\cite{GRV}, and is also used in the analyses by H1 and ZEUS.                                                                         
                                                                        
However, the more recent measurements of charm production at HERA                                                                         
\cite{H1C,ZEUSC} emphasise the importance of having a consistent theoretical                               
framework for heavy flavour production in deep inelastic scattering. Not only are                               
there more direct measurements of the charm structure function $F_2^c$, but     
the charm contribution could be                                                                         
20\% or more of the total $F_2$ at small $x$. Indeed, even the NMC data                                                                        
which contains only $5-10\%$ charm, but has rather smaller errors than the                                                                         
HERA data, is sensitive to the treatment of charm.  Hence, a modern global analysis                               
of structure functions must necessarily include a satisfactory description of $F_2^c$.                                                                      
                              
As a consequence there have been several recent theoretical studies                               
\cite{ACOT,MRRS,RT} to improve the treatment of heavy quark mass effects in deep                               
inelastic scattering.  For instance ref.~\cite{MRRS} proposes a simple procedure to                               
sum up the leading (and next-to-leading) log contributions of Feynman diagrams                               
including explicitly the $m_H \neq 0$ mass effects.  It is straightforward to generalize                               
this procedure to any order.  In this approach the natural scale to resolve charm quarks                               
in the proton is $Q^2 = (m_c^2 + k_T^2)/(z (1 - z)) \gapproxeq 4 m_c^2$, whereas                               
the conventional $\overline{\rm MS}$ scheme, which we adopt, requires the charm                               
threshold to be at $Q^2 = m_c^2$.  ($k_T$ and $z$ define the momentum of the                               
charm quark).  For this reason the procedure is difficult to reconcile with the                               
$\overline{\rm MS}$ scheme.                               
    
\medskip                                                                      
\noindent{\bf 5.1.  Theoretical procedure}                                                                        
                                                                        
In order to have a reliable treatment of massive quarks over the whole range of $Q^2$                                                                        
we must clearly use an approach which extrapolates smoothly from the FFNS at low                               
$Q^2$ to the massless evolution at high $Q^2$, maintaining the correct ordering in                               
both schemes.  To do this we use a method which has recently been developed by two                               
of the authors and is discussed in detail in \cite{RT}, and more briefly                                                                         
in \cite{RTLET}.  Since this treatment of charm is such a major change to our                               
previous analyses we present the method briefly here.  First we note that                                                                         
in the FFNS (\ref{eq:lof2cffns}) is valid up to corrections of                                                                        
${\cal O}(\Lambda^2/m_c^2)$ while the massless prescription                                                                         
is valid only up to corrections of ${\cal O}(m_c^2/\mu^2)$, i.e. threshold                                                                         
corrections. In order                                                                         
to improve the accuracy of the latter scheme we need to                                                                        
examine the connection between the parton densities in the two schemes.                                                                        
The connection between the $\msb$ parton densities for 3 and                                                                         
4 flavours takes the form                                                                         
\bea                                                                        
c_+(z,\mu^2,\mu^2/m_c^2) &=& A^{cg}(\mu^2/m_c^2) \:\otimes\:                                                                         
g_{n_f=3}(\mu^2) \nonumber \\                                                                        
g_{n_f=4}(z,\mu^2,\mu^2/m_c^2) &=& A^{gg}(\mu^2/m_c^2) \:\otimes\:                                                                         
g_{n_f=3}(\mu^2) \label{eq:partons}                                                                        
\eea                                                                         
at leading order,                                                                         
where the elements $A^{ba}$ which contain $\ln(\mu^2/m_c^2)$ terms,                                                                        
are, in general, part of a full 5$\times$4                                                                        
matrix which also connects the light quark flavours. Hence the charm                                                                         
distribution $c_+ \equiv c + \bar{c}$ is determined entirely in terms of the light                               
parton                                                                        
distributions, and it is the above equations which lead to the requirement                                                                         
of evolving from zero charm at $\mu^2=m_c^2$. Since we use the scale choice                                                                         
$\mu^2=Q^2$ for all the light partons we also take this simple choice                                                                         
for the heavy quark structure function. Thus, from now on we will always                                                                         
use $Q^2$ instead of $\mu^2$.                                                                         
                                                                        
For $Q^2 \gg m_c^2$, the equivalence of the FFNS and the massless                                                                        
scheme at all                                                                         
orders lead to the connections between the CF's in                                                                         
the two schemes                                                                        
up to ${\cal O}(m_c^2/Q^2)$ \cite{Buza}, in particular up to                                                                         
${\cal O}(\alpha_S^2)$                                                                        
\bea                                                                        
C_g^{\rm FF}(z,Q^2/m_c^2) &=& C_c^{n_f=4} \:\otimes\:                                                                        
A^{cg}(Q^2/m_c^2) \nonumber \\                                                                        
 &\:&+\:\: C_g^{n_f=4} \:\otimes\:                                                                        
A^{gg}(Q^2/m_c^2) \:+\: {\cal O}(m_c^2/Q^2).                                                                         
\label{eq:cfconn}                                                                        
\eea                                                                        
The details of the connection are fully worked out in \cite{Buza}. To improve                                                                        
the accuracy of (\ref{eq:cfconn}), where the uncertainty is reduced                                                                        
to ${\cal O}(\Lambda^2/m_c^2)$, requires defining `corrected' CF's,                                                                         
$C_i^{\rm VF}$ ($i=1,\ldots,4$), in another $n_f=4$ scheme                                                                         
-- the variable flavour number scheme (VFNS) --                                                                        
where one can write                                                                        
\bea                                                                        
F_2^c(x,Q^2,m_c^2) &=& C_c^{\rm VF}(Q^2/m_c^2) \:\otimes\:                                                                        
c_+(Q^2,Q^2/m_c^2) \nonumber \\                                                                        
 &\:&+\:\: C_g^{\rm VF}(Q^2/m_c^2)                                                                          
\:\otimes\: g_{n_f=4}(Q^2,Q^2/m_c^2) \:+\: {\cal O}(\Lambda^2/m_c^2),                                                                        
\label{eq:facttheha}                                                                        
\eea                                                                        
where the corrected CF's are related to the FFNS CF's by                                                                        
\begin{equation}                                                                         
C_i^{\rm FF}(z,Q^2/m_c^2) = C_j^{\rm VF}(Q^2/m_c^2)                                                                        
\:\otimes\: A^{ji}(Q^2/m_c^2),                                                                        
\label{eq:corrcoeff}                                                                        
\end{equation}                                                                        
the new $n_f$=4 CF's now being {\it exact} at all values of $Q^2$.                                                                        
                                                                        
Hence, our procedure is to use the FFNS                                                                        
for $Q^2 \leq m_c^2$ where it should be very reliable and switch to the                                                                         
VFNS for $Q^2 \geq m_c^2$.                                                                        
(The precise choice of the transition point is undetermined, however,                                                                          
taking $Q^2=m_c^2$ removes complications arising from                                                                        
$\ln(Q^2/m_c^2)$ terms in the matching conditions between the partons                                                                         
at threshold.) In order to define the VFNS                                                                        
one must solve (\ref{eq:corrcoeff}) for the $C_i^{\rm VF}$. Unfortunately                                                                        
the all-orders matching of $F_2^c$ in the two schemes, from which                                                                         
(\ref{eq:corrcoeff}) arose, is not sufficient since, for example,                                                                        
at low orders  the {\it single}                                                                         
quantity $C_g^{\rm FF}$ is expressed in terms of the {\it two} quantities                                                                        
$C_c^{\rm VF}$ and $C_g^{\rm VF}$. We stress that any choice satisfying                                                                         
(\ref{eq:corrcoeff}) is                                                                         
\lq correct' in the sense that it leads to the same all orders expression.                                                                         
Nevertheless,                                                                        
each choice leads to a different expression if one uses the usual rules of                                                                         
combining coefficient functions and parton distributions of a given order to                                                                        
obtain a fixed order in $\alpha_S$ expression for the structure functions.                                                                        
In order to remove this ambiguity we apply a sensible, physically                                                                         
motivated constraint and                                                                         
impose not only continuity of the structure function                                                                         
but also demand, in addition, order-by-order                                                                         
matching of the evolution of $F_2^c$ at threshold.     
    
The explicit form of ({\ref{eq:partons}) at                                                                        
${\cal O}(\alpha_S)$ is                                                                        
\bea                                                                        
c_+(z,Q^2,Q^2/m_c^2) &=& \frac{\alpha_S}{2\pi}\;                                                                        
\ln \biggl (\frac{Q^2}{m_c^2} \biggr )                                                                        
\;P_{qg}^0 \:\otimes\: g_{n_f=3} \nonumber \\                                                                         
g_{n_f=4}(z,Q^2,Q^2/m_c^2) &=& g_{n_f=3}(z,Q^2) \:-\:\frac{\alpha_S}{6\pi}                                                                        
\;\ln \biggl (\frac{Q^2}{m_c^2} \biggr )\; g_{n_f=3}.                                                                         
\label{eq:partdefcharm}                                                                        
\eea                                                                        
Inserting the implied expressions for the matrix elements                                                                         
$A^{cg}(z,Q^2/m_c^2)$ and $A^{gg}(z,Q^2/m_c^2)$                                                                        
into (\ref{eq:cfconn}) gives the relation (first seen in \cite{ACOT})                                                                        
\begin{equation}                                                                        
C_g^{(1)\;{\rm FF}}(z,Q^2/m_c^2) = C_g^{(1)\;{\rm VF}}(z,Q^2/m_c^2)                                                                         
\:+\: C_c^{(0)\;{\rm VF}}(Q^2/m_c^2) \:\otimes\: P_{qg}^0\;                                                                        
\ln \biggl (\frac{Q^2}{m_c^2} \biggr )                                                                        
\label{eq:cfgconnect}                                                                        
\end{equation}                                                                        
connecting the gluonic CF's in the FFNS and VFNS.  Let us now consider the                               
evolution of $F_2^c$.  From (\ref{eq:lof2cffns}) the                                                                         
LO expression in the FFNS for the $\ln Q^2$ derivative is simply                                                                        
\begin{equation}                                                                         
\frac{dF_2^c(x,Q^2,m_c^2)}{d\ln Q^2} = \frac{\alpha_S}{2\pi} \;                                                                        
\frac{d C_g^{(1)\;{\rm FF}}(Q^2/m_c^2)}{d\ln Q^2} \:\otimes\: g_{n_f=3}(Q^2).                                                                        
\label{eq:df2ffns}                                                                          
\end{equation}                                                                        
The corresponding expression obtained by differentiating the LO expression                                                                          
in the VFNS, for $Q^2$ just above $m_c^2$, is                                                                        
\bea                                                                         
\frac{dF_2^c(x,Q^2,m_c^2)}{d\ln Q^2} &=&                                                                         
\frac{dC_c^{(0)\;{\rm VF}}(Q^2/m_c^2)}{d\ln Q^2}\:\otimes\: c_+(Q^2) \nonumber                               
\\                                                                        
&+& \frac{\alpha_S}{2\pi}\;C_c^{(0)\;{\rm VF}}(Q^2/m_c^2)\:\otimes\:                                                                        
\biggl ( P_{qg}^0 \;\otimes\; g_{n_f=4}(Q^2) \:+\:                                                                        
P_{qq}^0 \;\otimes\; c_+(Q^2) \biggr ).                                                                        
\label{eq:df2vfns}                                                                        
\eea                                                                        
At $Q^2=m_c^2$, the                                                                        
terms in (\ref{eq:df2vfns}) involving $c_+$ vanish because of                                                                        
(\ref{eq:partdefcharm}) and so demanding continuity of the evolution                                                                        
across the transition point immediately leads, from                                                                         
(\ref{eq:df2ffns},\ref{eq:df2vfns}), to                                                                         
\begin{equation}                                                                        
C_c^{(0)\;{\rm VF}}(Q^2/m_c^2) \:\otimes\: P_{qg}^0 =                                                                         
\frac{dC_g^{(1)\;{\rm FF}}(z,Q^2/m_c^2)}{d\ln Q^2}.                                                                        
\label{eq:deflocf}                                                                        
\end{equation}                                                                        
Generalising this relation to be the definition of                                                                         
$C_c^{(0)\;{\rm VF}}(z,Q^2/m_c^2)$                                                                        
at {\it all} $Q^2$ guarantees a smooth passage                                                                         
for charm structure function from                                                                         
$Q^2 < m_c^2$ to $Q^2 > m_c^2$, by definition.                                                                        
It is also easy to see that in the limit $Q^2\to \infty$,                                                                         
\begin{equation}                                                                        
\frac{d C_g^{(1)\;{\rm FF}}(z,Q^2/m_c^2)}{d \ln Q^2} \rightarrow P^0_{qg}(z).                                                                        
\label{eq:limitp}                                                                        
\end{equation}                                                                         
Hence, from (\ref{eq:deflocf}), we see that                                                                         
$C_c^{(0)\;{\rm VF}}(z,Q^2/m_c^2)$ must indeed                                                                        
tend to the usual simple form $z\;\delta (1-z)$ in this limit.                                                                        
Also, since $C_g^{(1)\;{\rm FF}}(z,Q^2/m_c^2)$                                                                         
contains the                                                                        
factor $\theta(\hat W^2 - 4m_c^2)$ so does its $\ln Q^2$ derivative,                                                                        
thus ensuring the correct threshold behaviour in $W^2$ for $C_c^{(0)\;{\rm VF}}$                                                                        
and in turn for $F_2^c$ at LO. Furthermore (\ref{eq:deflocf}) allows                                                                        
the gluonic CF in the VFNS to be written as                                                                        
\begin{equation}                                                                        
C_g^{(1)\;{\rm VF}}(z,Q^2/m_c^2) = C_g^{(1)\;{\rm FF}}(z,Q^2/m_c^2) \:-\:                                                                        
\frac{dC_g^{(1)\;{\rm FF}}(z,Q^2/m_c^2)}{d \ln Q^2} \;                                                                         
\ln \biggl ( \frac{Q^2}{m_c^2} \biggr ),                                                                        
\label{eq:defnlogcf}                                                                        
\end{equation}                                                                         
and $C_g^{(1)\;{\rm VF}}$ also has the correct threshold behaviour as                                                                         
$Q^2/m_c^2 \rightarrow \infty$ and                                                                        
$C_g^{(1)\;{\rm VF}}(z,Q^2/m_c^2)$ tends to the correct asymptotic                                                                         
$\msb$ limit.                                                                         
The extension of this procedure to any arbitrary order,                                                                        
i.e. continuity of the derivative in the gluon sector, is described in                                                                        
full in \cite{RT}.                                                                        
    
The implementation of the charm coefficient function is also described in                                                                        
detail in \cite{RT}, and results in the relatively straightforward                                                                        
expression                                                                        
\bea                                                                        
C_c^{(0)\;{\rm FF}}(Q^2/m_c^2)\otimes c_+(Q^2)                                                                        
&=& - \int_{x}^{x_0} dz \, \frac{d C_g^{(1)\;{\rm FF}}(z,Q^2/m_c^2)}                                                                         
{d \ln Q^2}                                                                        
\biggl({x\over z}\biggr)^2 \: \frac{d c_+(x/z,Q^2)}                                                                        
{d (x/z)} \nonumber \\                                                                        
&+& 3\int_{x}^{x_0} dx \, \frac{d C_g^{(1)\;{\rm FF}}(z,Q^2/m_c^2)}                                                                         
{d \ln Q^2}                                                                        
\:{x\over z} \: c_+(x/z,Q^2) \nonumber \\                                                                        
&-& 2\int_{x}^{x_0} dz \, \frac{d C_g^{(1)\;{\rm FF}}(z,Q^2/m_c^2)}                                                                         
{d \ln Q^2}                                                                        
\int_{{x/z}}^1 d z'\, r(z') \: {x\over zz'}                                                                        
\: c_+(x/zz',Q^2) \nonumber \\ 
& &                                            
\label{eq:conv}                                                                         
\eea                                                                        
where $x_0=(1+4m_c^2/Q^2)^{-1}$ and $r(z)$ is given by                                                                        
\begin{equation}                                                                        
r(z) = z^{1 \over 2}\biggl[ \cos\Bigl({\sqrt 7 \over 2}\ln                                                                         
{1\over z}\Bigr) + {3\over \sqrt 7} \sin \Bigl({\sqrt 7 \over 2}\ln                                                                         
{1\over z}\Bigr)\biggr].                                                                         
\label{eq:defr}                                                                        
\end{equation}                                                                        
This general method can be applies at all orders, and                                                                        
the ${\cal O}(\alpha_S)$ charm coefficient function is determined                                                                        
by demanding continuity of $dF_2^c(x,Q^2,m_c^2)/d\ln Q^2$ (in the gluon                                                                        
sector) at ${\cal O}(\alpha_S^2)$, and is discussed in detail in \cite{RT}.                                                                        
However, in practice its contribution to the charm structure function is                                                                         
only at the level of a couple of percent at most, and it can be treated                                                                         
using a phenomenological approximate expression.                                                                         
                                                                        
Thus we can calculate the charm structure function at NLO.                                                                        
For $Q^2 < m_c^2$ we use the usual FFNS expression,                                                                         
i.e. using coefficient functions \cite{NLOCF}                                                                        
to ${\cal O}(\alpha_S^2)$\footnote{We are grateful to                                                                         
Steve Riemersma and Jack Smith for providing the                                                                         
program to compute the ${\cal O}(\alpha_S^2)$                                                                        
contributions.} and parton distributions with $3$ light quarks.                                                                        
For $Q^2 > m_c^2$ we use the VFNS coefficient functions to                                                                         
${\cal O}(\alpha_S)$ and the partons are evolved via the NLO DGLAP                                                                         
equations in $\msb$ scheme                                                                         
with $4$ massless quarks. Since the coefficient functions                                                                         
reduce to the usual massless expressions as $Q^2/m_c^2 \to \infty$, the                                                                         
structure functions approach the previous massless expressions in this                                                                         
limit.

We feel we should distinguish between this approach and previous                                                                         
implementations of a VFNS.                                                                        
In \cite{ACOT} (\ref{eq:cfgconnect}) served                                                                        
as the definition for $C_g^{(1)\;{\rm VF}}$ in terms of the PGF CF                                                                         
(\ref{eq:lochgffns}) with an assumed form of $C_c^{(0)\;{\rm VF}}$                                                                        
given by                                                                        
\begin{equation}                                                                        
\hat C_c^{(0)\;{\rm VF}}(z,Q^2/m_c^2)= z\;\delta(\hat x_0-z)\biggl(1+                                                                        
{4m_c^2\over Q^2}\biggr), \hskip 0.4in \hat x_0=\biggl(1+{m_c^2\over Q^2}                                                                        
\biggr)^{-1}                                                                        
\label{eq:acota}                                                                        
\end{equation}                                                                        
where the delta-function describes the tree-level diagram for                                                                        
a massive quark scattering from a photon                                                                        
and the modified argument of the delta-function follows from                                                                         
demanding that the massive quark is on-shell.\footnote{We note that the same                                                                         
definition of the zeroth order coefficient function is adopted in                                                                         
\cite{MRRS},                                                                        
although of course there are differences between this and \cite{ACOT}, notably the                                                                          
mass dependent evolution in the former.}     
We believe that this manner of determining the                                                                         
charm coefficient function does not reflect the true physics, i.e. that                                                                         
a real charm-anticharm pair must be generated via the photon scattering,                                                                        
leading to the physical threshold in $W^2$, and cancellation between terms                                                                         
is required to reflect this correct threshold. A full critique of this                                                                         
approach can be found in \cite{RT}, but we note here that our variable flavour                                                                        
number scheme is certainly very different to this alternative prescription                                                                        
(which is not yet implemented in their definition of NLO).                                                                         
                                                                        
The theoretical treatment of the bottom quark is essentially identical to                                                                         
that for the charm quark outlined above, and we use the FFNS below                                                                         
$Q^2=m_b^2$ and the VFNS above $Q^2=m_b^2$. As discussed in \cite{RT}, the                                                                        
procedure also generalizes to other processes in a simple manner.

\medskip                                                                      
\noindent{\bf 5.2  Implications for $F_2^c$ and the global analysis}                                                                        
                                                                        
In Fig.~\ref{fig:charm1} we show the result of the NLO calculation                                                                         
of $F_2^c$ for $x=0.05$.  The very                                                                         
smooth transition from the description at low $Q^2$ in terms                                                                        
of the FFNS to high $Q^2$ in terms of the massless prescription demonstrates the                               
success of our procedure.                                                                         
The result is qualitatively similar for all other $x$ values. However, the                                                                         
obvious discrepancy                                                                         
between the continuation of the FFNS curve and the zero-mass                                                                         
curve  for relatively low $Q^2$ (i.e. $\sim 5-20~\GeV^2$), diminishes as we                                                                         
go to lower $x$ as we are then further from the physical threshold in                                                                         
$W^2$, and mass-dependent effects become less important.                                                                         
                                                                        
As was demonstrated in \cite{RT}, the global fit to structure function data                                                                         
achieved using our prescription for charm was superior to that using                                                                         
either the FFNS or the massless prescription.  The former has too slow                                                                        
an evolution due to the lack of $\ln(Q^2/m_c^2)$ terms, and the latter gives a                                                                         
definite kink in $F_2(x,Q^2)$ at mass thresholds.  Thus our global fit incorporates the                               
best available treatment of heavy quark mass effects using NLO-in-$\alpha_S$                                                                        
QCD.                              
                              
One point to note is the influence of the charm prescription on  the                                                                         
optimum value of $\alpha_S(M_Z^2)$. In the most recent global                                                                         
analysis this value came out to be $0.113$, although an alternative set of                                                                         
partons was also given for $\alpha_S(M_Z^2)=0.120$ \cite{MRSR}. The fact                                                                        
that the value of $\alpha_S(M_Z^2)$ for the best fit has risen to $0.1175$                                                                        
is partially due to some of the new data in this fit, e.g.  the reanalysed                                                                         
CCFR data, and the final NMC data. However, a fit to the new data                                                                         
using the old massless charm prescription results in a value of                                                                         
$\alpha_S(M_Z^2)=0.116$. Hence, the effect of the new treatment of heavy                                                                        
quarks is to increase the value of $\alpha_S(M_Z^2)$ by about $0.002$. We also note                                                                         
that the prediction for $F_L(x,Q^2)$ is very different to previous                                                                         
analyses, i.e. smaller; the charm contribution being very suppressed near                                                                         
threshold since the gluon coefficient function behaves like                                                                        
$v^3$. This leads to a better description of the NMC data on $R$ \cite{NMC}, as                      
seen in Fig.~\ref{fig:NMCFL}.                                                                         
                                                                        
We can also look at the charm structure function directly.                                                                         
In Fig.~\ref{fig:charm2} we show the comparison of the charm structure function                                                                        
$F_2^c(x,Q^2)$ resulting from the MRST partons with all available                                                                        
data. The data at intermediate $x$ values come from EMC \cite{EMC}                                                                        
measurements of inclusive muons, while the new data from HERA are obtained by                               
measuring $D$ and $D^*$ cross sections \cite{H1C,ZEUSC}.  We show MRST  
curves together with those resulting from the MRST$(\gup)$ and                               
MRST$(\gdown)$ parton sets. In all cases the value of $m_c$                                                                         
is taken to be 1.35~GeV. (A comparison for different values of                                                                        
$m_c$ can be found in \cite{RT}.)                                                                         
The predictions using the central gluon agree well                                                                        
with the data. As we would expect the difference between the predictions                                                                         
is only significant for relatively high $x$. However, we see that for the                                                                         
two highest $x$ bins the curves for MRST($\gdown)$ tend to fall below                                                                         
the data. Thus, the EMC charm data are capable of acting as a weak                                                                         
constraint on the form of the gluon at high $x$, and it is clear that any                                                                         
parametrization which has an even smaller gluon at high $x$ than                                                                         
MRST($\gdown)$ would be inconsistent with these charm data.                                                                          
                                                                        
We should also justify our choice of $m_c=1.35$~GeV. This value is chosen                                                                         
somewhat as a compromise. As shown in \cite{RT} the charm data on their own                               
prefer a value in the region of $1.5$~GeV. However, the quality of the                                                                         
{\it global} fits obtained are slightly sensitive to the value of $m_c$ and                                                                         
we find that $m_c \sim 1.2$~GeV, or even lower, actually gives the best fit.                                                                         
Practically all this sensitivity to $m_c$ comes from the                                                                         
NMC $F_2$ data with $x<0.1$. As already discussed, these data grow with $Q^2$                               
more                                                                         
quickly than the theory predicts. If $m_c$ is lowered then the                                                                         
charm evolution is slowed down less by mass effects, and the slopes of the                                                                         
theory curves increase. This is seen clearly in Figs.~\ref{fig:NMCa} and                            
\ref{fig:NMCslope}                               
which demonstrate                                                                         
the improvement in the description, although it is clear that a problem remains. As                      
already                                                                        
mentioned there are other effects which could be responsible for the apparent                                                                         
discrepancy with the observed value of $dF_2/d\ln Q^2$                                                                        
in this region, and hence we do not take this as strong evidence for a                                                                         
low value                                                                         
of $m_c$. Indeed, as seen in \cite{RT}, the $F_2^c$ data themselves completely                                                                         
rule out such a low value of $m_c$. Hence we choose $m_c=1.35$~GeV as a                               
compromise between the                                                                        
values required by the best global fit and the best description of charm data.                                                                          
                                                                        
Of course there are as yet no data on the bottom quark contribution to the                                                                         
structure function, and because it contributes with a charge squared                                                                         
of $1/9$, and only at relatively high $Q^2$, it forms only a very                                                                        
small fraction of the total structure function.  Hence, the value of $m_b$ has                               
essentially no impact on the quality of the fit, and is always taken to be 4.3 GeV.                                                                         
Future data on the bottom structure function could act both as a verification                                                                        
of our treatment of heavy flavours and as a determination of the bottom quark mass.                                
The $b$ quark distribution at $Q^2 = 10^4$~GeV is shown in                             
Fig.~\ref{fig:MRSTpartsb}.   \\                                                                      
    
\noindent {\large \bf 6.  Drell-Yan production}                                                                                   
                                                                                   
The observation of Drell-Yan production in high energy $pN$ collisions offers a                                                                                    
valuable constraint on the sea quark distributions since the leading-order subprocess is                                                                                    
$q\overline{q} \rightarrow \gamma^* \rightarrow \ell^+ \ell^-$.  The small $x_F$ data                                            
of the E605                                                                                    
collaboration \cite{E605} are used in the global analysis and they constrain the sea in                                                                                    
the interval $0.15 \lapproxeq x \lapproxeq 0.4$.  The description of the E605 data is                                            
shown in                                                                                    
Fig.~\ref{fig:E605}.  The curves are obtained using NLO QCD, with the                                                                                    
factorization  and renormalization scales set equal to the invariant mass $M$ of the                                            
lepton pair, together with                                                                                    
an overall phenomenological normalization parameter which allows for possible                                                                                    
higher-order effects.                                                                                   
                                                                                   
The more recent measurements of the E772 collaboration \cite{E772} span a larger                                                                                    
kinematic range and, in principle, allow the sea quark distribution to be probed down                                                                                    
to $x \simeq 0.025$.  In general the agreement between the data and the MRST                                                                                    
prediction is quite satisfactory, see Fig.~\ref{fig:E772}.  However, we point out a                                            
discrepancy at high $x_F$ and low $\sqrt{\tau}$.  In this region the                                                                                    
dominant contribution is                                                                                   
\begin{equation}                                                                                   
d \sigma \; \simeq \; u (x_1) \left [ \bar{u} (x_2) \: + \: \bar{d} (x_2) \right                                                                                    
]                                                                                   
\label{eq:a12}                                                                                   
\end{equation}                                                                                   
where $x_1 \simeq x_F$ and $x_2 \simeq \tau/x_1$, and where the partons are                                                                                    
sampled at scales $Q^2 = M^2 = s\tau$ with $\sqrt{s} \simeq                                                                                    
40$~GeV.  If we assume that $u (x_1)$ is known, then there is a factor of two                                                                                    
discrepancy between the E772 data and ($\bar{u} + \bar{d}$) evaluated at                                                                                    
$x_2 \sim 0.03$.  We find that there is no way in which the global fit can remove this                                                                                    
discrepancy since, at such small $x$ values, ($\bar{u} + \bar{d}$) gives                                                                                    
approximately a third of the total contribution to $F_2$, and so is well determined by                                                                                    
the deep inelastic scattering data. \\                                                                                   
                                                                                   
\noindent {\large \bf 7.  Flavour decomposition of the sea}                                                                                   
                                                                                   
As seen in Figs.~\ref{fig:MRSTpartsa} and \ref{fig:MRSTpartsb} the sea quark                                         
distributions                                                                                    
$(\bar{u}, \bar{d}, \bar{s}, \bar{c}$ and $\overline{b})$ have an                                                                                    
interesting non-trivial structure.  Of course, as $Q^2$ increases, all these distributions                                                                                    
ultimately evolve to a common form, concentrated at small values of $x$, since they                                                                                    
are driven by $g \rightarrow q\overline{q}$ transitions.  However, at accessible                                                                                   
$Q^2$ values the different flavour sea quark distributions are quite distinct.  Due to                                                             
their heavy mass, the $\bar{c}$ and $\overline{b}$ can be treated perturbatively,                                                             
as discussed in                                                                                    
Section 5.  The $x$ and $Q^2$ dependence of these heavy quark densities are                                                                                    
therefore completely determined, with their mass being the only free parameter.                                                                                     
On the other hand for the light quark distributions, $\bar{u}, \bar{d}$ and                                                                                    
$\bar{s}$, we may use massless evolution, but here the distributions at the                                                                                    
starting scale have a non-perturbative origin and are determined by experiment.  For                                                                                    
the light quark distributions we use the parametric forms given in (\ref{eq:a3}),                                                                                    
(\ref{eq:a5})--(\ref{eq:a8}).                                                                                   
                                                                                   
\medskip                                                                                   
\noindent {\bf 7.1  The strange quark distribution}                                                                                   
                                                                                   
The observations of deep inelastic dimuon production indicate that $\bar{s}$ has                                                                                    
the same $x$ shape as ($\bar{u} + \bar{d}$) but with an overall suppression                                                                                    
of the order of 50\% at $Q^2 \simeq 4$~GeV$^2$, presumably due to the mass of the                                                                                    
strange quark.  We reflect this behaviour by using an $\bar{s}$ parametrization                                                                                    
which has the same shape as ($\bar{u} + \bar{d}$) at $Q_0^2$, but with a                                                                                    
single overall parameter (0.2 in (\ref{eq:a7})) which is adjusted to fit to the CCFR                                                                                    
dimuon data \cite{CCFRMM}.  Fig.~\ref{fig:strange} shows the MRST strange                                                                                    
quark distribution compared to the measurements determined by the CCFR                                                                                    
collaboration in a NLO analysis \cite{CCFRMM} of their dimuon production data.                                                                                     
We see that our minimal parametrization gives excellent agreement with their                                                                                    
experimental result, which supports the assumption that $\bar{s}/(\bar{u} +                                                                                    
\bar{d})$ is essentially independent of $x$.  As noted earlier, at small $x$, the                                                                                    
differences between $\bar{u}, \bar{d}$ and $\bar{s}$ decrease with                                                                                    
increasing $Q^2$, due to the dominance of the $g \rightarrow q\overline{q}$                                                                                    
subprocesses.  Finally we note that an independent measurement of $\bar{s}$                                                                                    
could, in principle, be made at HERA using the charged current subprocess $e^-                                                                                    
+ \bar{s} \rightarrow \nu + (\bar{c} \rightarrow \mu^-)$ provided that the                                                                                    
accelerator integrated luminosity was sufficiently high.                                                                                   
                                                                                   
\medskip                                                                                   
\noindent {\bf 7.2  Determination of the difference $\bar{u} - \bar{d}$}                                                                                   
                                                                                   
The structure function measurements (of $F_2^{\mu p}, F_2^{\mu n}, F_2^{\nu N}$                                                                                    
and $xF_3^{\nu N}$) determine $(\bar{u} + \bar{d})$, but not                                                                                    
$(\bar{u} - \bar{d})$.  The sum rules do give some information on the                                                             
integral                                                                                    
over $\bar{u} - \bar{d}$, which indicate that, on average, $\bar{d}$ is                                                                                    
greater than $\bar{u}$.                                                                                   
                                                                                   
For a direct determination of $\bar{u} - \bar{d}$ we must look elsewhere.                                                                                     
One proposal \cite{ES} is to measure the asymmetry of Drell-Yan production in $pp$                                                                                    
and $pn$ collisions                                                                                   
\begin{equation}                                                                                   
A_{\rm DY} \; \equiv \; \frac{\sigma_{pp} - \sigma_{pn}}{\sigma_{pp} +     
\sigma_{pn}} \; = \; \frac{1 - r}{1 + r},                                                            
\label{eq:a13}                                                                                   
\end{equation}                                                                                   
where $r = \sigma_{pn}/\sigma_{pp}$ and where $\sigma \equiv d^2                                                             
\sigma/dM dx_F$ with $M$ and $x_F$ being the invariant mass and                                                             
the Feynman $x$ of the produced lepton pair.  At leading order we have                                                            
\begin{equation}                                                            
r \; \equiv \; \frac{\sigma_{pn}}{\sigma_{pp}} \; = \; \frac{(4u_1 \bar{d}_2                                                             
+ d_1 \bar{u}_2 + 4 \bar{u}_1 d_2 + \bar{d}_1 u_2 + 2s_1                                                             
s_2 + 8 c_1 c_2) }{(4u_1 \bar{u}_2 + d_1 \bar{d}_2 + 4 \bar{u}_1 u_2 + \bar{d}_1                                          
d_2 + 2s_1 s_2 + 8c_1 c_2)}                                                           
\label{eq:b13}                                                           
\end{equation}                                                            
where the 1,2 subscripts indicate that the partons are to be evaluated at  
\begin{equation} 
x_1, x_2 = \frac{1}{2} \left(\pm x_F + \sqrt{x_F^2 + 4 \tau}\right), 
\label{eq:c13} 
\end{equation} 
with $\tau = M^2/s$.  We may rearrange the expression for $1 - r$,                                                                                    
and hence that for $A_{\rm DY}$, to show that it is dependent on the combinations                                                                                    
$(\bar{u}_1 - \bar{d}_1)$ and $(\bar{u}_2 - \bar{d}_2)$.                                                                                   
                                                                                   
The first experiment of this type was performed by the NA51 collaboration                                                                                    
\cite{NA51}.  They measured                                                                                   
\begin{equation}                                                                                   
R_{dp} \; \equiv \; \frac{\sigma_{pd}}{2 \sigma_{pp}} \; = \; \frac{1}{2} (1 + r)                                                                                   
\label{eq:a14}                                                                                   
\end{equation}                                                                                   
at $x_1 = x_2 = 0.18$ and found $A_{\rm DY} = -0.09 \pm 0.02 \pm 0.025$, which                                                                                    
corresponds to $\bar{d}/\bar{u} \simeq 2$.  Very recently the E866                                                                                    
collaboration \cite{E866} have measured $R_{dp}$ over a much wider range of                                                                                    
$M$ and $x_F$, which enables a study of the $x$ dependence of                                                                                    
$(\bar{u} - \bar{d})$ over the range $0.04 < x < 0.3$.  The continuous                                                                                    
curve in Fig.~\ref{fig:E866} shows our fit to these data\footnote{We calculate the                                                                                    
ratio $R_{dp}$ from the NLO QCD expression for $d^2 \sigma/dx_1 dx_2$, with                                                                                    
$x_1$ computed from the mean value of $x_2$ for each bin, and then we join the                                                                                    
resulting values of $R_{dp}$ to form a smooth curve.}.  The dotted curve shows the                                                                                    
values which would have been obtained for the ratio if we were to set $\bar{u}$                                                             
equal to                                                                                    
$\bar{d}$, that is if we were to take $A_\Delta = 0$ in (\ref{eq:a8}).  The                                                             
implications for                                                                                    
$\bar{d}/\bar{u}$ from our fit to the E866 data are shown in                                                                                    
Fig.~\ref{fig:ubarodbar}.  Interestingly the structure of                                                                                    
$\bar{d}/\bar{u}$ shows that, at the maximum value of $x$ that is                                                                                    
measured\footnote{Our parametric form should not be extrapolated to predict                                                                                    
$\bar{d}/\bar{u}$ at larger values of $x$, where both $\bar{u}$ and                                                                                    
$\bar{d}$ are very small.}, the ratio has decreased to give $\bar{d} \simeq                                                                                    
\bar{u}$.  Moreover we see that the NA51 measurement occurs at a value of $x$                                                                                    
for which $\bar{d}/\bar{u}$ is essentially at a maximum.  Nevertheless the                                                                                    
new data indicate a somewhat smaller value of $\bar{d}/\bar{u}$ at this                                                                                    
point, $x = 0.18$.  For comparison we also show the prediction for                                                                                    
$\bar{d}/\bar{u}$ obtained from the MRS(R2) set of partons                                                                                    
\cite{MRSR} --- partons which were obtained from a global fit which included the                                                                                    
NA51 measurement, but for which the E866 data were not available.                                                                                   
                                                                                   
Independent information on the $\bar{u}$,$\bar{d}$ flavour asymmetry is currently                                   
being obtained by the HERMES experiment \cite{HERMES} at HERA from the     
observation of semi-inclusive deep inelastic events.  By observing final state     
$\pi^\pm$ mesons, they measure the ratio                                  
\begin{equation}                                  
r (x, z) \; = \; \frac{\sigma (ep \rightarrow e \pi^- X) \: - \: \sigma (en \rightarrow e                                   
\pi^- X)}{\sigma (ep \rightarrow e \pi^+ X) \: - \: \sigma (en \rightarrow e \pi^+ X)}                                  
\label{eq:b14}                                  
\end{equation}                                  
where $z$ is the fractional energy of the fragmenting parton that is carried by the pion.                                    
The HERMES semi-inclusive data lie in the kinematic range $0.02 < x < 0.3$ and                                   
$0.3 < Q^2 < 10$~GeV$^2$.  At leading order $r (x, z)$ is a direct measure of                                   
$(\bar{u}-\bar{d})/(u - d)$ since                                  
\begin{equation}                                  
\frac{1 + r (x, z)}{1 - r (x, z)} \; = \; \frac{(u - d) + (\bar{u} - \bar{d})}{(u - d) -                                   
(\bar{u} - \bar{d})} \: \frac{3}{5} \left (\frac{1 + F}{1 - F} \right )                                  
\label{eq:c14}                                  
\end{equation}                                  
where $F (z)$ is the ratio of the disfavoured to favoured $u \rightarrow \pi$     
fragmentation functions, $F =                                   
D_u^{\pi^-}/D_u^{\pi^+}$.  
Fig.~\ref{fig:HERMES} compares\footnote{Note that     
we compare NLO MRST partons with a LO ratio extracted from the semi-inclusive     
data.  However the effect of the NLO corrections is expected to largely cancel in the     
ratio.} the MRST predictions                                   
with the preliminary HERMES measurements \cite{HERMES} of $(\bar{d} - \bar{u})/(u - d)$ 
as a function of $x$.                                    
The good agreement between the MRST curve and this independent measure of                                   
$\bar{u} - \bar{d}$ is confirmation that $\bar{d}/\bar{u}$ is now reliably known as a                                   
function of $x$ and $Q^2$ for $x < 0.3$.                                                                                   
%
%
%Fig.~\ref{fig:HERMES} shows the MRST predictions                                   
%in the kinematic region of the forthcoming HERMES 
%measurements of $(\bar{d} - \bar{u})/(u - d)$ as a function of $x$.                                    
%
                                                                                   
Historically the first indication of the $\bar{u} \neq \bar{d}$ flavour                                                                                    
asymmetry of the sea came from the evaluation of the Gottfried sum                                                                                   
\begin{equation}                                                                                   
I_{\rm GS} \; \equiv \; \int_0^1 \: \frac{dx}{x} \: \left ( F_2^{\mu p} - F_2^{\mu n}     
\right )                                                                                   
\label{eq:a15}                                                                                   
\end{equation}                                                                                   
by NMC \cite{NMCGS}.  The final NMC measurements \cite{NMCF} of                                                             
$(F_2^{\mu p}                                                                                    
- F_2^{\mu n})$ are compared with the values obtained from the MRST partons in                                                                                    
Fig.~\ref{fig:GS}.  The area under the MRST curve yields the value $I_{\rm GS} =                                                                                    
0.266$.  This is slightly larger than the value $I_{\rm GS} = 0.235 \pm 0.026$ found     
at $Q^2 = 4$~GeV$^2$ by NMC \cite{NMCF}.  The small discrepancy is induced in                                   
part by the requirement that the MRST partons also fit the E866 data. \\                                                                                   
                                                                                   
\noindent {\large \bf 8.  $W$ rapidity asymmetry}                                                                                   
                                                                                   
The $W^\pm$ charge asymmetry at the Fermilab $p\bar{p}$ collider,                                                                                   
\begin{equation}                                                                                   
A_W (y) \; = \; \frac{d\sigma (W^+)/dy - d\sigma (W^-)/dy}{d\sigma (W^+)/dy +                                                                                    
d\sigma (W^-)/dy},                                                                                   
\label{eq:a16}                                                                                   
\end{equation}                                                                                   
is a sensitive probe of the difference between $u$ and $d$ quarks in the $x \sim 0.1$,                                                                                    
$Q \sim M_W$ region \cite{BERGER}.  Because the $u$ quarks carry more momentum on average                                                                                    
than the $d$ quarks, the $W^+$ bosons tend to follow the direction of the incoming                                                                                    
proton and the $W^-$ bosons that of the antiproton, i.e. $A_W > 0$ for $y > 0$.  Thus                                                                                    
a precise measurement of the $W$ asymmetry serves as a valuable independent check                                                                                    
on the $u$- and $d$-quark distributions.  In practice it is the lepton asymmetry,                                                                                   
\begin{equation}                                                                                   
A (y_\ell) \; = \; \frac{\sigma (\ell^+) - \sigma (\ell^-)}{\sigma (\ell^+) + \sigma (\ell^-                                                                                   
)},                                                                                   
\label{eq:a17}                                                                                   
\end{equation}                                                                                   
which is measured, where $\sigma (\ell^\pm) \equiv d\sigma/dy_\ell$ are the                                                                                    
differential $p\bar{p} \rightarrow W^\pm X \rightarrow \ell^\pm \nu X$ cross                                                                                    
sections for producing $\ell^\pm$ leptons of rapidity $y_\ell$.  There is a direct                                                                                    
correlation between the lepton asymmetry and the {\it slope} of the $d/u$ ratio.  To                                                                                    
see this we first note that the dominant contribution to $W^+ (W^-)$ production                                                                                    
comes from the $u\bar{d}$ ($d\bar{u}$) annihilation process.  Thus                                                                                   
\begin{equation}                                                                                   
A_W (y) \; \simeq \; \frac{u (x_1) d (x_2) - d (x_1) u (x_2)}{u (x_1) d (x_2) + d (x_1)                                                                                    
u (x_2)},                                                                                   
\label{eq:a18}                                                                                   
\end{equation}                                                                                   
where the scale $Q = M_W$ is implicit for the parton distributions, and                                                                                   
\begin{equation}                                                                                   
x_{1,2} \; = \; x_0 \exp (\pm y), \quad x_0 \; = \; \frac{M_W}{\sqrt{s}}.                                                                                   
\label{eq:a19}                                                                                   
\end{equation}                                                                                   
If we introduce the ratio $R_{du} (x) = d(x)/u(x)$, then, for small $y$,                                                                                   
\begin{equation}                                                                                   
A_W (y) \; \simeq \; - x_0 y \: \frac{R_{du}^\prime (x_0)}{R_{du} (x_0)},                                                                                   
\label{eq:a20}                                                                                   
\end{equation}                                                                                   
where the prime denotes differentiation.  In reality, the situation is of course more                                                             
complicated --- it is the {\it lepton}                                                                                    
asymmetry which is measured, and there are subleading and higher-order corrections                                                                                    
to (\ref{eq:a18}).  Nevertheless, the correlation implied by (\ref{eq:a20}) is                                                                                    
evident in the full prediction.                                                            
                                                                                
The CDF collaboration \cite{CDF} have recently extended the range and improved                                                                                 
the precision of their measurements of the asymmetry $A_\ell$.  The new data extend                                                                                 
to higher values of lepton rapidity $y_\ell$ and the measured values are below the                                                                                 
extrapolation of the previous global fits. The fit\footnote{The curves                                                             
in Fig.~\ref{fig:WASY} are calculated using the next-to-leading-order program                                                             
DYRAD of \cite{DYRAD}.  We thank Nigel Glover for helping with these                                                             
calculations of the $W$ asymmetry.} to the lepton asymmetry data is shown in                                                             
Fig.~\ref{fig:WASY}.  For comparison we also show the result from a previous set of                                                             
partons, MRS(R2), which were fitted to earlier CDF asymmetry measurements.   
Recently the effects of soft gluon resummation on $A_\ell$ have been calculated  
\cite{BY}.  They increase the asymmetry slightly at the highest $y_\ell$ values of the  
data.  For example at the highest value, $y_\ell = 2.2$, the increase in $A_\ell$ is  
about 0.02 so that the MRST prediction would, as it happens, be raised to coincide  
exactly with the data point. 
 
In order to accommodate the new $A_\ell$                                                             
measurements the $d$ distribution increases with respect to the $u$ at $x \sim 0.3$ (in                                                             
comparison with our previous global analysis).  The change affects $F_2^n$ much                                                             
more than $F_2^p$.  The consequence is that the ratio $F_2^n/F_2^p$ is increased,                                                             
and the description of the new NMC data is improved relative to MRS(R2), see                                          
Fig.~\ref{fig:NMCratio}. \\                                                            
                                                            
\noindent {\large \bf 9. Implications for hadron collider cross sections }                                                            
                                                  
According to the QCD factorization theorem, `hard scattering' hadron collider                                                   
cross sections can be expressed in terms of parton distribution functions                                                   
convoluted with perturbatively calculable subprocess cross sections,                                                   
\beq                                                   
d\sigma_X\; =\; \sum_{ij}\; \int dx_1 dx_2\; f_i(x_1,\mu^2)                                                   
f_j(x_2,\mu^2)\; d\hat{\sigma}_{ij\to X}\; ,                                                   
\label{eq:factorization}                                                   
\eeq                                                   
with, for example, $X= W^{\pm},Z,Q\overline{Q}$, jets or Higgs.                                                   
Processes with well-measured final states, and for which the next-to-leading                                                   
order corrections to $\hat{\sigma}$ are known, can therefore provide important                                                    
cross-checks on the parton distributions. A prime example is the $W$ rapidity                                                   
asymmetry at the Tevatron which, as we have just seen, is used in our global analysis                                          
to constrain the $u$ and $d$ distributions.                                                   
Leading-order kinematics imply $x_{1,2} = M_X \exp(\pm y_X)/\sqrt{s}$.                                                   
In general, at colliders like the Tevatron and the LHC,                                                   
quark and gluon distributions are probed at $x$ values where they                                                   
are `measured' by deep inelastic scattering and prompt photon                                                   
experiments, but at significantly higher scales $\mu^2 \sim M_X^2$.                                                    
                                                   
A precise knowledge of parton distributions is absolutely vital                                                   
 for reliable predictions for signal and background cross sections                                                    
 at the LHC \cite{wjslhc}. Uncertainties can arise both from the starting distributions                                                   
 and from DGLAP evolution. A detailed assessment of these uncertainties                                                   
 illustrated with reference to various standard cross sections will be                                                   
 presented elsewhere \cite{mrstlhc}. In this section we focus on several                                                   
 standard cross sections measured at the Tevatron $p \bar p$ collider.                                                   
                                                    
\medskip                                                    
                                                    
\noindent {\bf 9.1  $W,Z$ production}                                                  
                                                   
While the $W$ (lepton) rapidity asymmetry probes the relative size of the $u$ and                                                   
$d$                                                   
distributions, the                                                    
 {\it total} cross sections for $W$ and $Z$ boson production                                                   
in $p \bar p$ collisions at $\sqrt{s} = 1.8\ \TeV$ provide an important                                                   
check of the overall magnitude of the                                                   
quark distributions in a region of $x\sim 0.05 - 0.1$ where they are constrained                                                   
at lower $\mu^2$  by deep-inelastic (in particular NMC) data.                                                   
Since the perturbative QCD subprocess cross section is known                                                   
to next-to-next-to-leading order \cite{WILLY} and the electroweak parameters                                                   
are precisely determined, there is very little theoretical                                                   
uncertainty in the predictions once the  parton distributions are                                                   
specified.                                                   
                                                   
We begin by displaying in Fig.~\ref{fig:WZsig} the $W$ and $Z$ total cross                                                   
sections                                                   
times the leptonic branching ratios\footnote{The Standard Model values                                                   
$B(W \to l \nu)  = 0.1084$ and $B(Z\to l^+l^-) = 0.03364$ are used. The electroweak                                                   
boson                                                   
masses are taken to be                                                    
$M_W = 80.43\; \GeV$ and $M_Z = 91.1887\; \GeV$.} as measured                                                   
by UA1 \cite{UA1SIGW}  and UA2 \cite{UA2SIGW}                                                    
 at $\sqrt{s} = 630\; \GeV$ and by CDF \cite{CDFSIGW}                                                    
and D0 \cite{D0SIGW} at                                                    
$\sqrt{s} = 1.8\; \TeV$, together with the predictions obtained                                                   
using our default MRST set.\footnote{Strictly speaking there is a slight
inconsistency from combining a subprocess cross section calculated to 
NNLO with NLO-evolved
partons. However the NNLO contribution is numerically very small.}
 Only the most recent (Run 1A)                                                   
Tevatron published measurements are included. The factorization and renormalization                                                   
scales are set equal to $M_W$ and $M_Z$ respectively. The overall agreement                                          
between theory                                                   
and experiment is excellent.                                                   
                                                   
To study this in more detail, we focus on the more precise Tevatron measurements                                                   
and show, in Fig.~\ref{fig:WZ5}, the predictions of the five MRST sets                                                   
(the default MRST set, together with  
$\gup, \gdown, \asup, \asdown$).  There is an overall spread  
of approximately $\pm 2\%$ about the default prediction,\footnote{A measure of the  
scale dependence of these predictions is                                                   
obtained by using instead $\mu = M_V/2$ and $2 M_V$. The effect is shown as error                                                   
bars on the default prediction and is evidently very small.} significantly                                                    
smaller than the current experimental errors.                                                   
The variations in the predictions are easily understood.                                                   
At these (small) $x$ values, the quark distributions {\it increase} with increasing                                                   
$\mu^2$. The larger the $\alpha_S$ the faster the increase, and so                                                   
$\sigma_V(\asdown) < \sigma_V({\rm MRST}) <                                   
\sigma_V(\asup)$.                                                     
The different gluon distributions also give rise to differences                                                   
in the $\sigma_V$ predictions.  In this $x \sim 0.05$ region, the ordering of the gluon                                                   
distributions is $\gup < g({\rm MRST}) < \gdown$, see Fig.~\ref{fig:WZ5}.                                                   
The larger the gluon the more rapid the DGLAP evolution, and so                                                    
$\sigma_V(\gup) < \sigma_V({\rm MRST}) < \sigma_V(\gdown)$. The gluon                                          
variation                                                   
is slightly smaller than the $\alpha_S$ variation.                                                   
                                                   
We may conclude from Fig.~\ref{fig:WZ5} that for a given set of                                                   
quark distributions fixed by DIS data at a relatively low $Q^2 \sim 10\;\GeV^2$                                                    
scale, the net uncertainty in the $\sigma_{W,Z}$ predictions at high $Q^2 \sim 10^4                                                   
\; \GeV^2$ coming from the gluon, $\alpha_S$ and unknown higher-order                                                   
corrections is of order $\pm 2\%$. The {\it true} theoretical uncertainty                                                   
has a significant additional component from the uncertainty in the absolute     
normalizations                                                   
of the ($u$ and $d$) distributions as determined by the normalization uncertainties                                                   
in the structure function data themselves.  In the relevant $x$ range for the Tevatron                                                   
$W$ and $Z$ cross sections, the main constraints on the quarks come from the                                                   
precise NMC $F_2^p$ and $F_2^d$ data. These have an overall systematic                                                   
(normalization)                                                   
error of approximately $\pm 2.5\%$, which would give a corresponding error of $\pm                                          
5\%$                                                   
on the weak boson cross sections, comparable to the current experimental errors  from                                                  
CDF and D0.  However this is almost certainly a large overestimate since the                                          
requirement of mutual consistency between the various deep inelastic data sets,                                          
together with the sum rule constraints, does not permit the possibility of changing the                                          
normalization of an individual data set over its full allowed range.  This point will be                                          
addressed in \cite{mrstlhc}.                                         
                                                   
A final point concerns the impact on the experimental cross sections of                                                   
the luminosity measurement and uncertainty. The latter ($\pm 3.6\%$ for CDF                                                   
and $\pm 5.4\%$ for D0 \cite{CDFSIGW,D0SIGW}) dominate the most precise                                                    
($W \to e \nu$) cross section errors. In addition, the value assumed                                                   
for the total $p \bar p$ cross section is slightly different for the two                                                   
experiments, and this may account in part for the systematically smaller                                                   
D0 cross sections displayed in Fig.~\ref{fig:WZ5}.                                                   
                                                   
\medskip                                                   
                                                   
\noindent {\bf 9.2  Top production}                                                   
                                                   
At the Tevatron, $t\bar t$ production occurs dominantly                                                   
($\sim$90$\%$) via $q\bar q$ annihilation. The cross section                                                   
therefore samples valence $u$ and $d$ quarks at $x\gapproxeq                                                   
2m_t/\sqrt{s} \sim 0.2 - 0.3 $ and $\mu^2 \sim m_t^2 \sim                                                    
10^{4-5}\; \GeV^2$. The leading order subprocess cross section                                                   
is proportional to $\alpha_S^2$, but the enhancement obtained by                                                    
increasing $\alpha_S$ is partially compensated by the decrease                                                   
in the parton distributions at large $x$ caused by more rapid                                                   
DGLAP evolution.                                                   
                                                   
The production cross section is known exactly at NLO \cite{TOPNLO}.                                                   
It is traditional to estimate the  residual theoretical scale dependent                                                   
uncertainty by varying the factorization and renormalization scales                                                   
in the range $m_t/2 < \mu < 2 m_t$ which, at the Tevatron, gives an                                                   
approximate $\pm 10\%$ variation about the $\mu = m_t$ prediction                                                    
for a fixed set of partons and fixed $\alpha_S$.                                                    
                                                   
Another important effect is the higher-order contributions from soft gluon                                                    
emission, which are expected to be large when the $t \bar t $                                                    
system is produced near threshold.                                                   
Techniques have recently  been developed for resumming the dominant                                                   
leading (LL) $\ln(1-4m_q^2/\hat{s})$ logarithms \cite{topresum}.                                                    
In Ref.~\cite{topbcmn} the NLL logarithms have been computed and used to                                                   
obtain an `improved' resummed cross section. The effects of resummation                                                   
are particularly large for large scale choices, whereas for $\mu = m_t/2$                                                   
the NLO cross section approximates the resummed cross section to better                                                   
than 1\% at Tevatron energies \cite{topbcmn}\footnote{For the choice                                                   
$\mu = m_t$ the effects of beyond-NLO resummation increase the NLO                                                    
cross section                                                   
by approximately 5\% \cite{topbcmn}.}. In our calculations we will therefore compute                                          
the top cross sections using NLO QCD, with $m_t                                                   
= 175\; \GeV$ and $\mu = m_t/2$.                                                    
                                                   
Figure~\ref{fig:top} shows predictions for the total $t\bar{t}$ cross section at the                                          
Tevatron using the five canonical MRST parton sets.  Data points from CDF                                   
\cite{topcdf}                                          
and D0 \cite{topdzero} are also shown.  The cross sections are slightly                                                   
larger for the sets  with the larger (large-$x$) gluon and larger                                                    
$\alpha_S$ (although for the latter, note that the variation is smaller                                                   
than the naive estimate of $\pm 10\%$ from the change                                                   
in the overall $\alpha_S^2$ value would suggest).                                                   
 The uncertainties due to the valence quarks, the gluons, and                                                    
$\alpha_S$ are all at the $\pm 5\%$ level. The resulting overall                                                   
parton distribution uncertainty in the $\sigma(t \bar t)$                                                   
prediction is therefore at the $\pm 10\%$ level, comparable                                                   
in magnitude to the scale dependence (combined with other higher-order effects                                                   
\cite{topbcmn}) uncertainty. The agreement with the current CDF and D0                                                   
measurements is entirely satisfactory.                                                  
                                                  
\medskip                                                   
                                                   
\noindent {\bf 9.3 Large $E_T$ jet production}          
          
The single jet inclusive $E_T$ distribution at the Tevatron is a particularly interesting           
observable.  Even though the measured cross section falls by more than six orders of           
magnitude when $E_T$ increases from 50 to 400~GeV, the NLO QCD predictions           
reproduce the data to well within the systematic error band.  Nevertheless the detailed           
shape of the spectrum, taking only the statistical errors into account, shows interesting           
features.  The spectrum has been measured by both the CDF and D0 collaborations           
\cite{CDFJ,D0J}.  The experimental spectra are compared with the        
NLO predictions\footnote{ We are grateful to Nigel Glover for help in performing       
these calculations.}           
obtained from five of our canonical sets of partons in Figs.~\ref{fig:J1}--\ref{fig:J4}.            
In each case we use the prediction of the MRST parton set as the base line for the           
comparison.  The predictions contain an overall normalization factor which is adjusted           
to give the optimum description of the data.  The value of the factor is shown on the           
plots for each set of partons.          
          
Figs.~\ref{fig:J1} and \ref{fig:J2} compare the CDF and D0 spectra with the           
predictions obtained from the three parton sets based on different gluons.  It is           
interesting to see that the set, MRST($\gup$), with the larger gluon at large $x$           
(that is the set in which the WA70 prompt photon data are fitted with partonic $\langle           
k_T \rangle = 0$) gives the best description of the shape of the observed spectrum and           
an overall normalization nearest to unity. The effect is particularly pronounced       
for the D0 data.  Figs.~\ref{fig:J3} and \ref{fig:J4} compare the CDF and D0 spectra      
with the predictions of parton sets corresponding to three different values of $\alpha_S           
(M_Z^2)$.  In this case the parton set, MRS($\asdown$), with           
the smallest $\alpha_S$ gives the best description.          
          
Note, however, that the comparisons between experiment and theory shown       
in Figs.~\ref{fig:J1}--\ref{fig:J4} should not be taken too literally. Aside from       
the large experimental systematic errors which have not been included,         
there are residual theoretical uncertainties, for example from scale dependence and       
the precise modelling of the experimental jet algorithm, which are important       
at the ${\cal O}(\pm 5\%)$ level (see for example the recent study in       
Ref.~\cite{D0study}). It is therefore premature to draw any definite conclusions       
about preferred gluons distributions and/or $\alpha_S$ values.       
       
A final point concerns the impact of `intrinsic' transverse momentum smearing       
on the jet $E_T$ distribution. Using the same procedure as implemented in the       
prompt photon studies of Section~4, we have investigated the effect of different       
choices of $\langle k_T \rangle$ on the shape of the distribution.          
Qualitatively, the effect is to {\it steepen} the distribution slightly at the        
low $E_T$ end of the spectrum, as for the $p_T(\gamma)$ distributions shown in        
Figs.~\ref{fig:WA70}--\ref{fig:E706b}. For example, for   $\langle k_T \rangle = 4\;       
\GeV$       
we find that the jet cross section is increased by $+3\%$ for $E_T = 75\; \GeV$, and       
by        
less than $+1\%$ for $E_T > 150\; \GeV$. The description of the CDF and D0 data       
by the canonical MRST parton set is not therefore improved.\\          
      
\noindent {\large \bf 10.  Conclusions}                                               
    
In order to examine the partonic structure of nucleons we have performed a    
global analysis of data on deep inelastic scattering and related hard     
scattering processes using NLO-in-$\alpha_S$ QCD. This present treatment     
represents a significant improvement over our previous analyses for a number of     
reasons. First, there is the availability of a number of updated, or     
completely new sets of data. These have all been discussed in the introduction.    
However, we note that those experiments which have had a major impact     
on the changes to the parton distributions include those that produce    
a relatively small number of data points:     
i.e. the new data from the E866 and HERMES experiments, 
%i.e. the new data from the E866 experiment, 
which constrain the value of $(\bar u - \bar d)$; the extended      
data on the $W$ rapidity asymmetry, which constrain $d/u$; and the new    
prompt photon data from the E706 experiment, which probe the gluon at high    
values of $x$.     
    
There are also new features in the way in which we perform the    
analysis itself. For the     
first time we incorporate intrinsic parton $k_T$     
when examining those data which are    
sensitive to it, which in practice means the prompt photon data. We also     
use a new, much more sophisticated treatment of the charm (and bottom)    
contribution to structure functions. This new procedure naturally includes     
both a smooth behaviour in the threshold region and the summation of large    
logarithms at high $Q^2$, and gives a parameter free (up to the charm mass)    
prediction for the charm contribution. For the first time we have also     
examined the sensitivity of the partons sets obtained to cuts in both     
$Q^2$ and $x$ imposed on the data, letting the former vary between $2~\GeV^2$    
and $10~\GeV^2$ and the latter up to $x_{\rm min}=0.01$. We have also     
investigated the uncertainties which exist for our determination of the     
strong coupling constant $\alpha_S$ in a more systematic manner than in     
previous analyses, obtaining a quantitative estimate of the allowed variation     
about our central value of $\alpha_S(M_Z^2)=0.1175$. Finally, for the     
first time we have also     
made a thorough investigation of the uncertainty in the gluon distribution    
obtained from our analysis, producing two sets of partons which represent     
two extremes as well as the central, preferred    
set.      
    
\medskip    
    
Both the new data sets and the new theoretical procedures have led to     
significant differences between the MRST partons and those produced by     
previous analyses. In order to exhibit these differences we can look at    
Fig.~\ref{fig:MRSTvsR2},    
which shows the comparison between the MRST partons and the preferred set from     
our previous global analysis, i.e. MRS(R2). First, looking at    
the comparison at $Q^2=20~\GeV^2$ (upper plot),     
most partons show significant differences. The new gluon is much smaller     
for $x\gapproxeq 0.2$, since the inclusion of intrinsic $k_T$ means a smaller gluon     
is required  to fit the WA70 prompt photon data. The form of the gluon at     
small $x$ is constrained by the HERA data, and therefore the momentum sum    
rule allows a larger gluon at $x\sim 0.05$. The form of the charm quark     
distribution is clearly very different to our previous MRS(R) analysis \cite{MRSR},  
where the     
evolution of the charm quark in the $\msb$ scheme took place from     
$Q^2=1~\GeV^2$ rather than the correct value of $Q^2=m_c^2$,     
and a phenomenological damping factor was used. It is     
this damping factor which led to the previous small charm distribution at     
small $x$.    
However, the charm distribution is driven by the gluon distribution, and    
one can see that the difference in shape between the new and old charm     
distribution mirrors that of the gluon. It is this which causes the MRS(R2)    
charm quark to be larger at large $x$, overcoming the effect of      
the damping factor which is a function of $Q^2$ only.    
The light quark distributions are all a little larger for the MRS(R2)     
partons at small $x$,    
compensating for the smaller charm distribution. At high $x$ the $u$ quark is     
essentially unchanged, the $d$ quark is larger in the MRST partons in     
order to accommodate the extended $W$ asymmetry data, and the strange quark    
(which effectively represents the sea quark distribution) is a little     
smaller, presumably due to the requirements of fitting the Drell-Yan data with    
a slightly larger valence quark distribution. Also examining the lower
plot of  Fig.~\ref{fig:MRSTvsR2} we see     
that the systematic differences between the light partons are of the same     
form as at the lower $Q^2$ value, but have been washed out somewhat    
(in particular the small $x$ form of the partons is almost identical in     
each set). However, the charm distribution is still very different.    
Again, the shape difference mirrors that of the gluon, but the MRS(R2)    
charm distribution is larger. This is because the damping factor is     
now unity at such a high value of $Q^2$, but the fact that the evolution     
of the MRS(R2) distribution began at lower $Q^2$ results in a     
constant difference between the two distributions, i.e. the charm     
generated in the evolution from $Q^2=1~\GeV^2$ to $Q^2=m_c^2$. Although this    
difference disappears in the ratio asymptotically, it is still     
significant at $Q^2 =10^4~\GeV^2$.       
    
We can also compare the new MRST partons with the preferred set of the most     
recent CTEQ analysis, i.e. CTEQ4M. This comparison is displayed in  
Fig.~\ref{fig:CTEQpart}.     
We first compare the partons at $Q^2=10~\GeV^2$.     
In the high $x$ region there is very little difference between the CTEQ4M     
partons and the MRS(R2) partons. The CTEQ4M strange (and therefore sea)    
distribution becomes much larger at $x\gapproxeq 0.4$, but this is beyond     
the range of the E605 Drell-Yan data. Hence, all differences     
between quarks in     
this region are presumably due to the new data included in the MRST analysis.    
Even the gluon is similar to MRS(R2)    
in this range, being constrained mainly by the high-$E_T$ jet data, which,    
as we have already noted, require a high $x$ gluon similar to MRST($\gup)$.    
However, the CTEQ4M gluon becomes significantly larger than the MRST gluon    
at small $x$, leading to a stronger growth of $F_2(x,Q^2)$ with $Q^2$ in this    
region. This is partly due to the $Q^2$ cut of $4~\GeV^2$ imposed by    
CTEQ, and a similar trend is seen for our gluon when similar cuts are     
imposed, as demonstrated in Fig.~\ref{fig:partoncut}. Since the evolution of the charm  
parton     
distribution is the same for CTEQ4M as for MRST (the treatment of the     
coefficient function being very different), differences in the     
charm are largely due to differences in the gluon, and indeed the shape     
of the charm ratios mirrors that of the gluon ratios. The light quark     
distributions at small $x$ are a little smaller than those for the MRST    
partons, but as with the gluon differences this can be attributed     
mainly to the different cuts, the same qualitative     
effect again being seen in Fig.~\ref{fig:partoncut}. However, the     
slightly different ratio between the $u$ and $d$ quarks and the     
strange quarks,     
even at $x=10^{-4}$, is due to the CTEQ requirement that $2\bar s/(\bar u + \bar  
d)=0.5$ at $Q^2=2.56~\GeV^2$ rather    
than $1~\GeV^2$, which results in the strange quark being a little    
smaller. In the intermediate $x$ region the CTEQ4M $u$ and $d$ quarks are    
both slightly larger than those for MRST, compensating for the smaller    
strange and charm quarks. As with the previous comparison, differences     
between the parton sets tend to be washed out at very high $Q^2$.    
    
In summary the main differences between the CTEQ4M and    
MRST partons arise from (i) our inclusion of new data, (ii) the different treatment of  
the gluon at large $x$ and (iii) the choice of cuts in $Q^2$. However, there are some    
small additional systematic differences between the two analyses. The normalizations     
imposed on the data sets are a percent or two higher for CTEQ than MRST.  If we  
compare CTEQ4M partons with those obtained in our analysis using a comparable  
$Q^2$ cut ($Q^2 > 5$~GeV$^2$) then the partons are more similar, but significant  
differences remain.  These may be due to the     
systematic difference between the output of the evolution programs, as noted in     
\cite{partoncomp}. With the quality of data now available such differences    
are becoming significant.     
    
Let us briefly justify our confidence in our new canonical set of partons    
(MRST). As we have seen they give an excellent overall description of the     
data. There are certain conflicts: as in our previous analyses we     
omit the (new) CCFR neutrino $F_2$ and $xF_3$ data for $x<0.1$ from our fit as  
explained in Section~3;    
and we have also noted a discrepancy with the E772 Drell-Yan data for large     
$x_F$ and small $\tau$. There is also a slight systematic discrepancy    
with the measured value of $\partial F_2(x,Q^2)/\partial \ln Q^2$ for the     
NMC data below $x=0.1$ and for some of the HERA data. The quality of the     
fit to these data is clearly very satisfactory, but this observed effect     
may be a sign that NLO-in-$\alpha_S$ DGLAP evolution is not sufficient     
at small $x$, and that theoretical corrections are required.\footnote{The     
discrepancy     
with the E772 data occurs when one of the partons has $x\sim 0.03$, but it seems     
unlikely that such a large discrepancy can be cured by leading     
$\ln (1/x)$ effects at such a high value of $x$.} It will be interesting    
to see whether further HERA data confirms the observed trend, and measurements    
at high $y$ may be particularly important since they probe both $F_2(x,Q^2)$    
and $F_L(x,Q^2)$.  We see from Fig.~\ref{fig:alpha}    
that if we vary $\alpha_S$ away from our     
central value of $0.1175$ in either direction, then the quality of the fit to     
one or more of the data sets becomes rather poor very quickly. Hence, variation of     
$\alpha_S$ allows very little variation in our partons.  We believe that our default set  
of partons, denoted simply by MRST, gives the most reliable treatment of the gluon at  
high $x$.  The E706 prompt photon data have made it clear that intrinsic parton  
$k_T$ is     
required, as one might expect from theoretical arguments and from the     
evidence already observed in Drell-Yan data. It is also     
very plausible that the amount of intrinsic $k_T$ increases with     
centre of mass energy, and this hypothesis has recently received support from  
an approximate soft-gluon resummation calculation~\cite{SMEAR}.
In the absence of a fully rigorous theoretical treatment 
of the broadening of the prompt photon $p_T$ distribution we have adopted
an experimentally motivated Gaussian form for the  parton intrinsic
$k_T$ smearing.
The     
MRST($\gup$) gluon obtained from assuming zero intrinsic $k_T$ for     
the fit to WA70 data is clearly at one extreme, while the MRST($\gdown$)    
gluon obtained from allowing the partons to have the maximum $\langle k_T    
\rangle$ in     
the fit to WA70 data is at the other extreme by construction. The default MRST     
gluon has $\langle k_T \rangle$ intermediate to these extremes.  While     
we cannot be absolutely sure of the most appropriate values of     
$\langle k_T \rangle$ to     
take for the two prompt photon experiments they cannot be much different to those     
employed in MRST. Since MRST($\gup$) and MRST($\gdown$)     
represent such extreme variation in parton $\langle k_T \rangle$ we feel     
that the variation in the gluon they encompass    
is also sufficient to account for the smaller effect of scale dependence in    
prompt photon production. We note that there is no dependence on intrinsic    
parton $k_T$ in any of the other data sets we analyse except for the high     
$E_T$ jet data from CDF and D0, where the effect is very small, and not    
helpful.              
    
We now  have the most complete determination of the proton     
parton distributions yet achieved, both in terms of the theoretical treatment     
and in terms of the data analysed. However, in both there are possible     
improvements to be made. In terms of the theoretical procedure used in the analysis     
there are two major points. At present we use a leading twist     
NLO-in-$\alpha_S$ calculation. However, there are both    
leading $\ln(1/x)$ and higher twist corrections which should have a     
significant effect in certain kinematic domains (we impose a cut    
in $W^2$ in order to remove the sensitivity to higher twist effects at    
high $x$ and low $Q^2$), and a hint of this is perhaps seen in the results         
of our analysis when different cuts are applied. In order to be truly confident of     
our results over the whole kinematic domain we need either proof that these     
possible corrections are very small, or a reliable method of including them     
in our analysis. Also, our present treatment of intrinsic parton $k_T$     
is phenomenologically inspired. While we believe it is a good     
representation of the true effects we await an improved     
theoretical understanding of the origin of both perturbative and     
nonperturbative intrinsic $k_T$, hopefully pinning down the actual amount    
in a quantitative manner.     
    
There are a number of improvements in data which would also be very important    
in increasing the accuracy of determinations of parton distributions.    
Obviously, any increase in the precision of existing data and/or a     
widening of the kinematic range would be useful.  It would be     
advantageous to have further sources of neutrino structure function data as a     
consistency check, since at present    
the analysis relies on a single high precision experiment.    
Of particular use would be an increase in both the amount and precision    
of data on the charm structure function, as this would help     
differentiate between different prescriptions for heavy flavour effects,    
be a direct probe of the gluon, and possibly be important in     
determining small $x$ corrections. We anticipate data of this type from HERA     
in the near future. Likewise, data on the bottom structure function     
would be useful for the same reasons, but since the mass is higher, it     
would be a cleaner probe of perturbative QCD. Finally, it would be     
very interesting to have a direct measurement of $F_L(x,Q^2)$ at     
small $x$, as this is an excellent test of different approaches to     
incorporating small $x$ corrections.     
    
In summary we have used all the deep inelastic and hard scattering data in order to  
obtain what we believe is the most precise determination of parton distributions of the  
proton to date.  On the theoretical side this illuminates the short-distance structure of  
the proton, while at the same time it provides an essential ingredient for extracting  
new physics from current and forthcoming hadron colliders. \\ 
    
\noindent {\large \bf Acknowledgements}    
    
We thank Arie Bodek, Chuck Brown, Antje Bruell, Albert De Roeck, Jerry Garvey,  
Joey Huston, Vladimir Shekelyan,
Mike Vetterli and Manuella Vincter for valuable discussions and  
information concerning the data.
This work was supported in part by the EU Fourth Framework
Programme `Training and Mobility of Researchers', Network
`Quantum Chromodynamics and the Deep Structure of
Elementary Particles', contract FMRX-CT98-0194 (DG 12 - MIHT).

\newpage                                                                                     
                                                                                     
\noindent TABLE I. The numerical values of the parameters of the starting                                                                                
distributions of three parton sets which differ in the value of the initial                                                                                
state partonic $\left<k_T\right>$ used to describe the prompt photon data.                                                                                
The three columns correspond to $\left<k_T\right>=$ 0.64, 0.4 and 0~GeV                                                                                
for WA70 respectively. Note that $A_g$ is fixed by the momentum sum rule and that                                                                                
$A_u, A_d$ are fixed by flavour sum rules, and are not therefore free                                                                                
parameters.                                                                              
                                                                               
\vspace{1cm}                                                                               
\begin{center}                                                                               
\begin{tabular}{|cccc|}\hline                                                                               
                & Lower gluon & Central gluon & Higher gluon \\                                                                               
       & MRST($\gdown$)  & MRST          & MRST($\gup$)\\ \hline                                                                               
$(A_g)$         & 89.32       & 64.57         & 152.1        \\                                                                               
$(A_u)$         & 0.8884      & 0.6051        & 0.7763       \\                                                                               
$(A_d)$         & 0.05950     & 0.05811       & 0.06015      \\                                                                                
$\lambda_g$     & $-$1.082    & $-$0.9171     & $-$1.035     \\                                                                                 
$\eta_g$        & 6.124       & 6.587         & 7.451        \\                                                                               
$\epsilon_g$    & $-$2.409    & $-$3.168      & $-$4.341     \\                                                                               
$\gamma_g$      & 1.562       & 3.251         & 5.251        \\                                                                               
                &             &               &              \\                                                                               
$\eta_1$        & 0.4710      & 0.4089        & 0.4398       \\                                                                               
$\eta_2$        & 3.404       & 3.395         & 3.427        \\                                                                               
$\epsilon_u$    & 1.628       & 2.078         & 1.152        \\                                                                               
$\gamma_u$      & 9.628       & 14.56         & 12.36        \\                                                                               
$\eta_3$        & 0.2736      & 0.2882        & 0.2694       \\                                                                               
$\eta_4$        & 3.902       & 3.874         & 3.941        \\                                                                               
$\epsilon_d$    & 29.78       & 34.69         & 27.96        \\                                                                               
$\gamma_d$      & 35.09       & 28.96         & 38.35        \\                                                                               
                &             &               &              \\                                                                               
$A_S$           & 0.2699      & 0.2004        & 0.1786       \\                                                                               
$\lambda_S$     & 0.2410      & 0.2712        & 0.2819       \\                                                                               
$\eta_S$        & 7.549       & 7.808         & 8.212        \\                                                                               
$\epsilon_S$    & 0.2062      & 2.283         & 3.725        \\                                                                               
$\gamma_S$      & 18.35       & 20.69         & 21.80        \\                                                                               
                &             &               &              \\                                                                               
$A_\Delta$      & 0.1494      & 1.290         & 1.260        \\                                                                               
$\eta_\Delta$   & 0.6440      & 1.183         & 1.157        \\                                                                               
$\gamma_\Delta$ & 42.94       & 9.987         & 9.778        \\                                                                               
$\delta_\Delta$ & $-$100.8    & $-$33.34      & $-$30.83     \\ \hline                                                                               
\end{tabular}                                                                               
\end{center}                                                                               
\vspace{1cm}

\newpage                                                                                     
                                                                               
\noindent TABLE II. The fractions of the total momentum of the proton carried by                                                                               
the various partons in the MRST set.                                                                               
\vspace{0.5cm}                                                                               
\begin{center}                                                                               
\begin{tabular}{|ccccccccc|}\hline                                                                               
$Q^2$~(GeV$^2$) & $u_v$  & $d_v$  & $2\bar{u}$ & $2\bar{d}$ & $2\bar{s}$ &                                                                                 
$2\bar{c}$ & $2\bar{b}$ & $g$ \\ \hline                                                                               
2             & 0.310 & 0.129 & 0.058 & 0.075 & 0.037 & 0.001 & 0.000 & 0.388 \\                                                                               
20            & 0.249 & 0.103 & 0.063 & 0.077 & 0.046 & 0.017 & 0.000 & 0.439 \\                                                                               
200           & 0.216 & 0.090 & 0.066 & 0.078 & 0.052 & 0.026 & 0.012 & 0.456 \\                                                                               
$2\times10^3$& 0.194 & 0.080 & 0.068 & 0.079 & 0.056 & 0.032 & 0.020 & 0.466 \\                                                                               
$2\times10^4$& 0.178 & 0.074 & 0.070 & 0.080 & 0.058 & 0.036 & 0.026 & 0.472 \\                                                                               
$2\times10^5$& 0.165 & 0.068 & 0.072 & 0.081 & 0.061 & 0.040 & 0.030 & 0.477 \\                                                                               
\hline                                                                               
\end{tabular}                                                                               
\end{center}                                                                               
\vspace{2cm}

\newpage                                                                                    
\noindent TABLE III. Processes studied in the global analysis 
($^*$ indicates data fitted).                                                                                    
\vspace{0.5cm}                        
                                                                                                                                             
\centering                                                                                     
\begin{tabular}{|l|l|l|}    \hline                                                                                        
%                   &                           &                           \\                                                                                       
{\bf Process/}     &     {\bf Leading order}   & {\bf Parton behaviour probed}\\                                                                                       
{\bf Experiment}   &  {\bf subprocess}         &                           \\                                                                                       
%                   &                           &                  \\                       
\hline                                                                                       
&\hfill \raisebox{-0.5ex}[0.5ex][0.5ex]{$\btop$}&                      \\                                                                                       
{\bf DIS} $\mbox{\boldmath $(\mu N \rightarrow \mu X)$}$ &  $\gamma^*q                                                                                       
\rightarrow q$ \hfill {\arrayrulewidth=1pt\vline}\hspace*{4pt}&  \\                                                                                       
$F^{\mu p}_2,F^{\mu d}_2,F^{\mu n}_2/F^{\mu p}_2$                                                                                       
& \hfill {\arrayrulewidth=1pt\vline}\hspace*{4pt}&                                                                                         
Four structure                                                                                       
functions $\rightarrow$  \\                                                                                       
(SLAC, BCDMS,& \hfill {\arrayrulewidth=1pt\vline}\hspace*{4pt}& \hspace*{1cm}                                                                                        
$u + \bar{u}$  \\                                                                                       
NMC, E665)$^*$& \hfill {\arrayrulewidth=1pt\vline}\hspace*{4pt}&                                                                                       
\hspace*{1cm} $d + \bar{d}$   \\                                                                                       
      &\hfill $\bmid$ & \hspace*{1cm}  $\bar{u} + \bar{d}$  \\                                                                                      
{\bf DIS} $\mbox{\boldmath $(\nu N \rightarrow \mu X)$}$ & $W^*q \rightarrow                                                                                       
q^{\prime}$  \hfill {\arrayrulewidth=1pt\vline}\hspace*{4pt}& \hspace*{1cm}                                                                                         
$s$ (assumed = $\bar{s}$),\\                                                                                       
$F^{\nu N}_2,xF^{\nu N}_3$    &\hfill {\arrayrulewidth=1pt\vline}\hspace*{4pt}&                                                                                      
but only                                                                                       
$\int xg(x,Q_0^2)dx \simeq 0.35$ \\                                                                                       
(CCFR)$^*$     &\hfill {\arrayrulewidth=1pt\vline}\hspace*{4pt}&                                                                                       
and $\int(\bar{d}-\bar{u})dx\simeq 0.1$  \\                                                                                      
                &\hfill \raisebox{0.5ex}[1.5ex][1.5ex]{$\bbot$} &         \\                                                                                       
{\bf DIS (small $x$)}    &   $\gamma^* (Z^*)q \rightarrow q$   &   $\lambda$     \\                                                                                       
$F^{ep}_2$ (H1, ZEUS)$^*$   &                  & $(x\bar{q} \sim x^{-\lambda_S},                                                                                      
\ xg \sim  x^{-\lambda_g})$    \\[2mm]                                                                                      
%& & \\                       
{\bf DIS ($\mbox{\boldmath $F_L$}$)}    &   $\gamma^* g \rightarrow q\bar{q}$                     
&   $g$     \\                                                                                       
NMC, HERA   &                  &     \\[2mm]                                 
%                      &                         &                     \\                                                                                        
%{\bf DIS (GSR)}    &   $\gamma^*q \rightarrow q$                                                                                        
%& $\int_{0.004}^{0.8} (\bar u -\bar d)dx$ \\                                                                                       
%$\int (F^{\mu p}_2 - F^{\mu n}_2 )dx/x$    &    &   \\                                                                                       
%  (NMC)              &                            &                           \\                                                                                       
%               &                            &                           \\                                                                                       
$\mbox{\boldmath $\ell N \rightarrow c \bar{c} X$}$ &                                                                                      
  $\gamma^*  c\rightarrow  c$    & $c$ \\                                                                                       
 $F_2^{c}$  (EMC; H1, ZEUS)$^*$            &     &     $  \quad (x \gapproxeq 0.01;\                                                                                       
x \lapproxeq 0.01 )$    \\[2mm]                                                                                      
%               &                            &                           \\                                                                                       
$\mbox{\boldmath $\nu N \rightarrow \mu^+\mu^-X$}$ &  $W^* s \rightarrow                                                                                        
c$    & $s \approx \frac{1}{4} (\bar{u} + \bar{d}) $ \\                                                                                       
(CCFR)$^*$            &   $\;\;\;\;\;\;\;\;\;\;\;\;\;\hookrightarrow \mu^+$     &         \\[2mm]                                                                                      
%                      &                         &                     \\                                                                                        
$\mbox{\boldmath $p N \rightarrow \gamma X$}$  &  $qg \rightarrow \gamma q$  &                                                                                      
$g$ at $x \simeq 2 p_T^\gamma/\sqrt{s} \rightarrow$ \\                                                                                       
(WA70$^*$, UA6, E706, \ldots)    &      &  \hspace*{.5cm}   $x \approx 0.2 - 0.6  $            
\\[2mm]                                                                                      
 %          &                   &                          \\                                                                                       
$\mbox{\boldmath $pN \rightarrow \mu^+\mu^- X$}$   &  $q\bar{q} \rightarrow                                                                                       
\gamma^*$  &  $\bar{q} = ...(1-x)^{\eta_S}$ \\                                                                                       
(E605, E772)$^*$          &                     &                             \\[2mm]                                                                                      
%               &                     &                                 \\                                                                                       
$\mbox{\boldmath $pp, pn \rightarrow \mu^+\mu^- X$}$ & $u\bar{u},d\bar{d}                                                                                       
\rightarrow \gamma^*$   & $\bar{u} - \bar{d}~~(0.04 \lapproxeq x \lapproxeq 0.3)$                    
\\                                                                                       
(E866, NA51)$^*$             &  $u\bar{d},d\bar{u} \rightarrow \gamma^*$   &     \\[2mm]                                                                                     
%                 &                   &                               \\                                                                                       
$\mbox{\boldmath $ep, en \rightarrow e \pi X$}$    &  $\gamma^* q \rightarrow q$         
with   &   $\bar{u} - \bar{d}~~(0.04 \lapproxeq x  \lapproxeq 0.2)$     \\                                                                                       
(HERMES)   & $q = u, d, \bar{u}, \bar{d}$ &     \\[2mm]                                
%     &     &    \\                        
$\mbox{\boldmath $p\bar{p} \rightarrow WX(ZX)$}$    &  $ud \rightarrow W$                                                                                       
&  $u,d$ at $x \simeq M_W/\sqrt{s} \rightarrow$  \\                                                                                       
(UA1, UA2; CDF, D0)          &   & \hspace*{.5cm}  $x \approx 0.13;~0.05$ \\[2mm]                              
%&    &   \\                        
$\;\;\mbox{\boldmath $\rightarrow \ell^{\pm}$}$ {\bf asym} (CDF)$^*$   &         &                                                                                       
slope of $u/d$ at $x \approx 0.05 - 0.1$  \\[2mm]                                                                                      
%                 &                   &                               \\                                                                                      
$\mbox{\boldmath $p\bar{p} \rightarrow t\bar{t} X$}$ & $q\bar{q}, gg \rightarrow          
t\bar{t}$ & $q, g$ at $x \gapproxeq 2m_t/\sqrt{s} \simeq 0.2$ \\         
(CDF, D0) &                       & \\[2mm]         
% & &           
$\mbox{\boldmath $p\bar{p} \rightarrow ${\bf jet}$\, + \, X$}$    &                                                                                       
 $gg,qg,qq\rightarrow 2j$  &  $q,g$                                                                                       
at $x \simeq 2 E_T/\sqrt{s} \rightarrow$  \\                                                                                       
(CDF, D0)         &   & \hspace*{.5cm}  $x \approx 0.05 -  0.5$ \\ \hline                                                                                      
%                         &                         &          \\                                                                                      
%\hline                                                                                         
\end{tabular}

\newpage                                                                                    
        
\noindent TABLE IV. The $\chi^2$ values for the DIS data included in the three                                                                                
global fits which resulted in the parameter values listed in Table I.                                                                               
\vspace{0.5cm}                                                                               
\begin{center}                                                                               
\begin{tabular}{|ccccc|}\hline                                                                               
Data set          & No. of   & MRST & MRST($\gup$)&                                                                               
MRST($\gdown$)\\                                                                               
                  & data pts &      &          &           \\ \hline                                                                               
H1 $ep$           & 221      & 164  & 166      & 161       \\                                                                               
ZEUS $ep$         & 204      & 269  & 273      & 258       \\                                                                               
BCDMS $\mu p$     & 174      & 248  & 239      & 264       \\                                                                               
NMC $\mu p$       & 130      & 141  & 148      & 142       \\                                                                               
NMC $\mu d$       & 130      & 101  & 107      & 104       \\                                                                               
SLAC $ep$         & 70       & 119  & 104      & 135       \\                                                                               
E665 $\mu p$      & 53       & 59   & 54       & 56        \\                                                                               
E665 $\mu d$      & 53       & 61   & 62       & 61        \\                                                                               
CCFR $F_2^{\nu N}$& 66       & 93   & 102      & 92        \\                                                                               
CCFR $F_3^{\nu N}$& 66       & 68   & 69       & 67        \\                                                                               
NMC \it{n/p}      & 163      & 186  & 192      & 174       \\ \hline                                                                               
\end{tabular}                                                                               
\end{center}                                                                               
{\footnotesize                                                                               
\begin{tabular}{lrp{5in}}                                                                               
Notes: & (i)  & The relative normalizations of the data sets are taken to be                                                                               
                unity, except that the BCDMS $\mu p$ data is normalized down                                                                                
                by 2\% and the SLAC $ep$ data up by 2.5\%. \\                                                                               
       & (ii) & The CCFR data are corrected for heavy (iron) target effects                                                                                
                using the information obtained from the muon-nucleus                                                                                
                measurements. Since the $x<0.1$ corrected CCFR data are in                                                                                
                disagreement with the NMC data, they are omitted from the fit.                                                                                
                Only statistical errors together with a 1.5\% uncertainty (to                                                                               
                represent uncertainty in the heavy target correction) are                                                                                
                included in the $\chi^2$ for the                                                                                
                CCFR data since no overall systematic errors are given. \\                                                                               
       & (iii)& All deuterium data are corrected for shadowing effects using                                                                                
                the method of Badelek and Kwiecinski \cite{BK}. \\                                                                               
\end{tabular}}

\newpage 
         
\begin{figure}[H]
%\vspace{-0.5cm}                                                                
\begin{center}     
\epsfig{figure=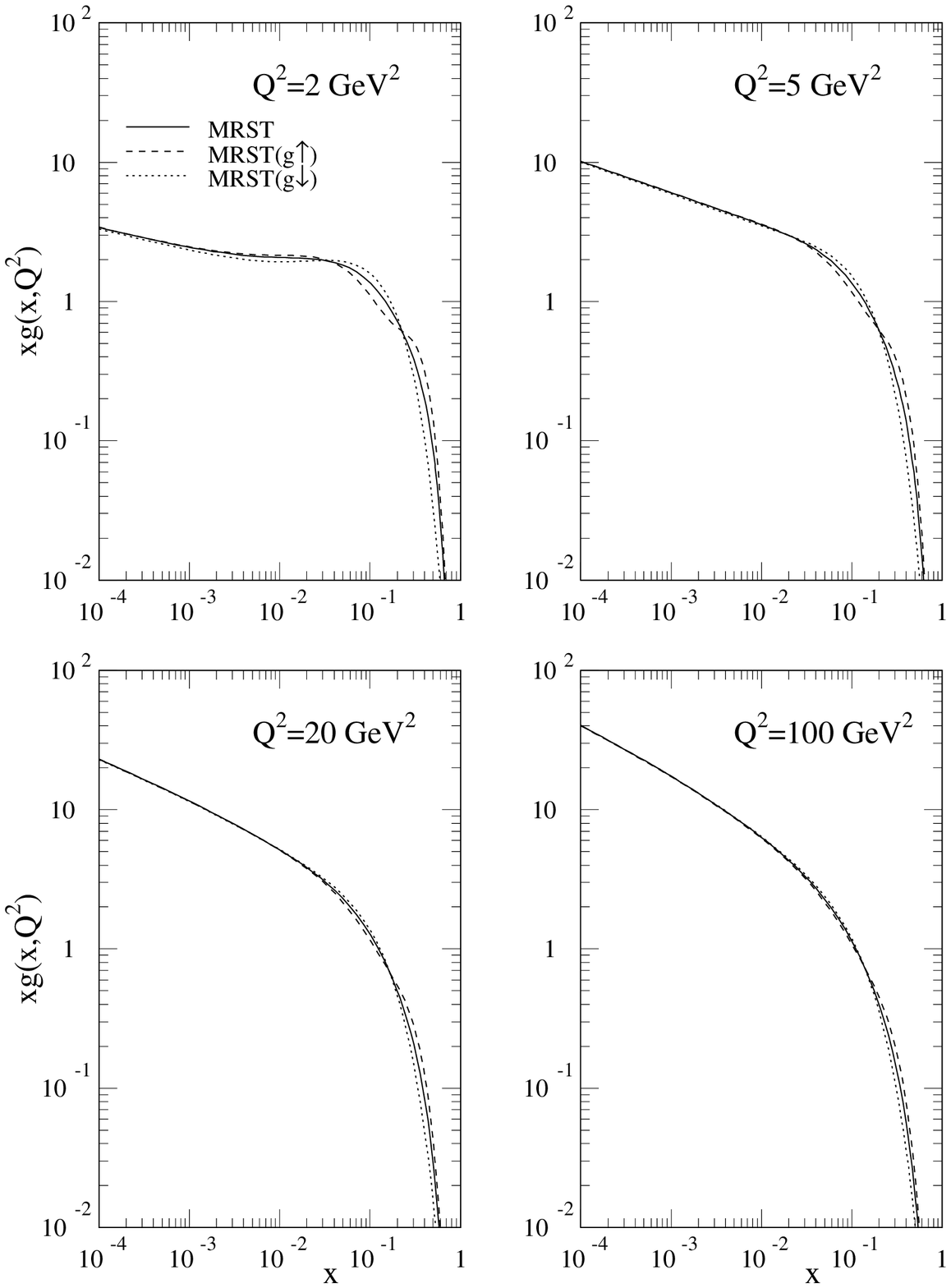,height=20cm}    
\end{center}    
\caption{The gluon distributions at $Q^2$ = 2, 5, 20 and 100~GeV$^2$    
corresponding to the     
MRST, MRST($\gup$) and MRST($\gdown$) sets of partons with, respectively,     
the central, larger and smaller gluon at large $x$.    
We take MRST as the default set of partons throughout the paper.}    
\label{fig:gluon}
\end{figure}                                                                 
\newpage    
                 
\begin{figure}[H]
%\vspace{-0.5cm}                                                                
\begin{center}     
\epsfig{figure=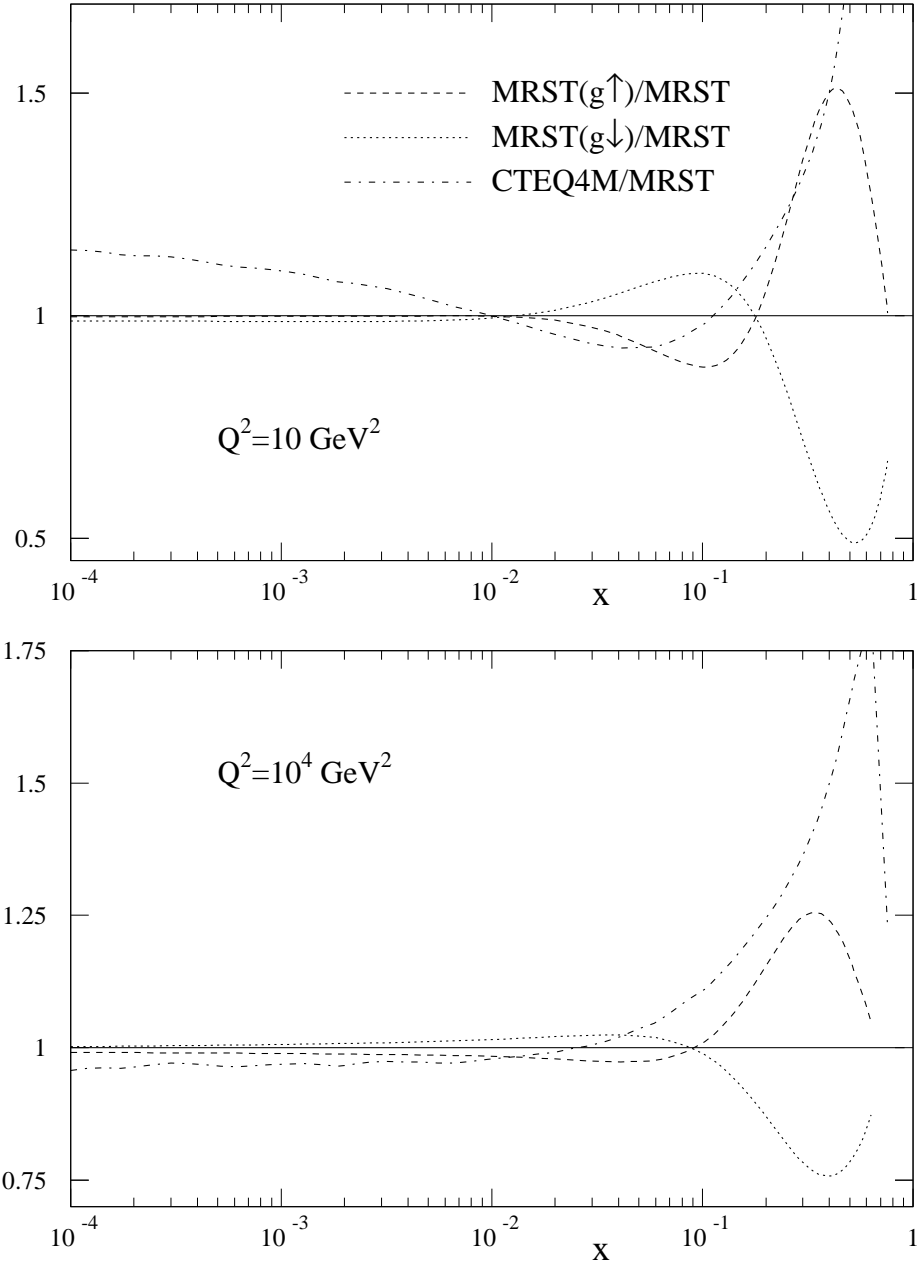,height=20cm}    
\end{center}    
\caption{The ratios of the $\gup$ and $\gdown$ gluons to the    
`central' gluon (MRST) at $Q^2$ = 10 and 10$^4$~GeV$^2$. For comparison the    
ratio of the CTEQ4M \protect\cite{CTEQ4M} gluon to our central gluon is     
also shown.}     
\label{fig:gluonratio}    
\end{figure}                                                                    
\newpage    
                                                                 
\begin{figure}[H]                                                               
%\vspace{-0.5cm}                                                                
\begin{center}     
\epsfig{figure=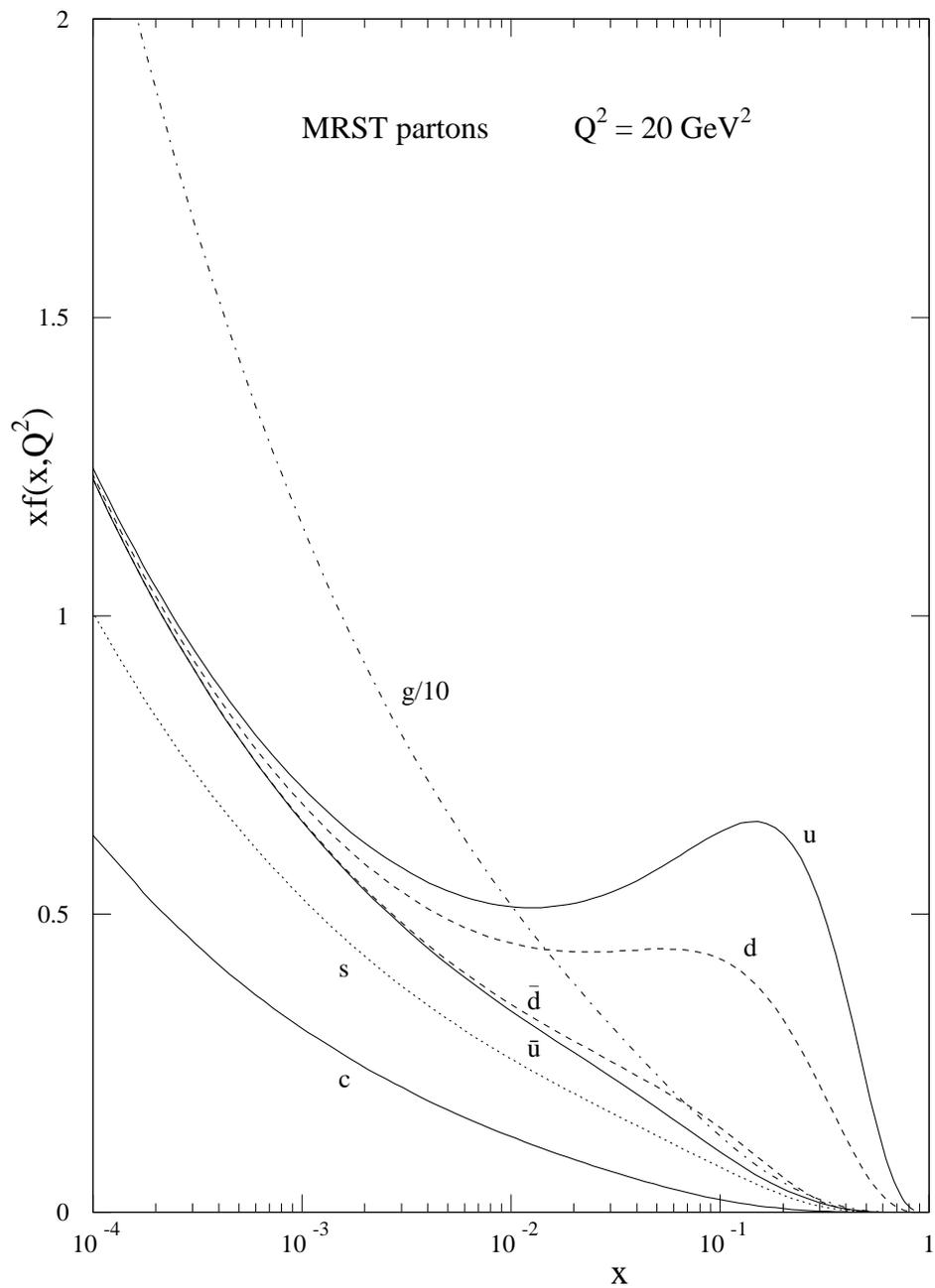,height=20cm}    
\end{center}    
\caption{MRST partons at $Q^2 = 20$~GeV$^2$.}    
\label{fig:MRSTpartsa}                            
\end{figure}                                                                    
\newpage    
    
\begin{figure}[H]                                                               
%\vspace{-0.5cm}                                                                
\begin{center}     
\epsfig{figure=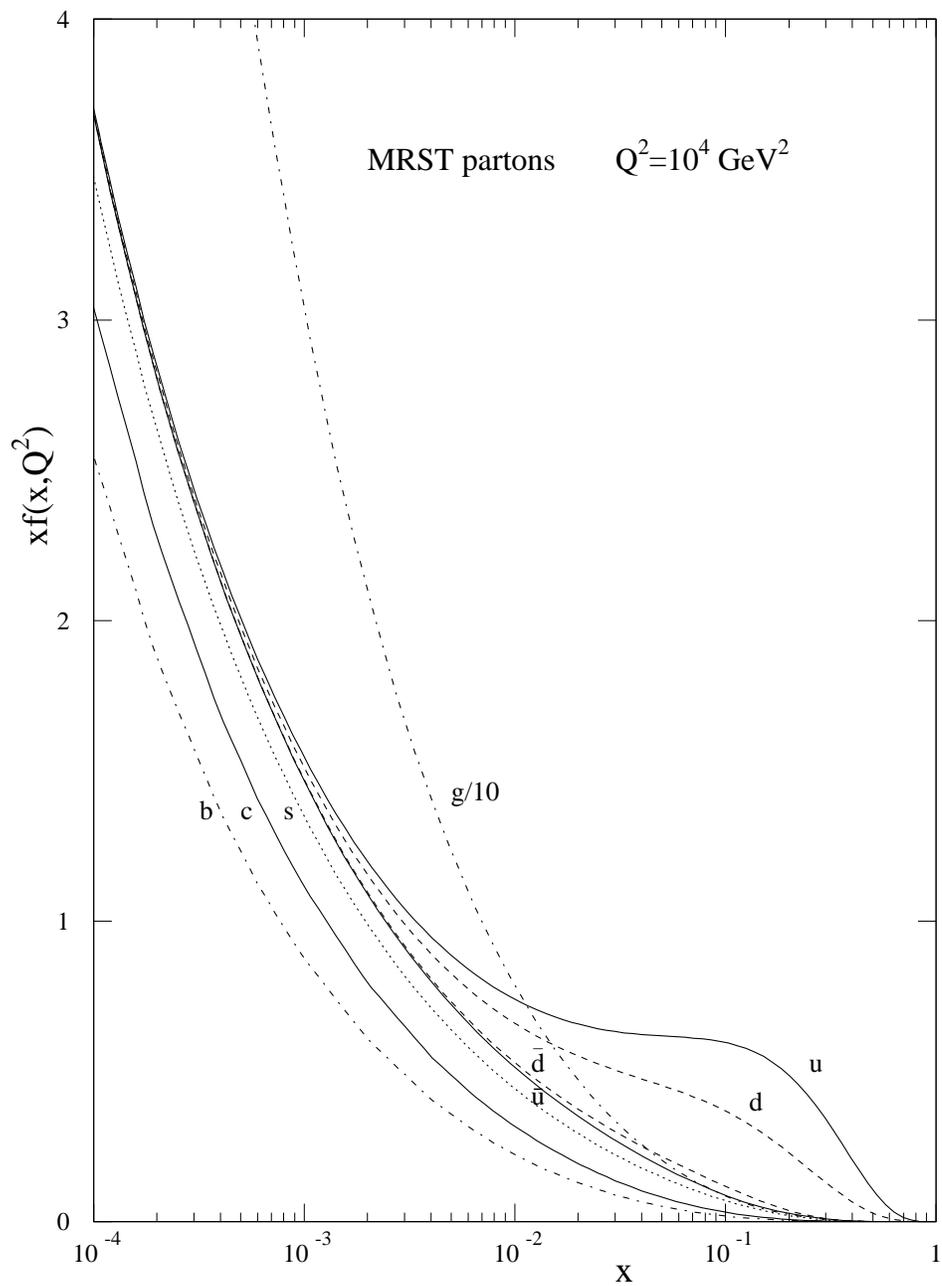,height=20cm}    
\end{center}    
\caption{MRST partons at $Q^2 = 10^4$~GeV$^2$.}    
\label{fig:MRSTpartsb}                            
\end{figure}                                                                    
\newpage 
   
\begin{figure}[H]                                                               
%\vspace{-0.5cm}                                                                
\begin{center}     
\epsfig{figure=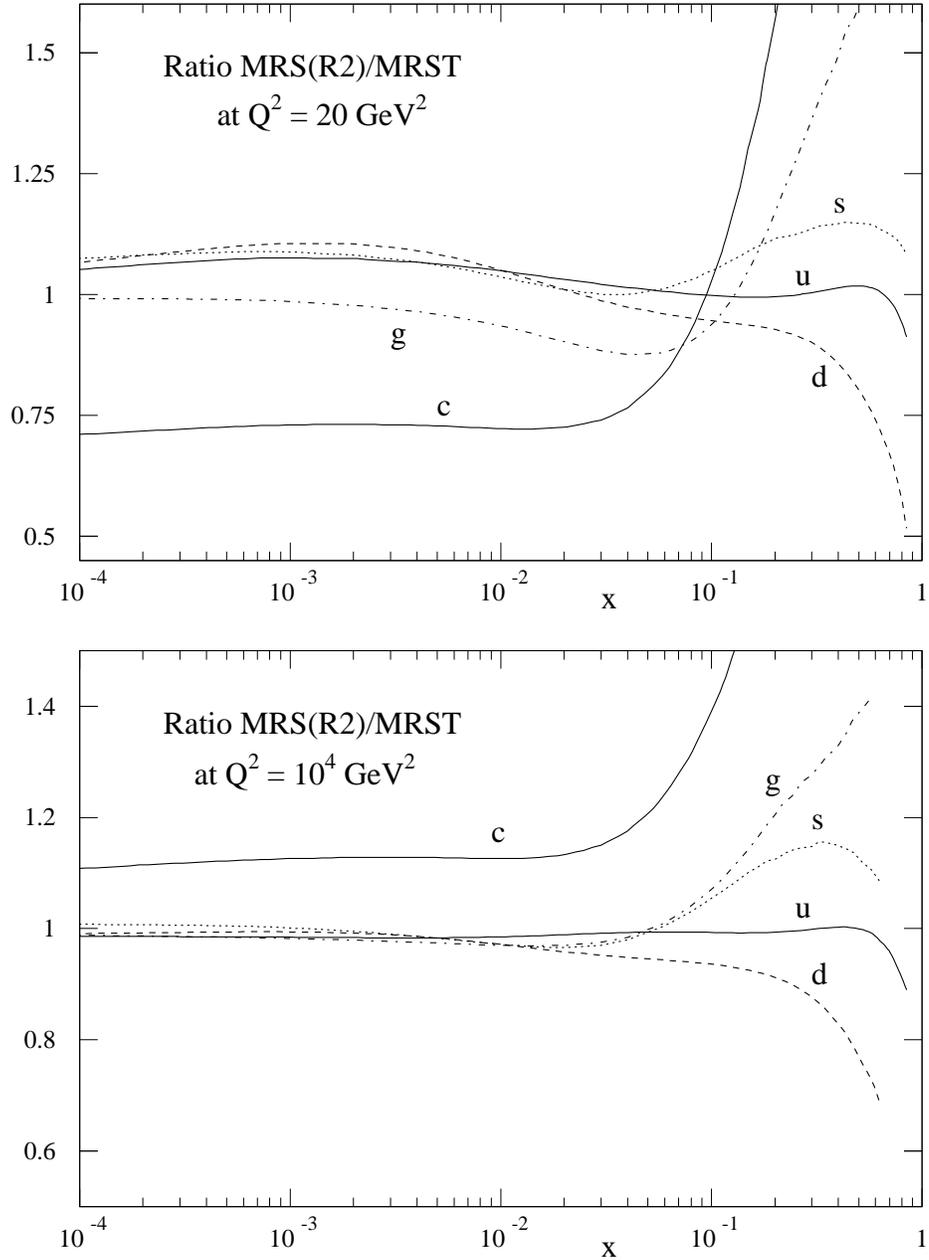,height=20cm}    
\end{center}    
\caption{Comparison of the MRST partons
with those of the previous MRS(R2) set at 
$Q^2 = 20$~GeV$^2$ (upper plot) and
$Q^2 = 10^4$~GeV$^2$ (lower plot).}  
\label{fig:MRSTvsR2}                            
\end{figure}                                                                    
\newpage    
                                                                 
\begin{figure}[H]
%\vspace{-0.5cm}                                                                
\begin{center}     
\epsfig{figure=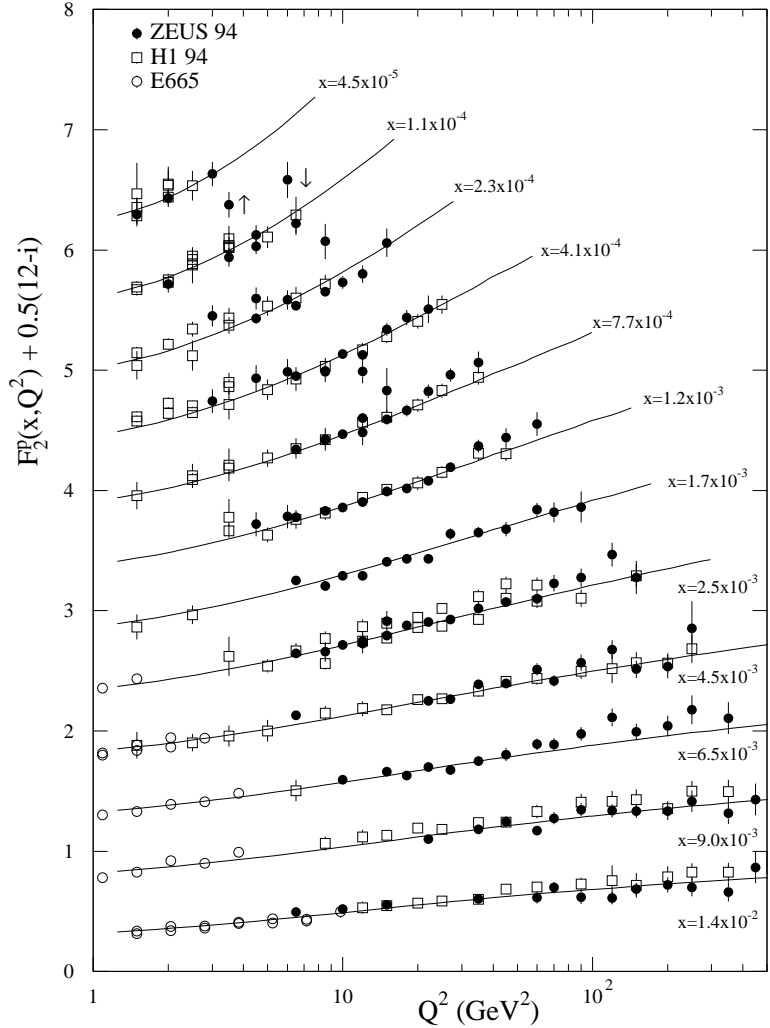,height=16cm}    
\end{center}    
\vspace{-0.5cm}                                                                
\caption{The description of the $F_2^p$ data at small $x$ by the MRST set of partons.
The comparison is made at twelve values    
of $x$ chosen to be the most appropriate for the new HERA data.    
For display purposes we add $0.5(12-i)$ to $F_2^p$ each time    
the value of $x$ is decreased, where $i=1,12$.    
The experimental data are assigned to the $x$ value which is    
closest to the experimental $x$ bin. Thus the ZEUS data \protect\cite{ZEUS}    
are shown in groupings with $x$ values     
$( 3.5,6.3,6.5     \times  10^{-5})$,    
$( 1.02,1.20     \times  10^{-4})$,    
$( 1.98,2.53      \times  10^{-4})$,    
$( 4.0,4.5     \times  10^{-4})$,    
$( 6.32,8.00     \times  10^{-4})$,    
$(1.02,1.20      \times  10^{-3})$,    
$(1.612     \times  10^{-3})$,    
$(2.53,2.60      \times  10^{-3})$,    
$(4.00     \times  10^{-3})$,    
$( 6.325     \times  10^{-3})$,    
$(1.02     \times  10^{-2})$,    
$(1.612      \times  10^{-2})$,    
and the H1 data \protect\cite{H1} in groupings with $x$ values    
$( 3.2,5.0     \times  10^{-5})$,    
$( 0.80,1.30     \times  10^{-4})$,    
$( 2.0,2.5      \times  10^{-4})$,    
$( 3.2,5.0     \times  10^{-4})$,    
$( 6.3,8.0     \times  10^{-4})$,    
$(1.3     \times  10^{-3})$,    
$(1.585      \times  10^{-3})$,    
$(2.0,2.5,3.2      \times  10^{-3})$,    
$(3.98,4.0,5.0     \times  10^{-3})$,    
$( 6.3     \times  10^{-3})$,    
$(8.0     \times  10^{-3})$,    
$(1.3      \times  10^{-2})$.    
The E665 data \protect\cite{E665}, which are shown on the curves with the    
five largest $x$ values, are measured at $x=    
 (2.46     \times  10^{-3})$,    
$(3.698,5.2    \times  10^{-3})$,    
$( 6.934     \times  10^{-3})$,    
$(8.933     \times  10^{-3})$,    
$(1.225,1.73      \times  10^{-2})$.}    
\label{fig:DISsmallx}        
\end{figure}                                             
\newpage    
                            
\begin{figure}[H]
%\vspace{-0.5cm}                                                                
\begin{center}     
\epsfig{figure=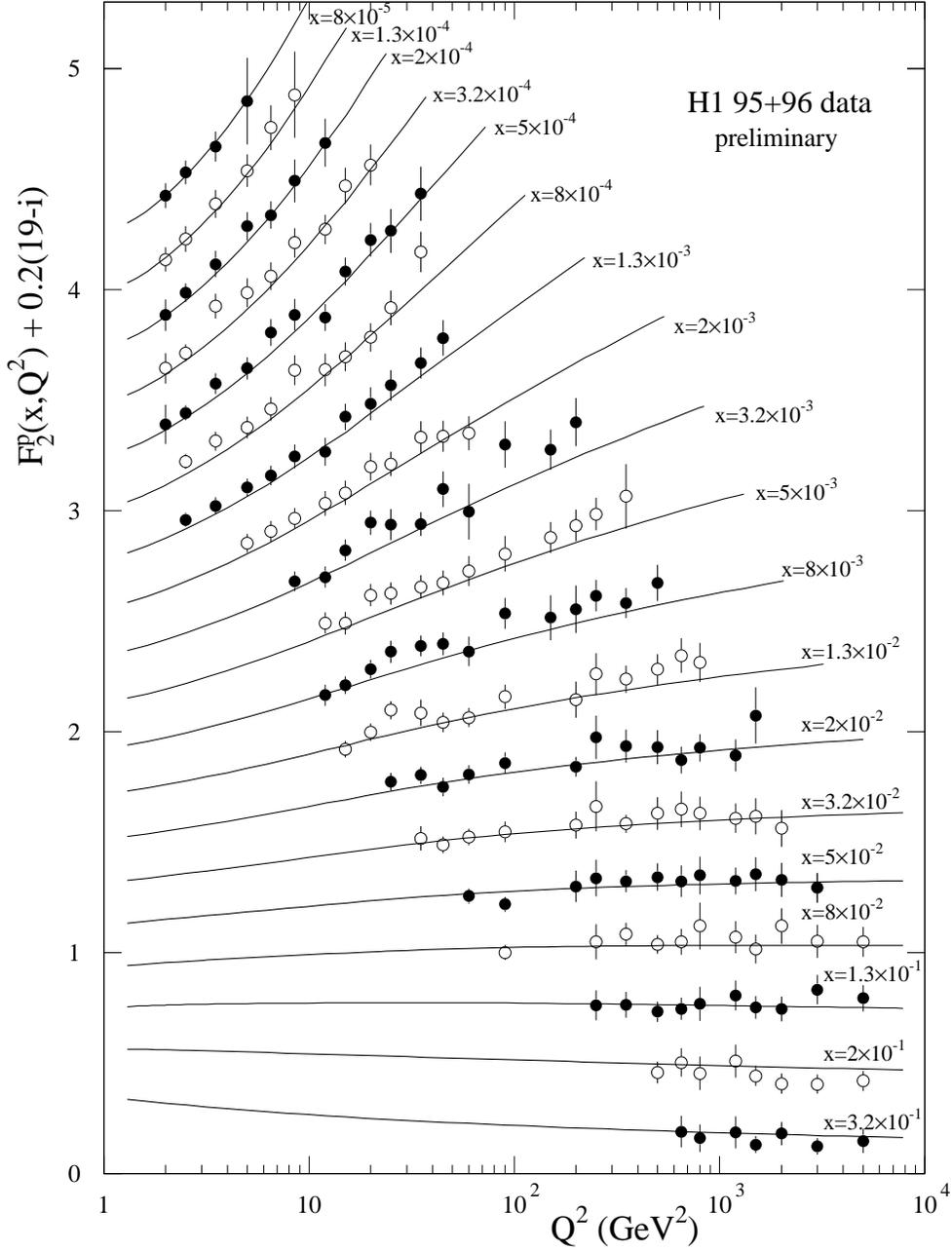,height=20cm}    
\end{center}    
\caption{Comparison of the $F_2^p$ predictions of the MRST partons with the     
preliminary
1995 and 1996 nominal vertex data of H1 \protect\cite{F2H196}.    
For display purposes we add $0.2(19-i)$ to $F_2^p$ each time    
the value of $x$ is decreased, where $i=1,19$. These data are not used in    
the global analysis.}    
\label{fig:F2H196}
\end{figure}                                                                   
\newpage    
                                                                 
\begin{figure}[H] 
%\vspace{-0.5cm}                                                                
\begin{center}     
\epsfig{figure=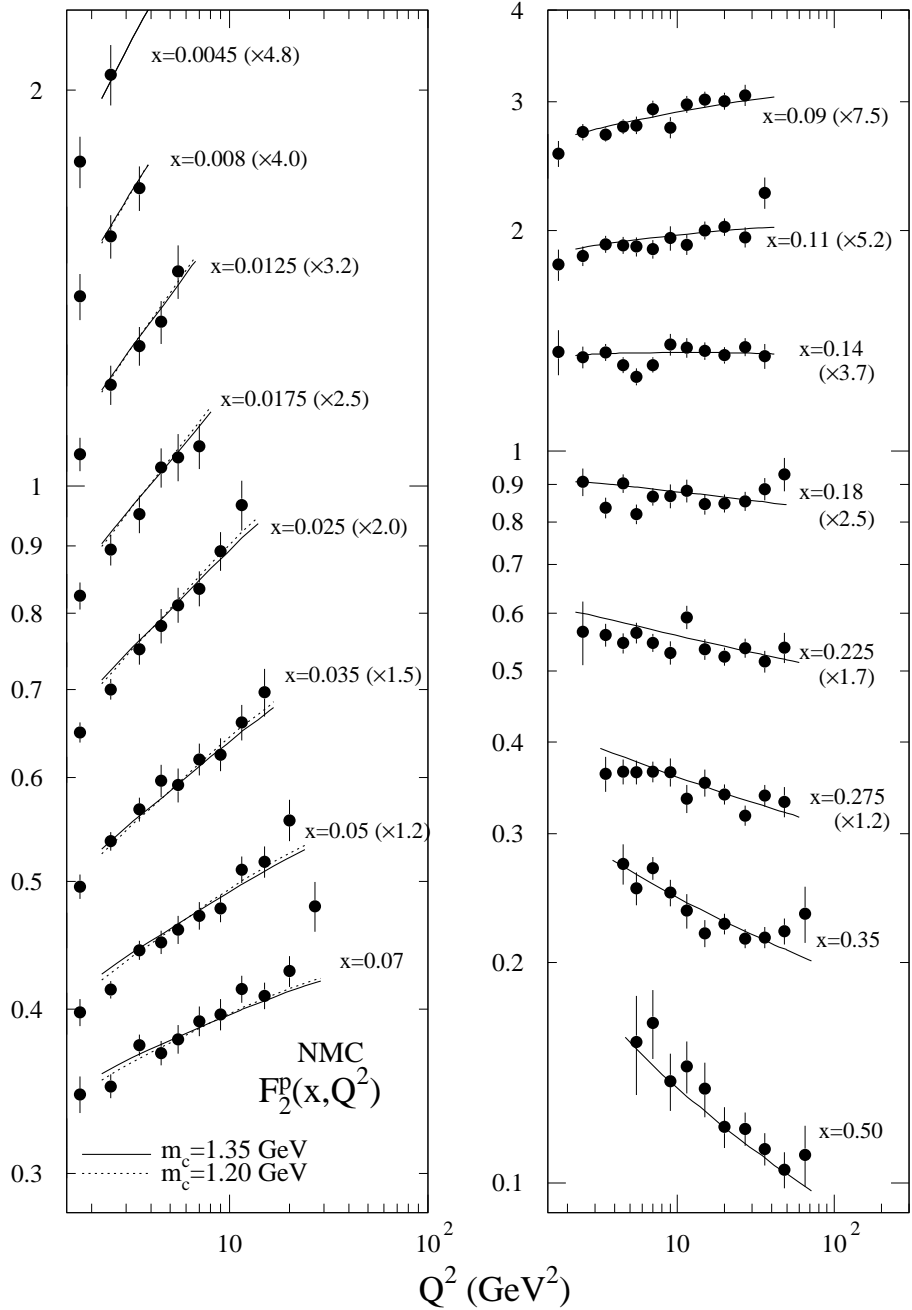,height=20cm}    
\end{center}    
\caption{Description of the NMC  $F_2^p$ data \protect\cite{NMC}    
 by the MRST partons. The effect of lowering the charm     
quark mass from 1.35 to 1.20 GeV is shown by the dotted curve. For display    
purposes we have multiplied $F_2^p$ by the numbers shown in brackets.}    
\label{fig:NMCa}  
\end{figure}                                                                 
\newpage    
                                                                 
\begin{figure}[H] 
%\vspace{-0.5cm}                                                                
\begin{center}     
\epsfig{figure=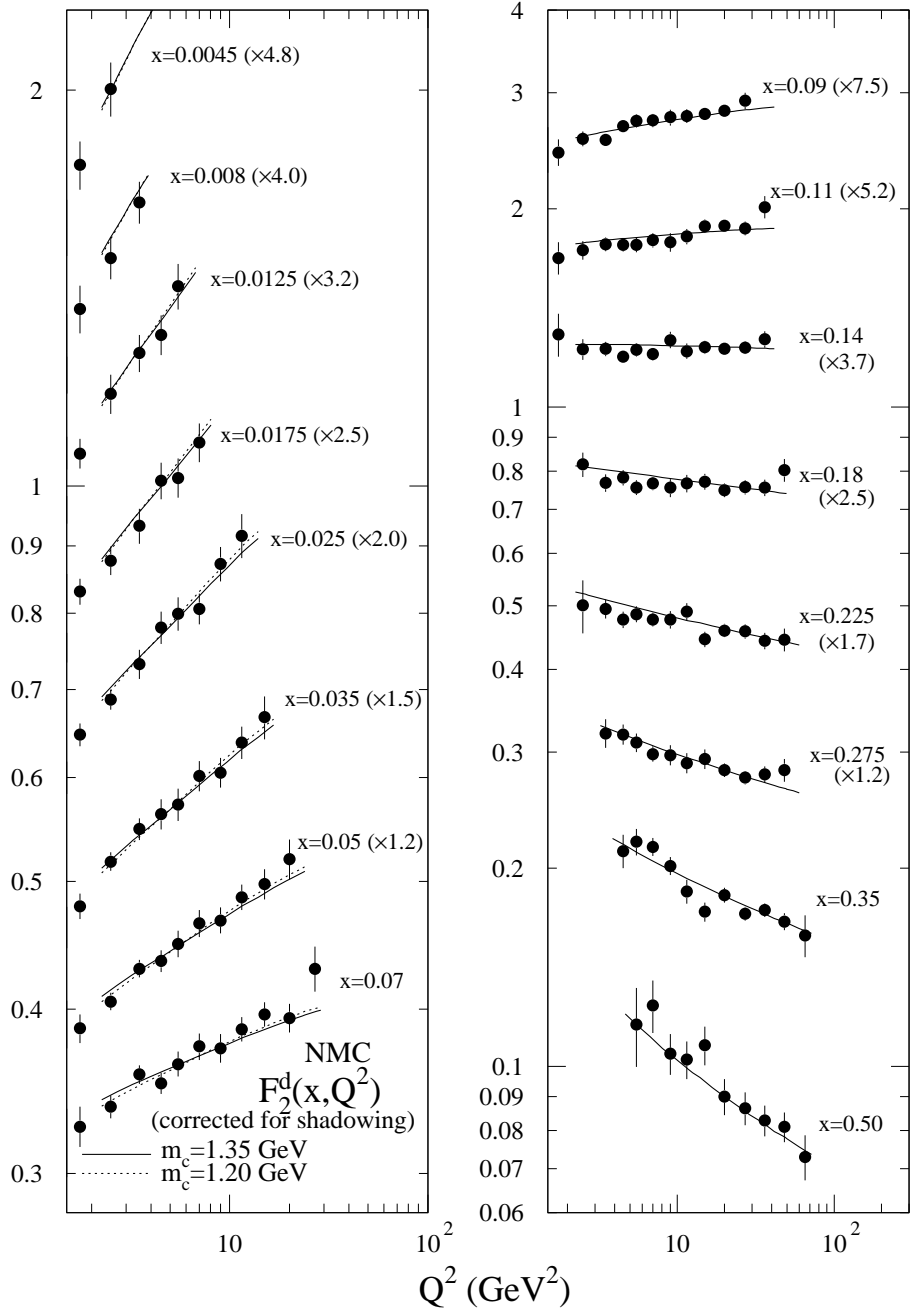,height=20cm}    
\end{center}    
\caption{Comparison of $F_2^d$ from the MRST set with the     
NMC deuteron data~\protect\cite{NMC}. The effect of lowering the charm     
quark mass from 1.35 to 1.20 GeV is shown. For display    
purposes we have multiplied $F_2^d$ by the numbers shown in brackets.}    
\label{fig:NMCb}  
\end{figure}                                                                 
\newpage    
                                                                 
\begin{figure}[H] 
%\vspace{-0.5cm}                                                                
\begin{center}     
\epsfig{figure=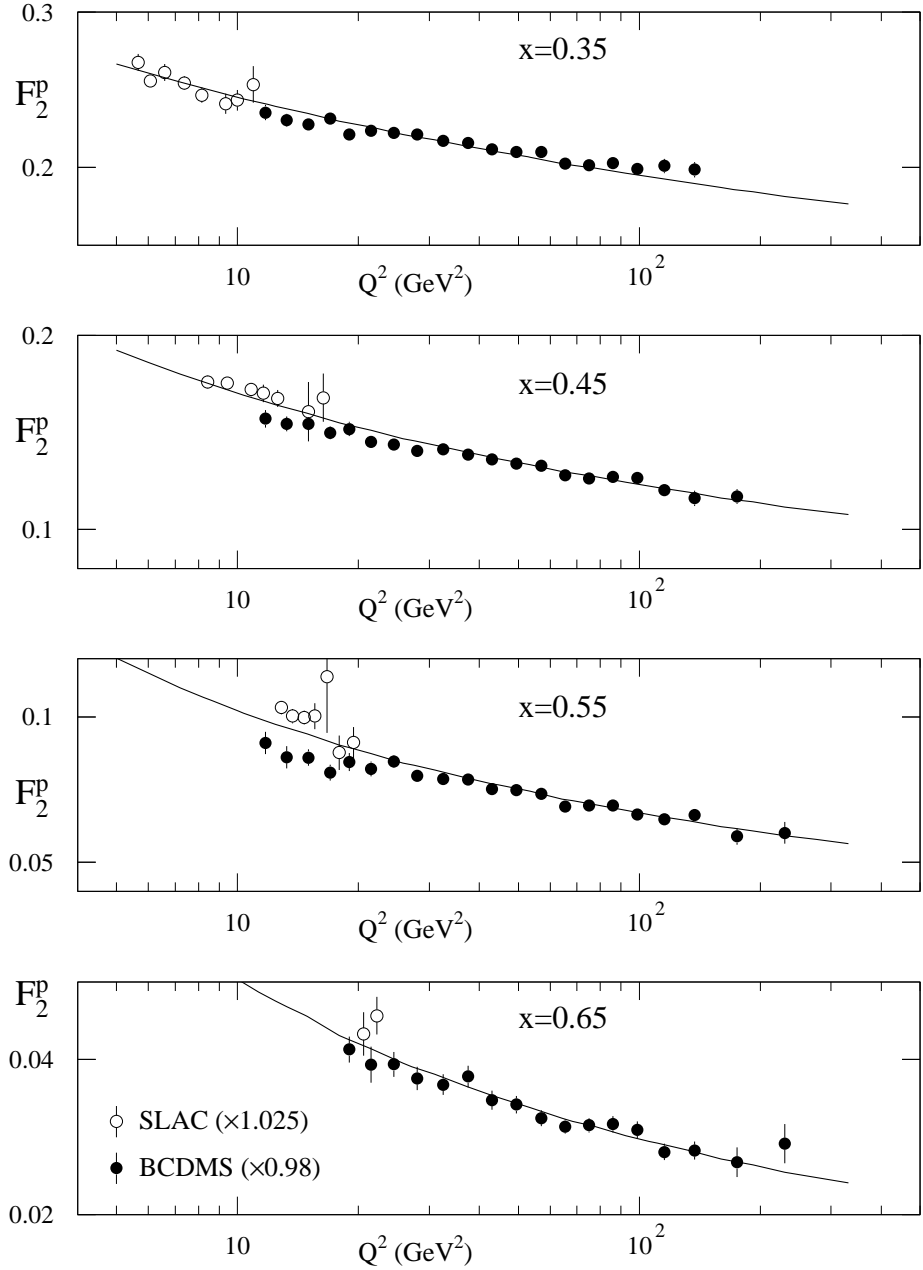,height=20cm}    
\end{center}    
\caption{The description of the large $x$ BCDMS~\protect\cite{BCDMS}     
and SLAC~\protect\cite{SLAC} measurements of $F_2^p$ by the MRST partons.}      
\label{fig:BCDMS} 
\end{figure}                                                                 
\newpage    
                                                                 
\begin{figure}[H] 
%\vspace{-0.5cm}                                                                
\begin{center}     
\epsfig{figure=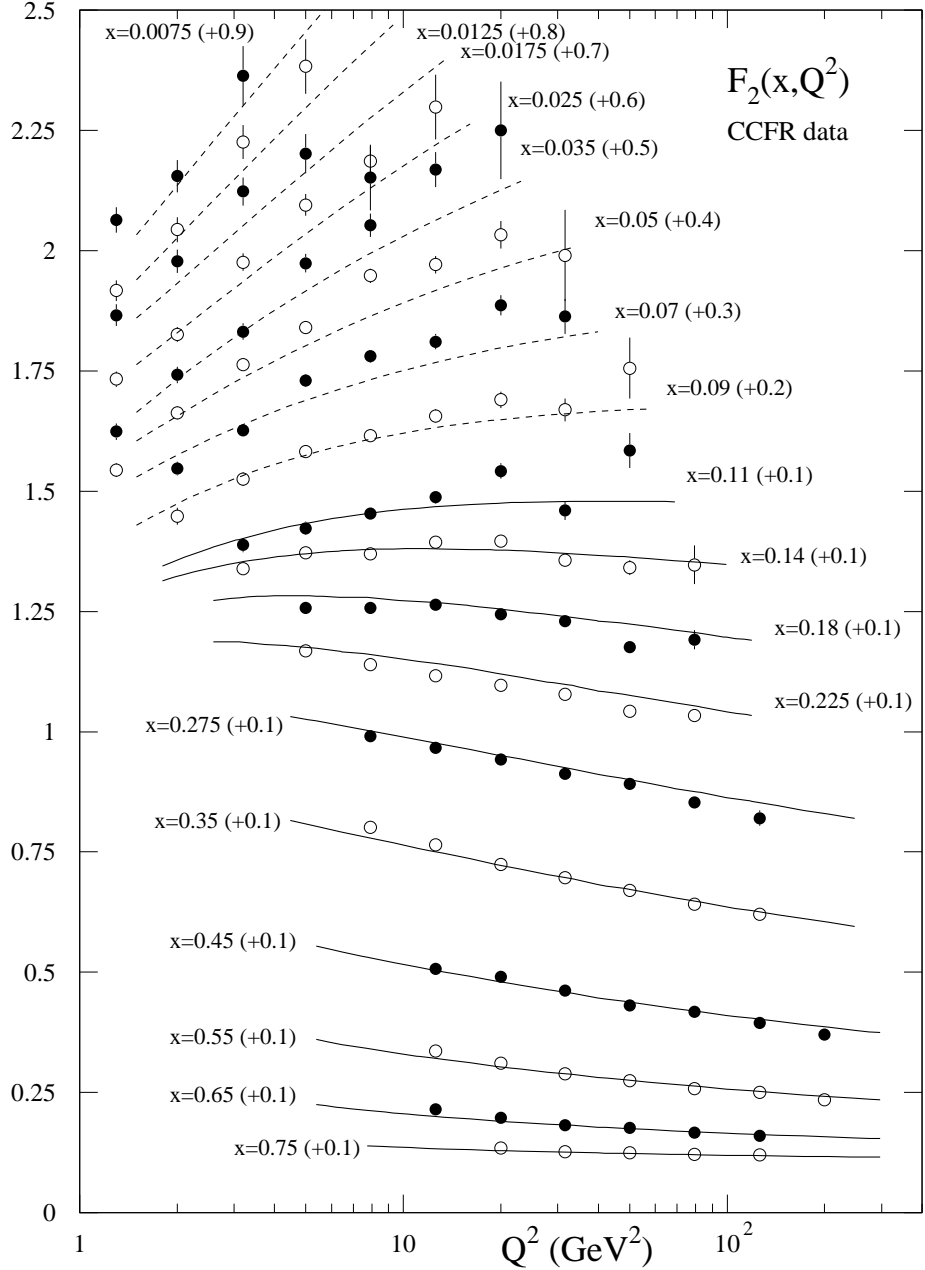,height=20cm}    
\end{center}    
\vspace{-0.5cm}                                                                
\caption{The description of the CCFR~\protect\cite{CCFR2} measurements    
of $F_2^{\nu N}$ by the MRST partons.  Only the data with $x > 0.1$ are     
included in the global fit. An $x$-dependent heavy target correction has     
been applied.    
The errors shown correspond to the quoted statistical errors together with a 
1.5\% uncertainty to represent the uncertainty of the heavy target correction.
 For display    
purposes we have added  to $F_2^{\nu N}$  the numbers shown in brackets.}    
\label{fig:CCFRa}
\end{figure}                                 
\newpage    
                                             
\begin{figure}[H]                            
%\vspace{-0.5cm}                                                                
\begin{center}     
\epsfig{figure=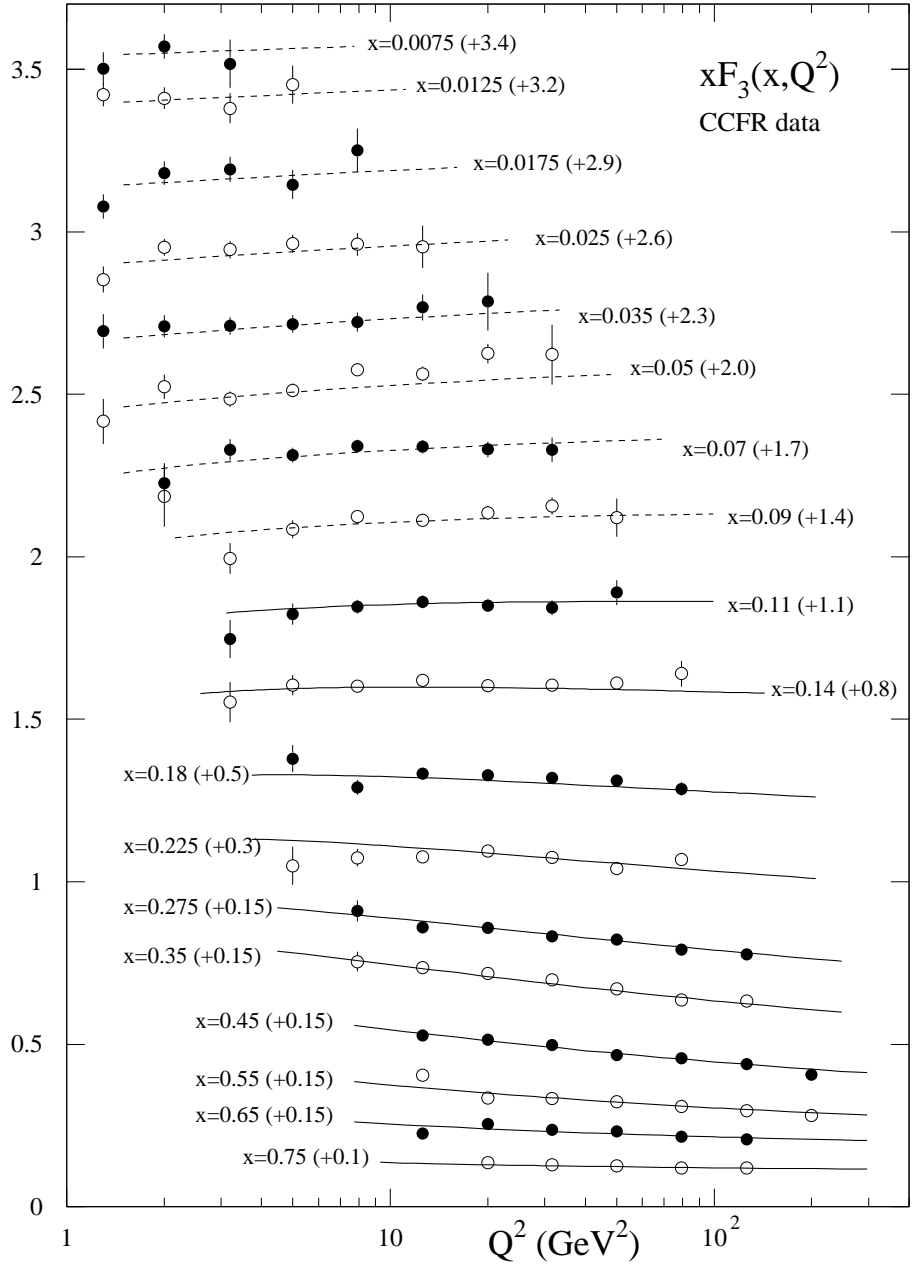,height=20cm}    
\end{center}    
\caption{The same as Fig.~\protect\ref{fig:CCFRa} but for the    
structure function $xF_3^{\nu N}$.}    
\label{fig:CCFRb}
\end{figure}                                                                 
\newpage    
                                                                 
\begin{figure}[H]
%\vspace{-0.5cm}                                                                
\begin{center}     
\epsfig{figure=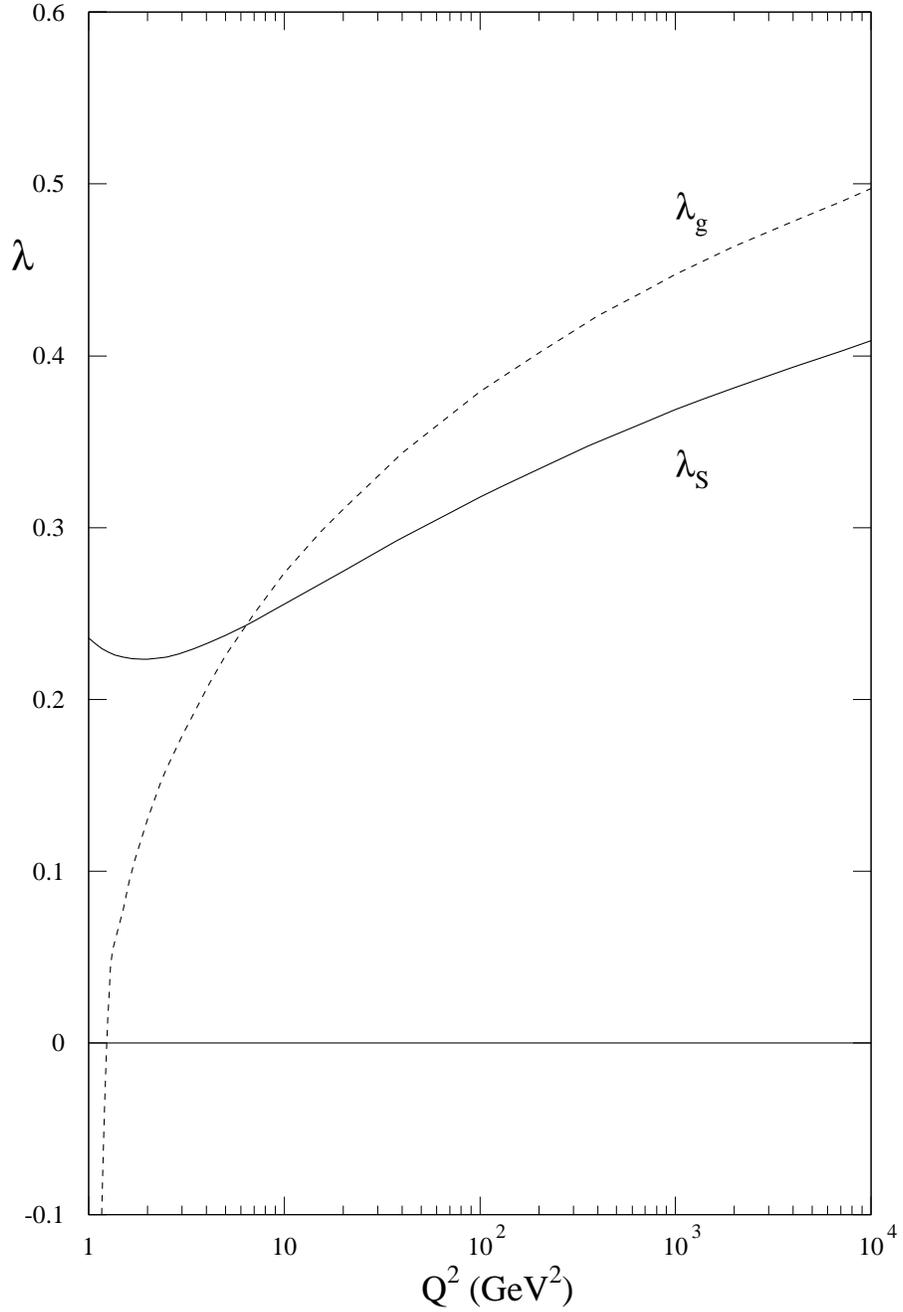,height=20cm}    
\end{center}    
\caption{The effective exponents $\lambda_S, \lambda_g$ of the     
small $x$ behaviour of the sea-quark and gluon distributions     
of the default MRST partons versus $Q^2$, defined such that    
$xS \sim x^{-\lambda_S}$ and $xg \sim x^{-\lambda_g}$.}     
\label{fig:lambda}
\end{figure}                                                                 
\newpage    
                                                                 
\begin{figure}[H] 
\vspace{-0.5cm}                                                                
\begin{center}     
\epsfig{figure=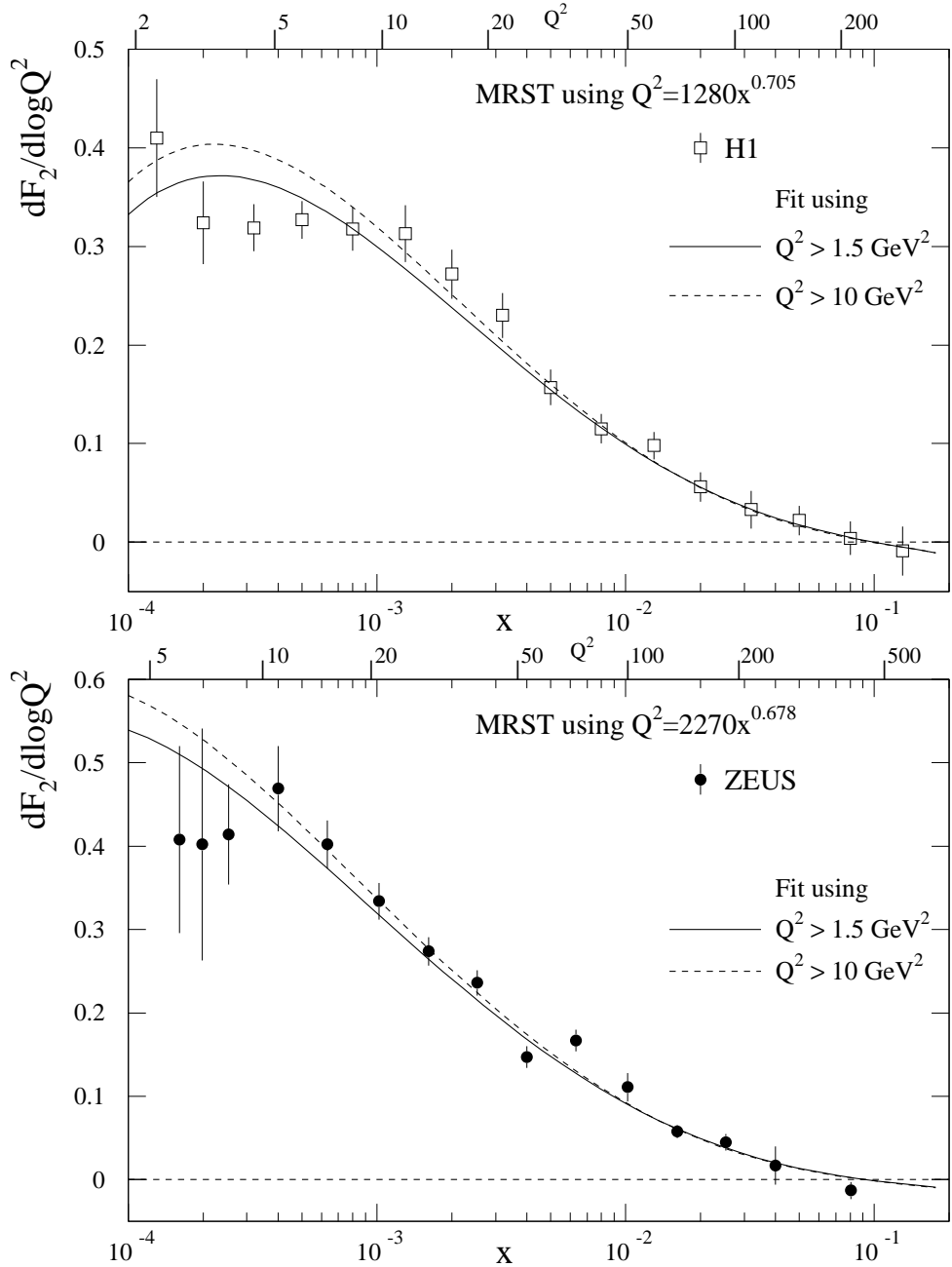,height=20cm}    
\end{center}    
\vspace{-0.5cm}                                                                
\caption{The description of the slopes $\partial F_2/\partial     
\ln Q^2$ versus $x$ for the HERA data. The experimental points were 
computed from linear fits to the data in Fig.~\protect\ref{fig:DISsmallx}.    
The solid curve corresponds to the MRST set while the dashed    
line indicates the effect of removing low $Q^2$ data    
(with $Q^2 < 10\; \GeV^2$) from the fit. The values of $Q^2$ appropriate    
to the data are given (in GeV$^2$) approximately by the formulae shown and the
corresponding scale is indicated on the upper edge of each plot.}    
\label{fig:slope}
\end{figure}                                  
\newpage    
                            
\begin{figure}[H]
\vspace{2cm}                                                                
\begin{center}     
\epsfig{figure=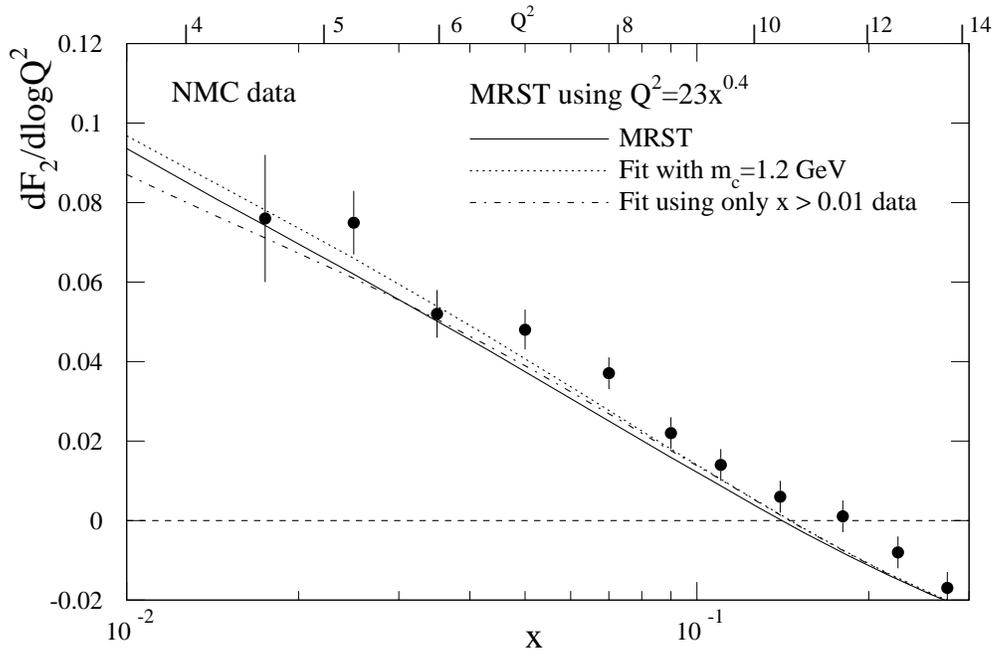,height=20cm}    
\end{center}    
\vspace{-9cm}                                                                
\caption{The continuous curve is the     
description of the slopes $\partial F_2/\partial     
\ln Q^2$ of the NMC proton data. The experimental points were     
computed from linear fits to the data in Fig.~\protect\ref{fig:NMCa}.    
We also show the effect of modifying the charm quark mass and  of excluding    
low $x$ data (with $x < 0.01$) from the global  fit.}    
\label{fig:NMCslope}    
\end{figure}                                          
\newpage    
                            
\begin{figure}[H]
%\vspace{-0.5cm}                                                                
\begin{center}     
\epsfig{figure=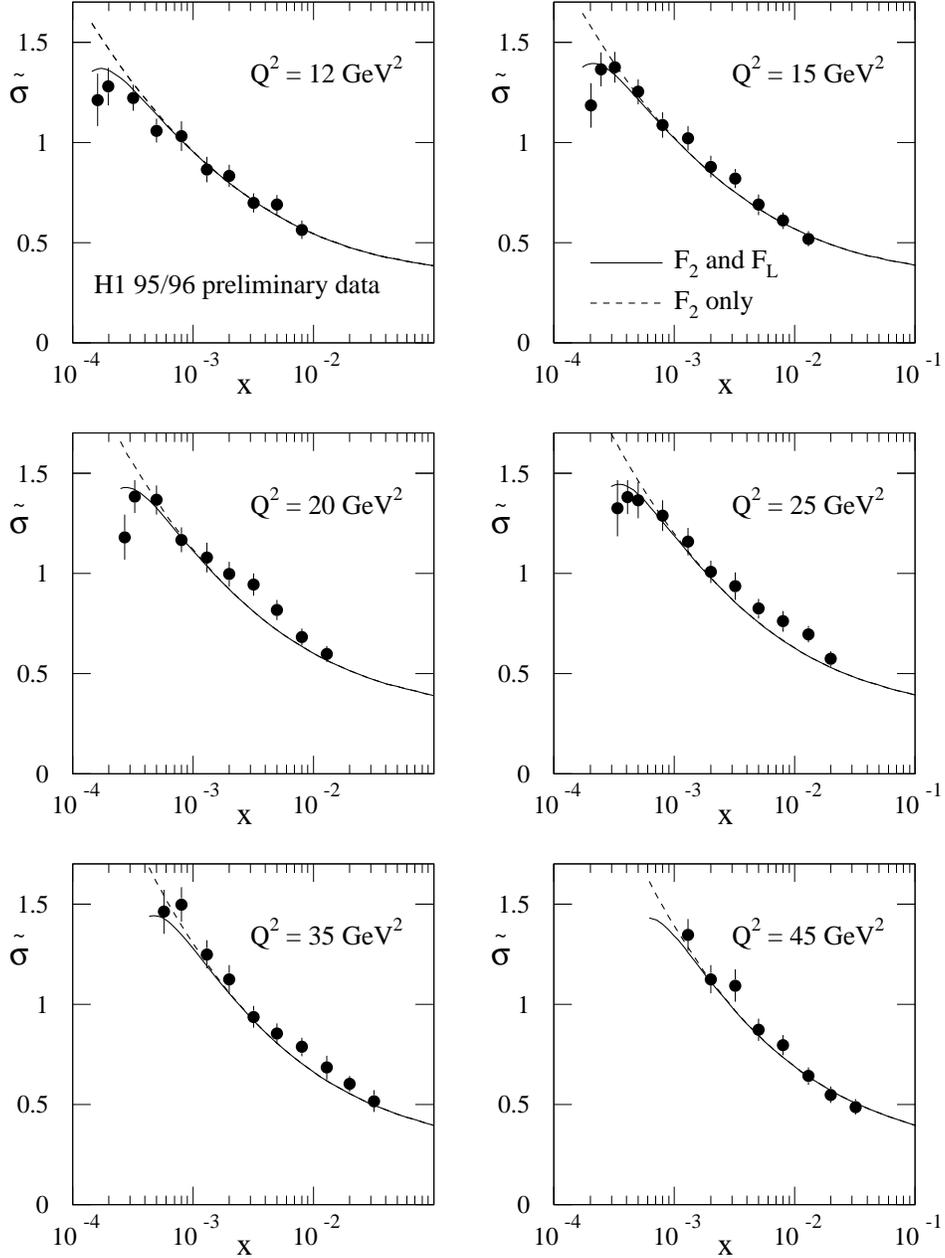,height=20cm}    
\end{center}    
\caption{The description of the quantity $\tilde \sigma = F_2 -    
{y^2}/{[1 + (1-y)^2]}F_L$ compared with the preliminary
H1 1995/96~\protect\cite{F2H196}    
data. Also shown (dashed curve) is the contribution to $\tilde \sigma $    
of $F_2$ only.}    
\label{fig:sigFL}                                   
\end{figure}                                    
\newpage    
                            
\begin{figure}[H]                                   
%\vspace{-0.5cm}                                                                
\begin{center}     
\epsfig{figure=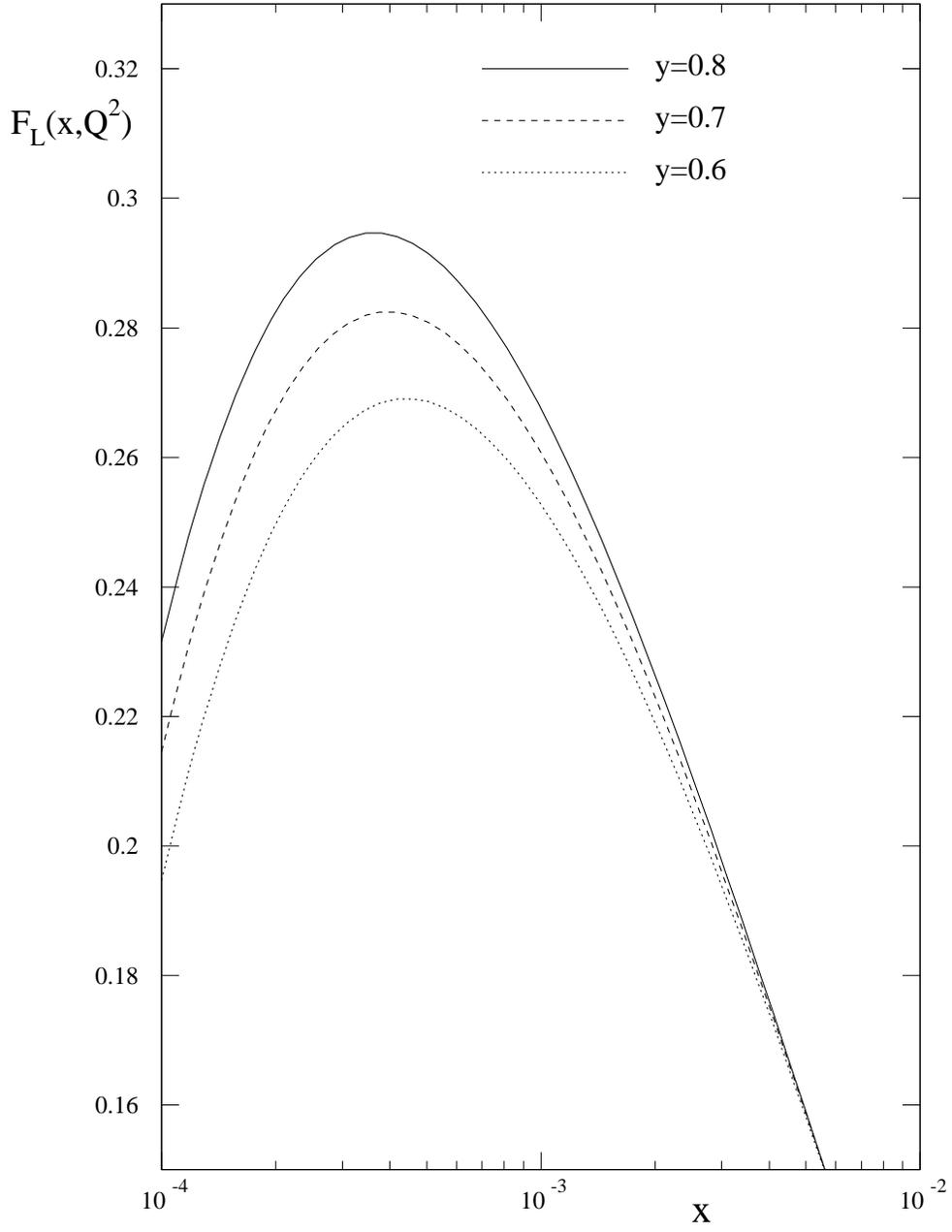,height=20cm}    
\end{center}    
\caption{Prediction for $F_L$ at HERA versus $x$ for $y = 0.6, 0.7, 0.8$.
Note that   
$F_L$ has an implicit $y$ dependence via the relation $Q^2 = xys$.}    
\label{fig:predFL}
\end{figure}                                                                  
\newpage    
                               
\begin{figure}[H] 
\vspace{2cm}                                                                
\begin{center}     
\epsfig{figure=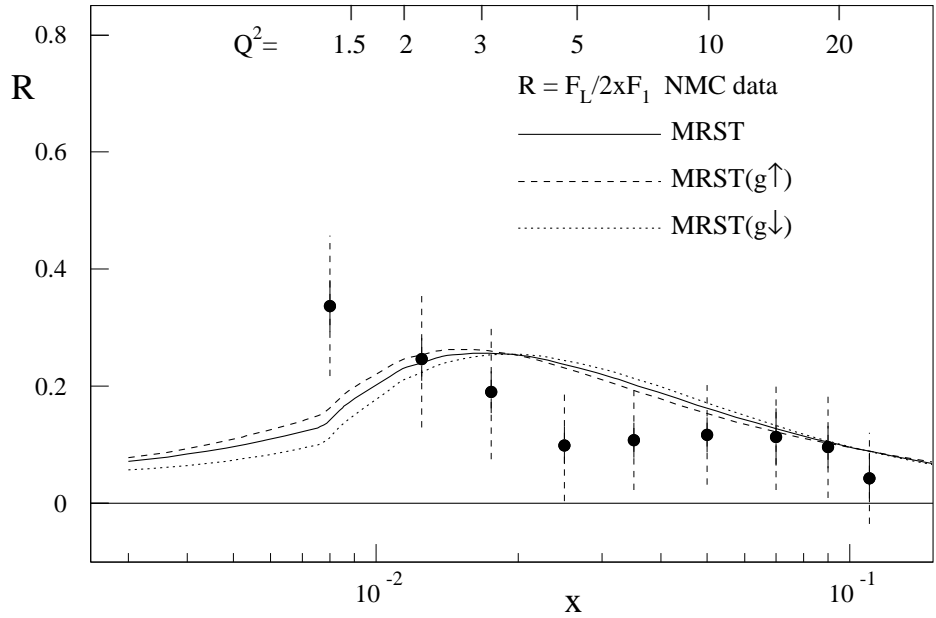,height=20cm}    
\end{center}    
\vspace{-9cm}                                                                
\caption{Comparison of the predictions for  $R=F_L/2xF_1$  with the     
NMC~\protect\cite{NMC} data. The values of $Q^2$ appropriate    
to the data are given (in GeV$^2$) approximately by the formula 
$Q^2 = 262 x^{1.09}$,  and the
corresponding scale is indicated on the upper edge of the plot.}                                                
\label{fig:NMCFL} 
\end{figure}                                      
\newpage    
                            
\begin{figure}[H] 
%\vspace{-0.5cm}                                                                
\begin{center}     
\epsfig{figure=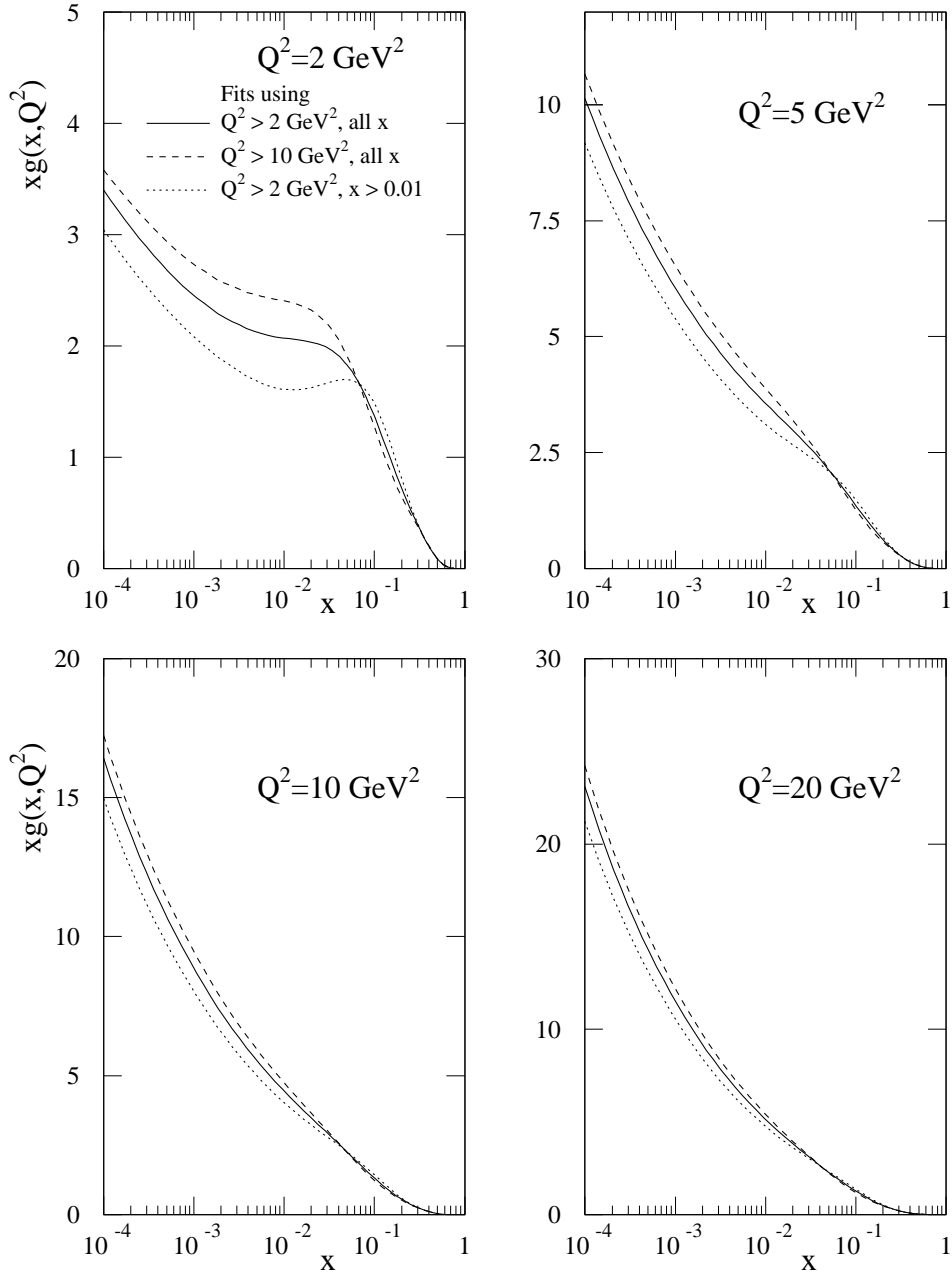,height=20cm}    
\end{center}    
\caption{The gluon distributions which result from making different cuts    
in $Q^2$ and $x$ to the data included in the fits. The cuts which we use
are specified  in the first plot.}    
\label{fig:gluoncut}    
\end{figure}            
\newpage    
    
\begin{figure}[H] 
%\vspace{-0.5cm}                                                                
\begin{center}     
\epsfig{figure=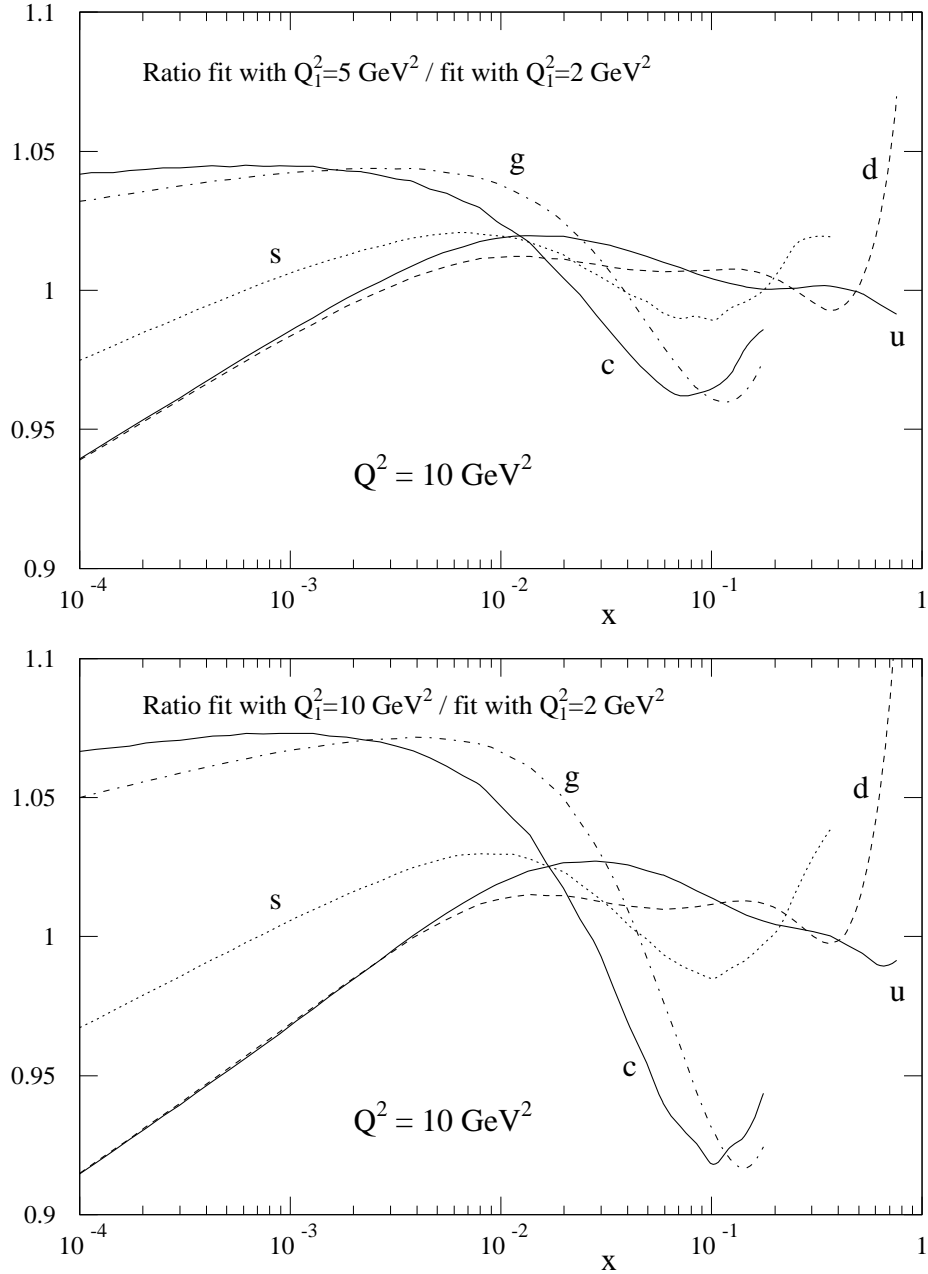,height=20cm}    
\end{center}    
\caption{The parton distributions at $Q^2 = 10\; \GeV^2$, compared    
with the default MRST partons, obtained by making $Q_1^2 = 5\; \GeV^2$ and     
$Q_1^2 = 10\; \GeV^2$  cuts    
in $Q^2$ to the data included in the fits.}    
\label{fig:partoncut}    
\end{figure}             
\newpage    
                                                              
\begin{figure}[H] 
\vspace{-0.5cm}                                                                
\begin{center}     
\epsfig{figure=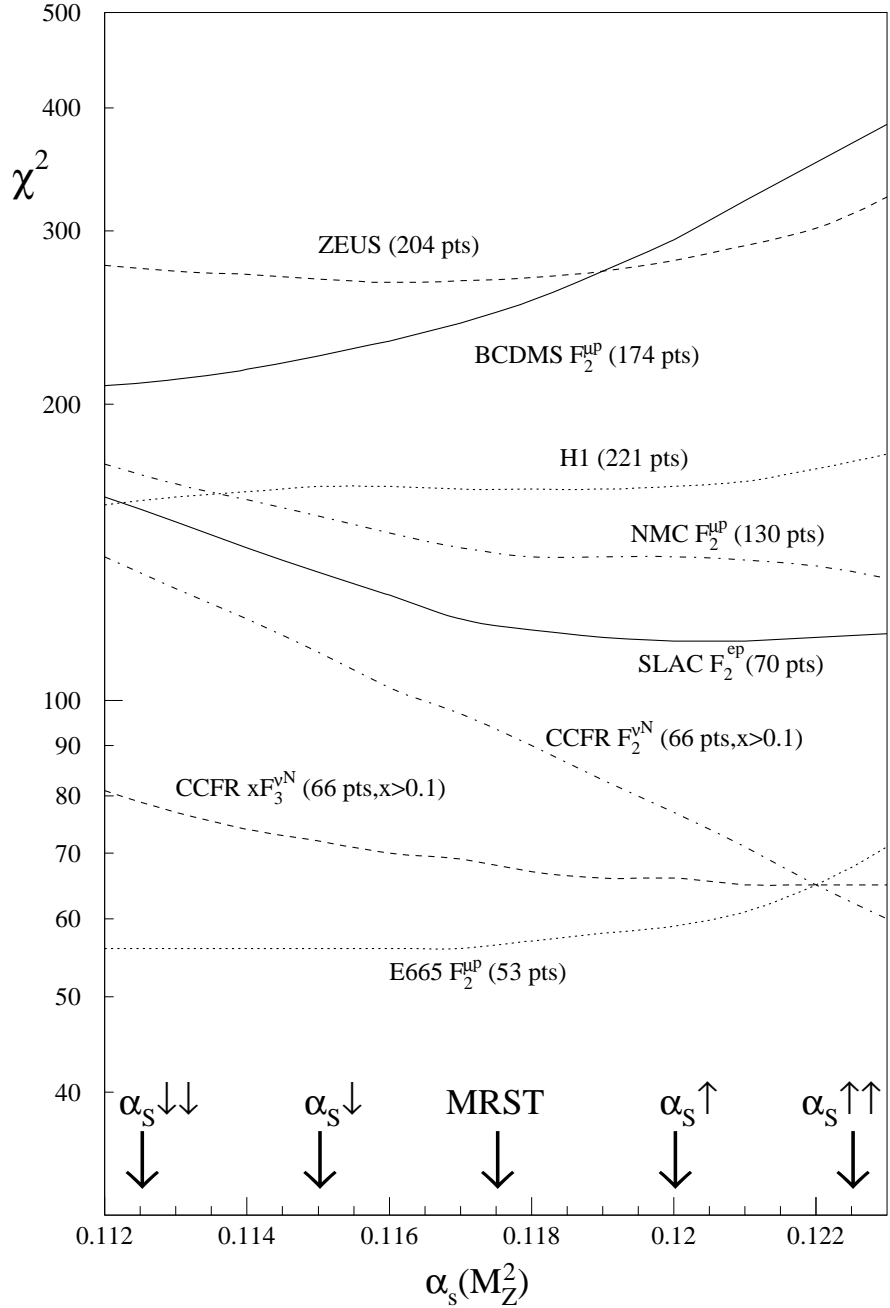,height=20cm}    
\end{center}    
\vspace{-0.5cm}                                                                
\caption{The contributions to the total global fit $\chi^2$ from the various
data sets as a function of $\alpha_S$.  The parton set corresponding to the     
optimum value $\alpha_S = 0.1175$ is denoted simply MRST and is the default
set of partons used throughout the paper. The four other sets, which correspond
to the adjacent values of $\alpha_S$ indicated by arrows, are used
for comparison purposes.  The $\chi^2$ values
for the CCFR data are obtained from the statistical error and an additional    
$1.5\%$ `systematic' error added in quadrature.}
\label{fig:alpha}                               
\end{figure}                                      
\newpage    
    
\begin{figure}[H] 
%\vspace{-0.5cm}                                                                
\begin{center}     
\epsfig{figure=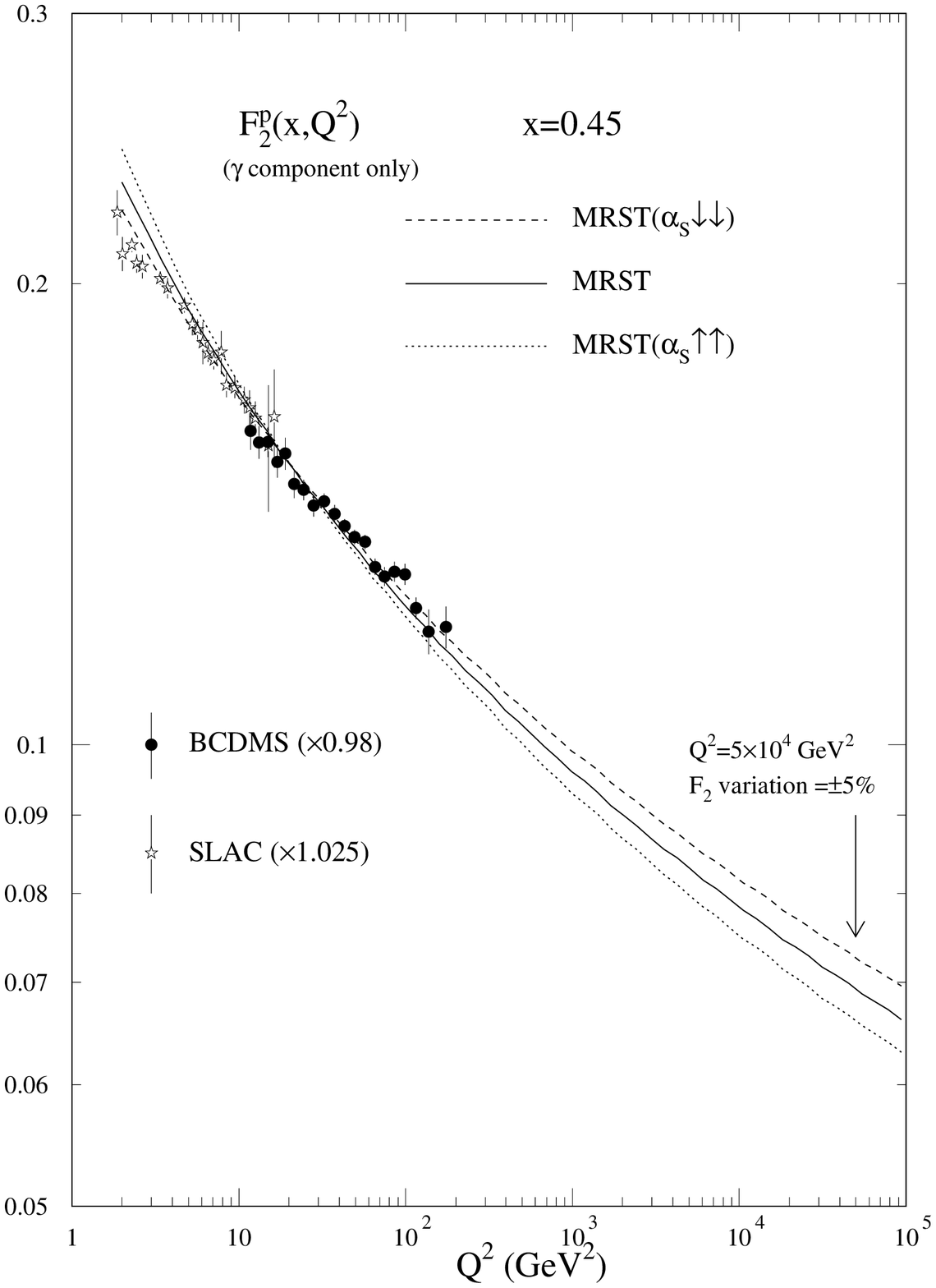,height=20cm}    
\end{center}    
\caption{The extrapolation of the fits at $x=0.45$ to high $Q^2$ using  the    
MRST, MRST($\asup$) and MRST($\asdown$) set of partons.}    
\label{fig:x45plot}    
\end{figure}             
\newpage    
                            
\begin{figure}[H]                               
%\vspace{1cm}                                                                
\begin{center}     
\epsfig{figure=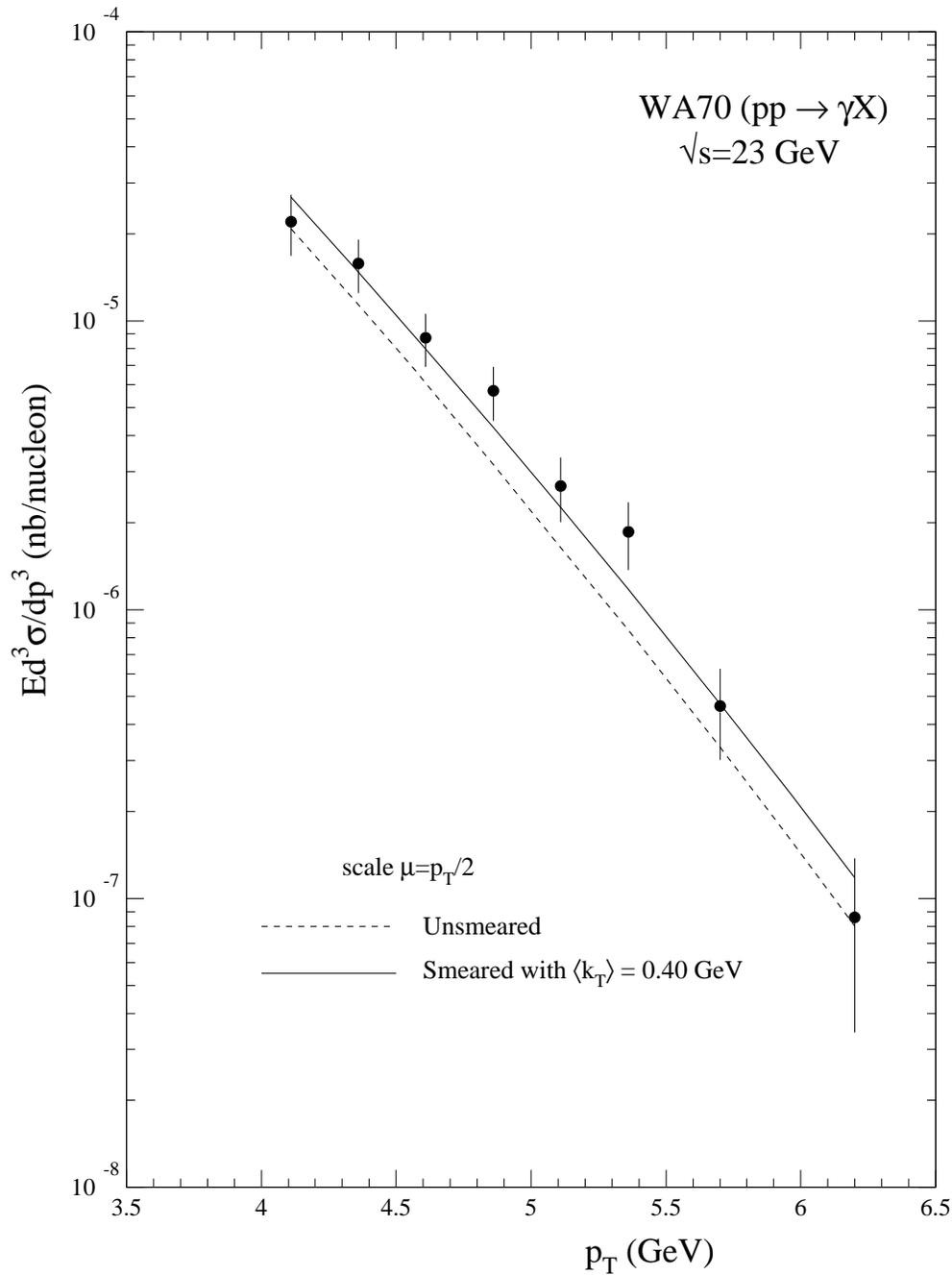,height=20cm}    
\end{center}    
\caption{Comparison of the WA70~\protect\cite{WA70} data with the MRST parton
set with and without smearing in transverse momentum.}                       
\label{fig:WA70}                                                             
\end{figure}                                                                 
\newpage    
                            
\begin{figure}[H]                                                            
%\vspace{1cm}                                                                
\begin{center}     
\epsfig{figure=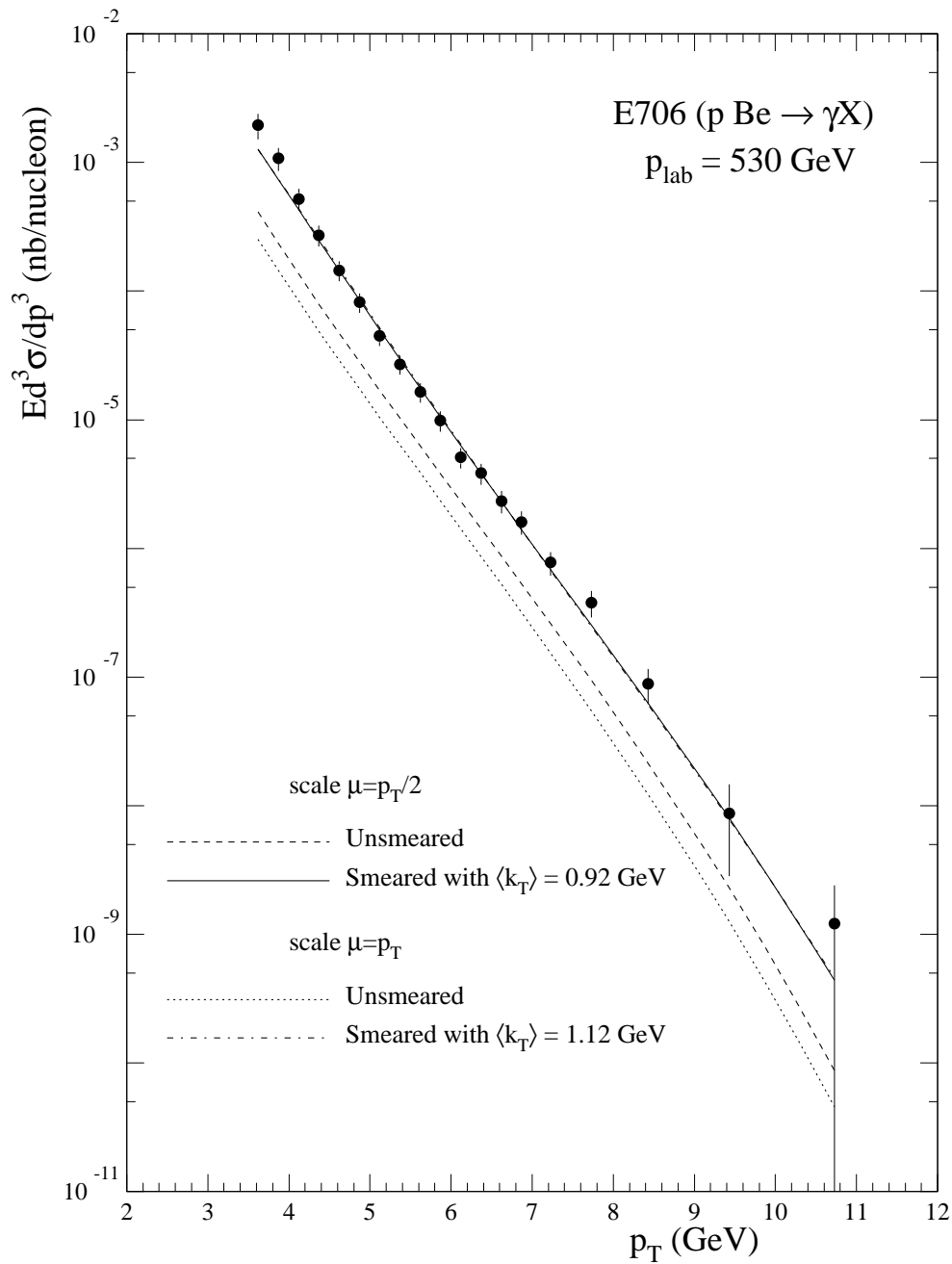,height=20cm}    
\end{center}    
\caption{Comparison of the E706~\protect\cite{E706} data at 530 GeV with the    
MRST parton set. The results for     
two choices of the factorization scale are shown, together    
with the improvement of the description obtained by     
including parton transverse momentum in each case.  
The two `improved' curves lie on    
top of each other.}           
\label{fig:E706a}             
\end{figure}                                  
\newpage    
                           
\begin{figure}[H]             
%\vspace{1cm}                                                                
\begin{center}     
\epsfig{figure=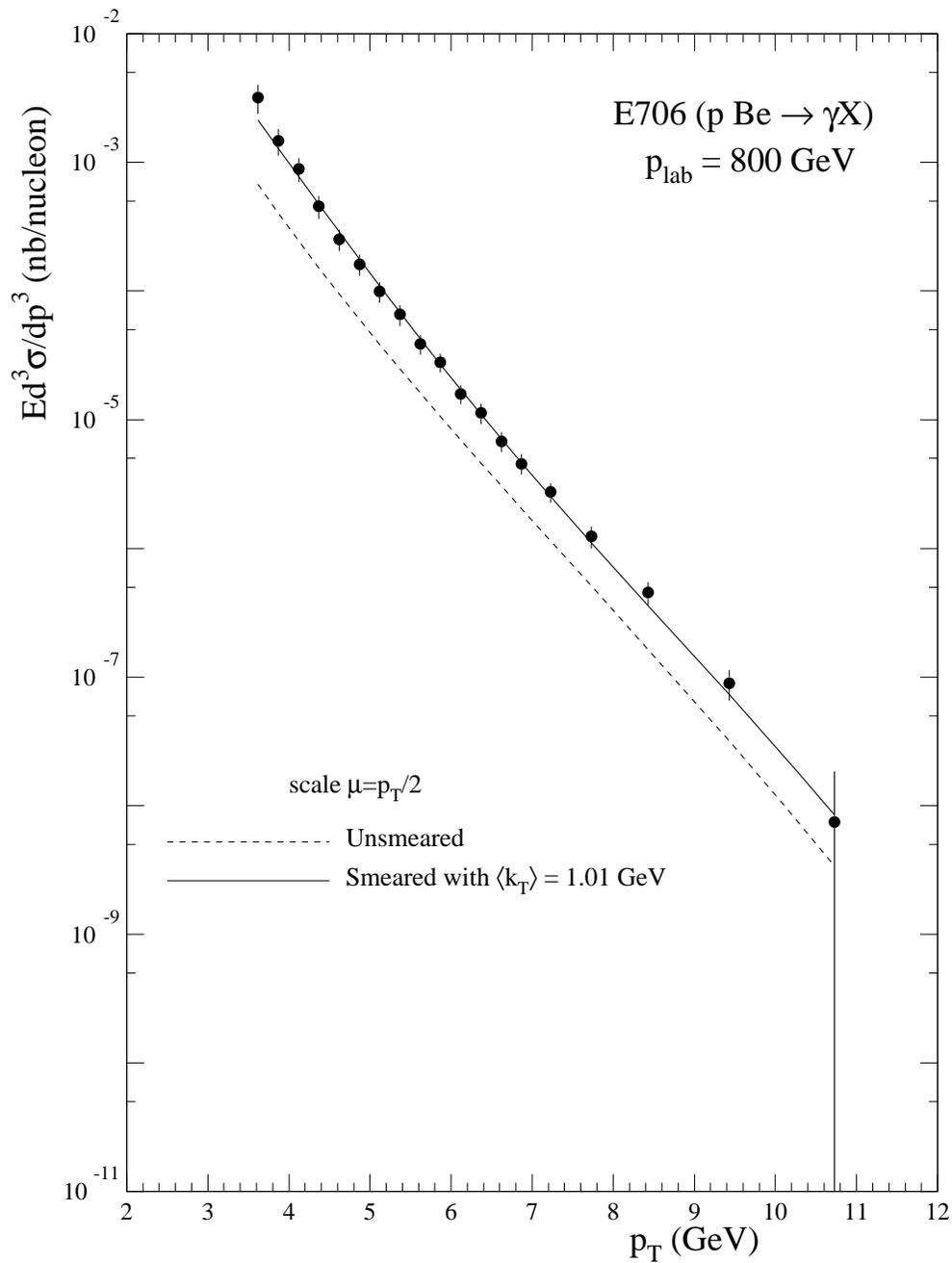,height=20cm}    
\end{center}    
\caption{Comparison of the E706~\protect\cite{E706} data at 800~GeV with the    
MRST parton set. The scale is chosen to be $p_T/2$ and the effect     
of including parton transverse momentum is shown.} 
\label{fig:E706b}                                  
\end{figure}                                       
\newpage    
                                                    
\begin{figure}[H]                                                               
%\vspace{1cm}                                                                
\begin{center}     
\epsfig{figure=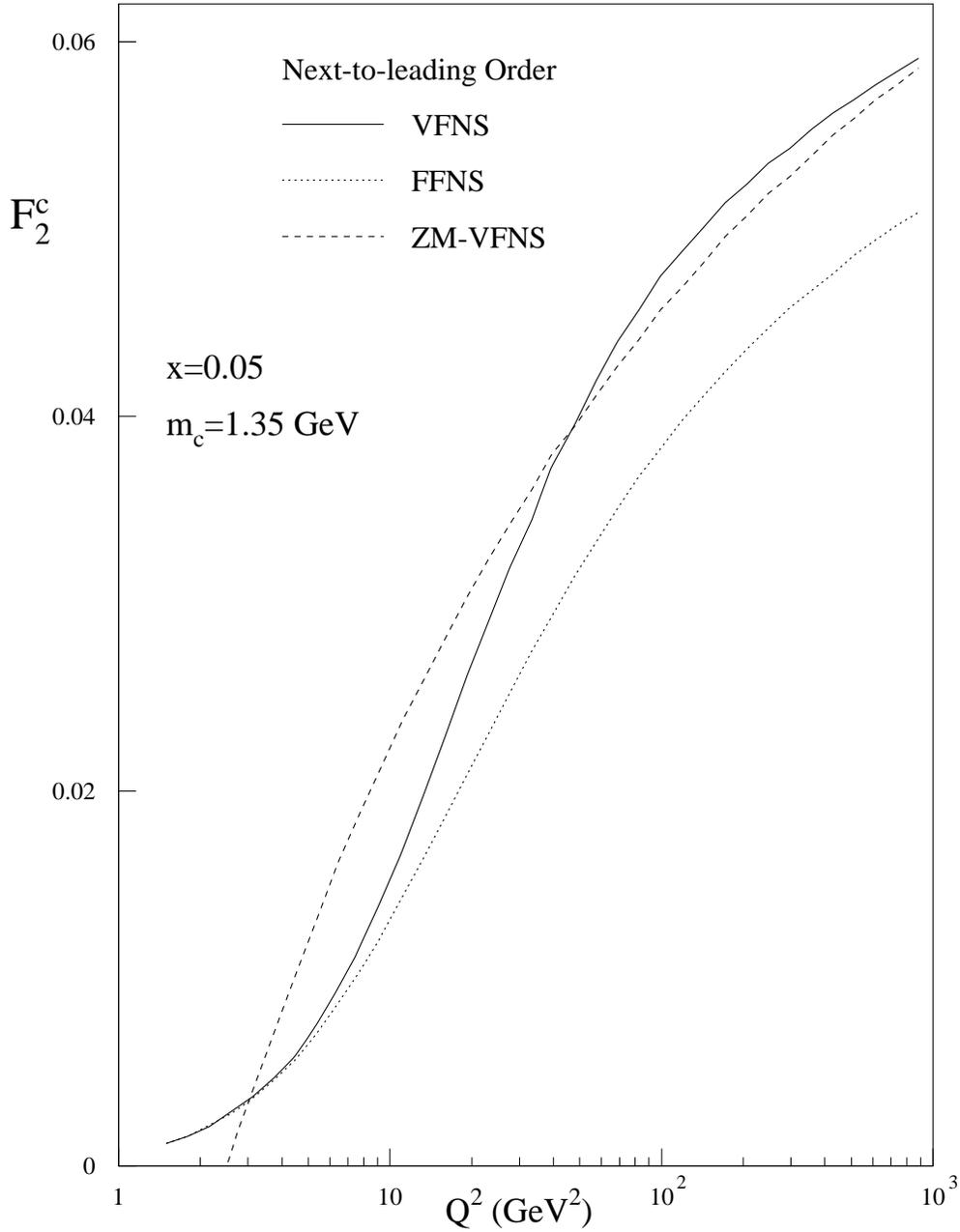,height=20cm}    
\end{center}    
\caption{The continuous curve is the MRST     
charm structure function at $x=0.05$ obtained using the NLO    
prescription and NLO evolution (VFNS). For comparison, the prescriptions    
of the fixed flavour number scheme (FFNS) and the zero mass scheme (ZM-VFNS)    
are also shown.}                                   
\label{fig:charm1}                                 
\end{figure}                                                                 
\newpage    
                                                                 
\begin{figure}[H]                                                               
%\vspace{1cm}                                                                
\begin{center}     
\epsfig{figure=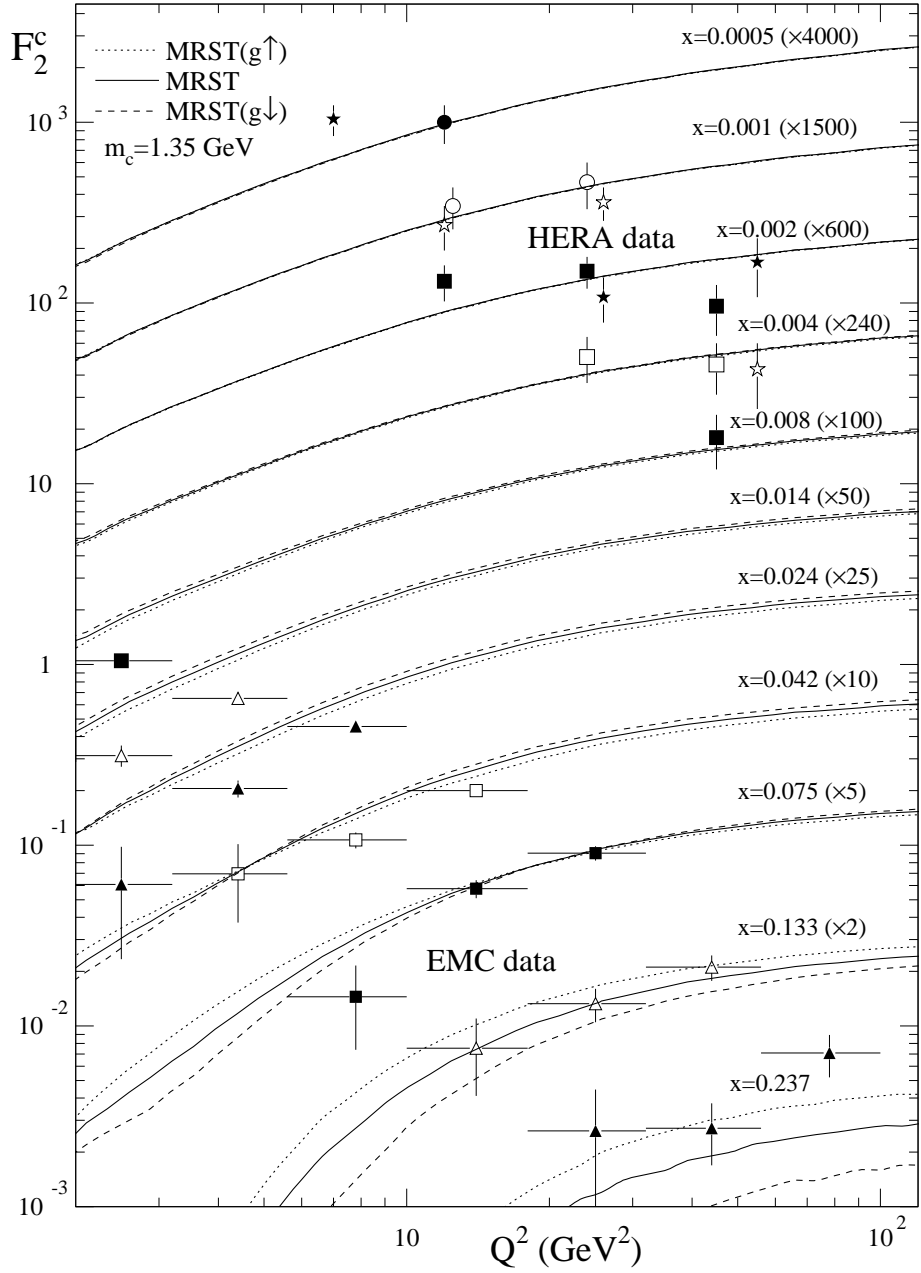,height=20cm}    
\end{center}    
\caption{The prediction for the charm structure function using the three    
MRST parton sets characterized by the different large $x$ gluon behaviour.    
The HERA data at small $x$ are from H1~\protect\cite{H1C} and      
ZEUS~\protect\cite{ZEUSC}, and the large $x$ data are  from     
EMC~\protect\cite{EMC}.}                           
\label{fig:charm2}                                 
\end{figure}                                                                 
\newpage    
                                                                 
\begin{figure}[H]                                                               
%\vspace{1cm}                                                                
\begin{center}     
\epsfig{figure=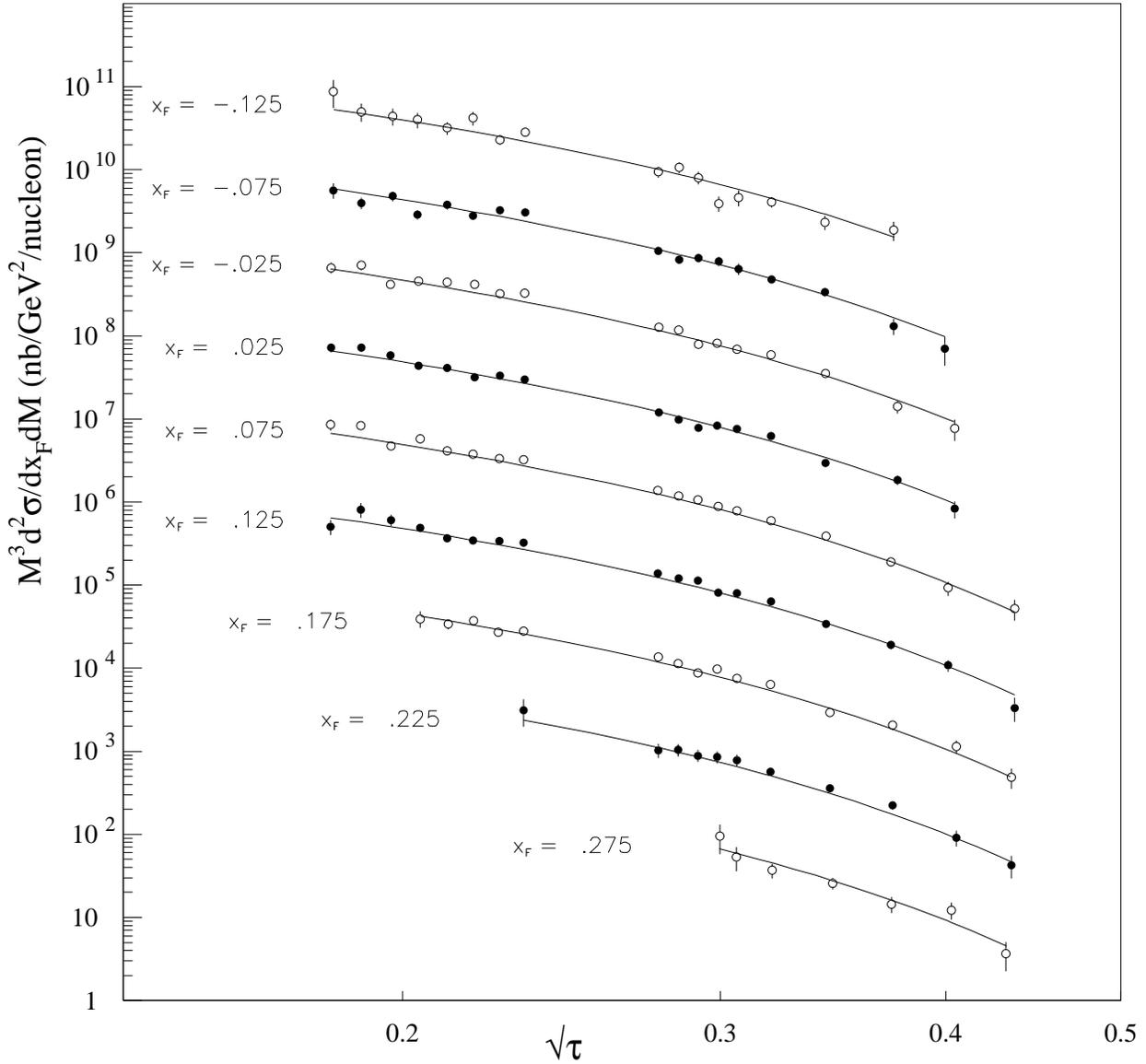,height=18cm}    
\end{center}    
\caption{Hadroproduction of dileptons computed from the MRST parton set    
compared with the E605 data~\protect\cite{E605}. The theory curves    
include an additional  $K'$ factor of 0.9. No correction for the heavy target    
has been made. The scale on the left--hand axis is appropriate
for the theory and data at $x_F = -0.125$. For display    
purposes  the normalization is then decreased by a factor of ten
for each step upwards in $x_F$.}                                    
\label{fig:E605}                                   
\end{figure}                                                       
\newpage    
                                                   
\begin{figure}[H]                                                               
%\vspace{1cm}                                                                
\begin{center}     
\epsfig{figure=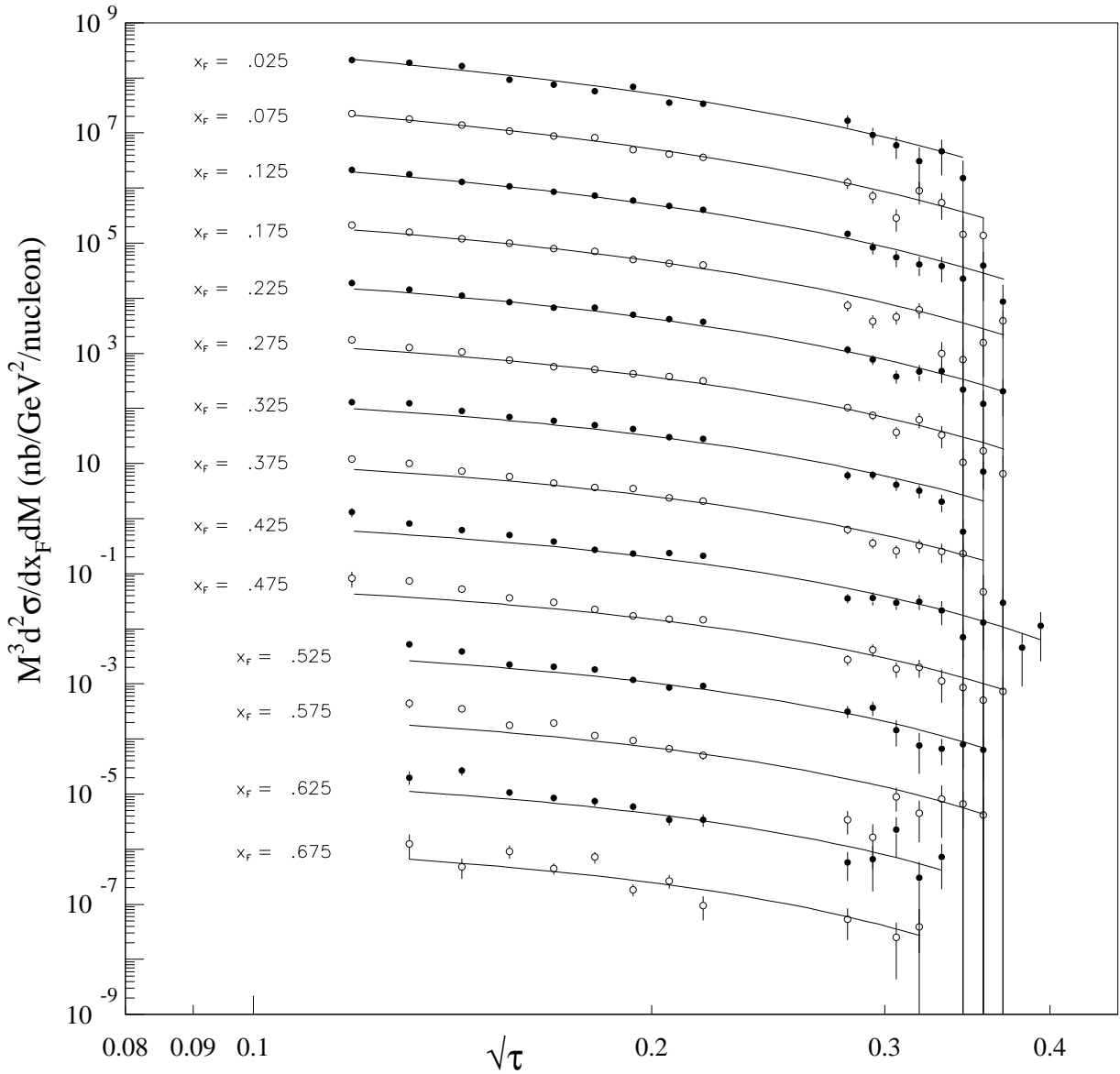,height=18cm}    
\end{center}    
\caption{Hadroproduction of dileptons computed from the MRST parton set    
compared with the E772 data~\protect\cite{E772}. The theory curves    
include an additional $K'$ factor of 0.96. The scale on the left--hand axis is appropriate
for the theory and data at $x_F = 0.025$. For display    
purposes  the normalization is then decreased by a factor of ten
for each step upwards in $x_F$.}                    
\label{fig:E772}                                   
\end{figure}                                                                 
\newpage    
                                                      
\begin{figure}[H]                                  
%\vspace{1cm}                                                               
\begin{center}     
\epsfig{figure=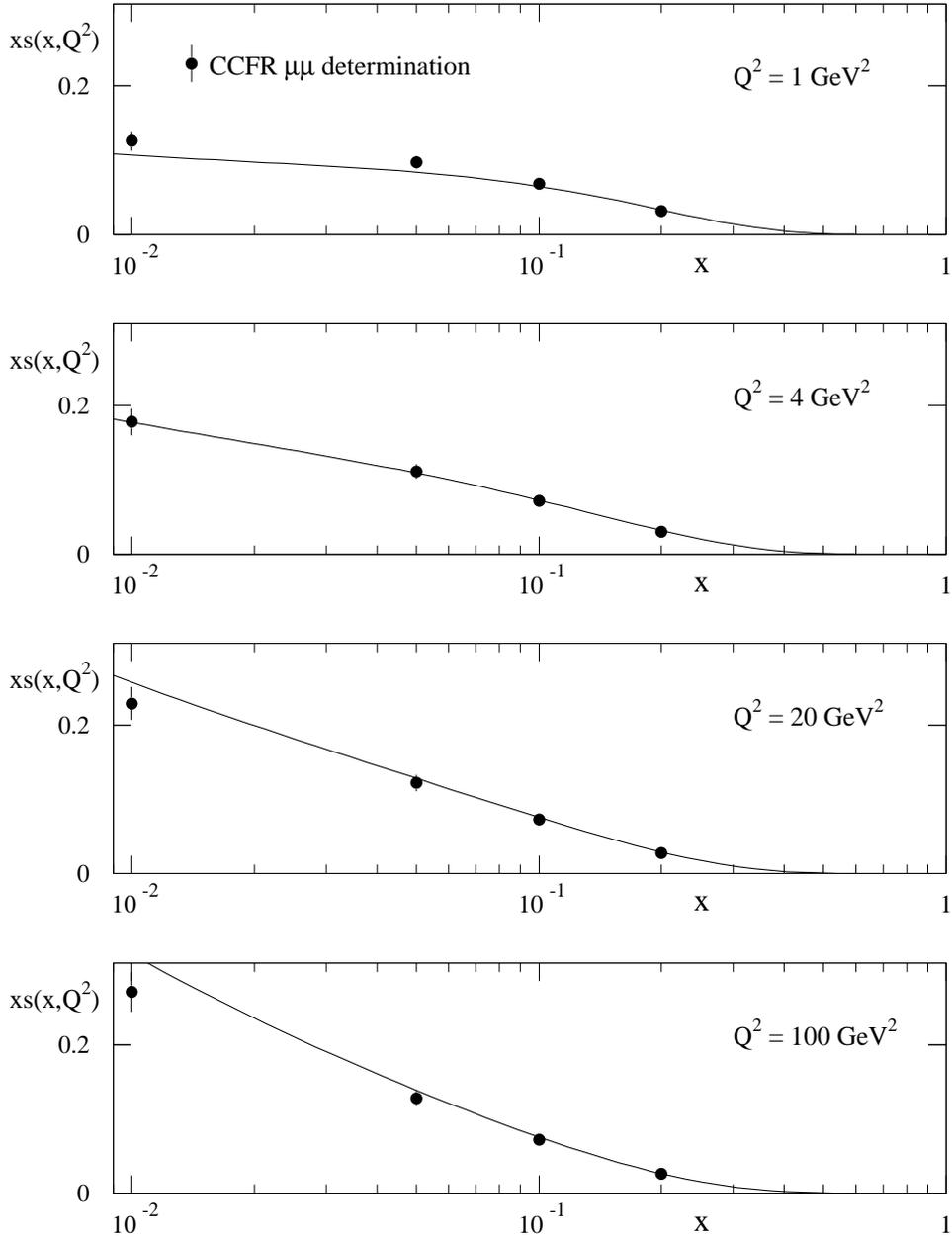,height=20cm}    
\end{center}    
\caption{Comparison of the strange quark distribution from the MRST set    
of partons compared with the determination of the strange sea obtained by     
 the CCFR collaboration~\protect\cite{CCFRMM}  in a    
NLO analysis of their data on the neutrino production of dimuons.}
\label{fig:strange}                                               
\end{figure}                                                                    
\newpage    
                                                                
\begin{figure}[H]                                                               
%\vspace{1cm}                                                                
\begin{center}     
\epsfig{figure=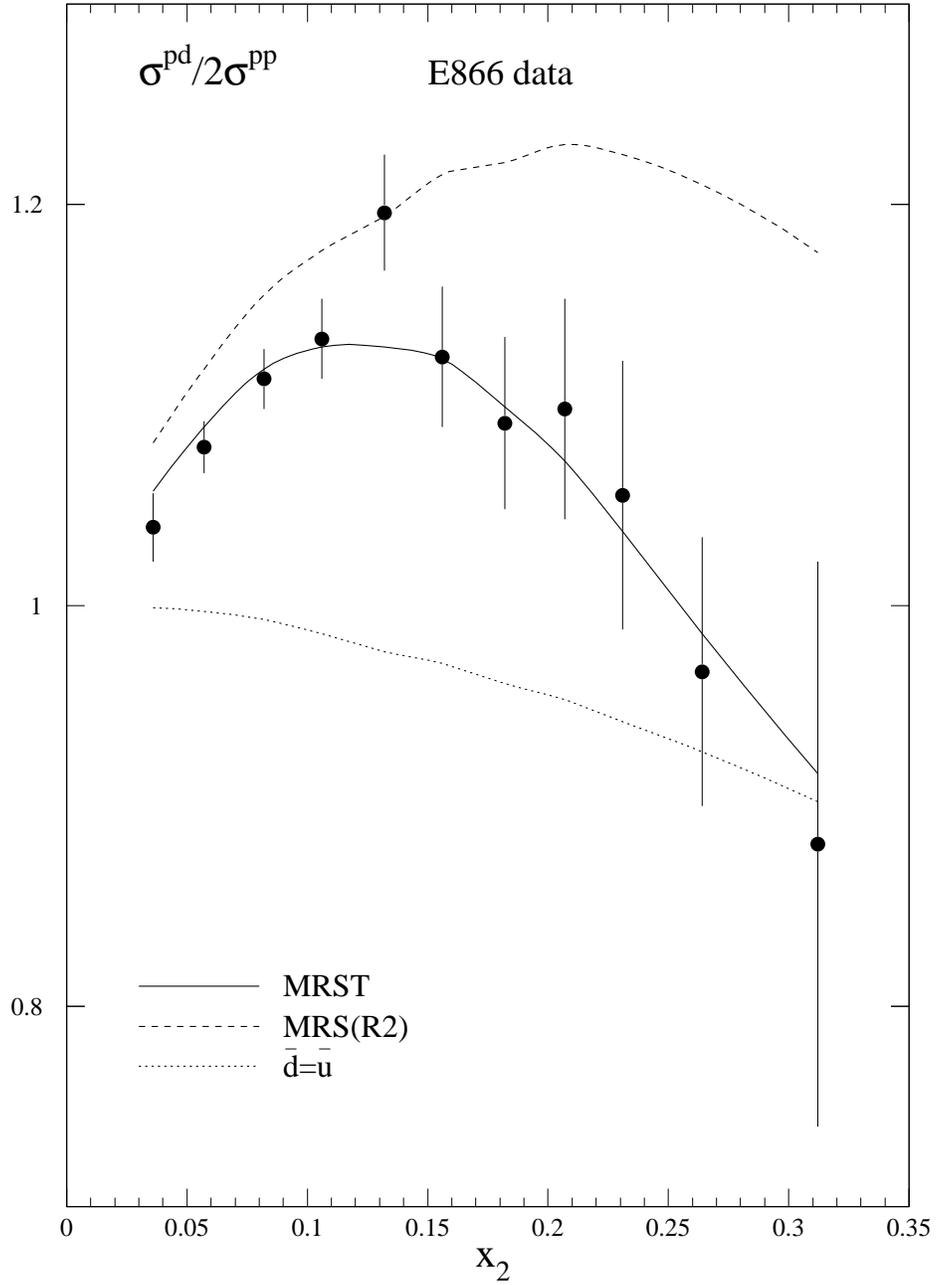,height=20cm}    
\end{center}    
\caption{The continuous curve is the MRST description of the     
E866~\protect\cite{E866} data for the ratio of the cross sections     
for hadroproduction of dileptons    
for proton and deuterium targets versus $x_2$, the fractional momentum     
of the parton in  the target. The other curves are for comparison only.} 
\label{fig:E866}                                                         
\end{figure}                                                                 
\newpage    
                                                                 
\begin{figure}[H]                                                               
%\vspace{1cm}                                                                
\begin{center}     
\epsfig{figure=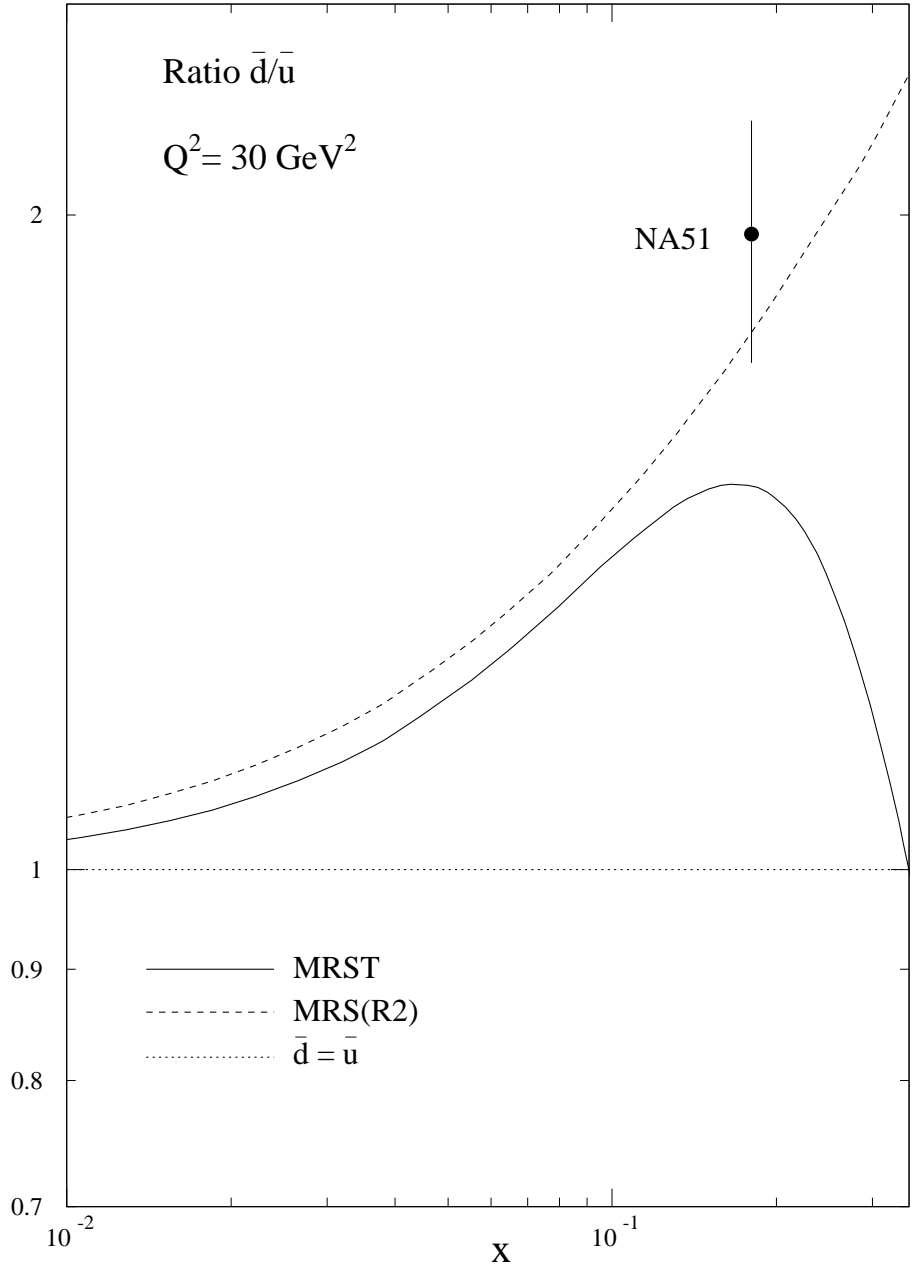,height=20cm}    
\end{center}    
\caption{The ratio of the parton distributions $\bar d / \bar u$ at $Q^2$    
= 30 GeV$^2$ for the MRST and MRS(R2) parton sets compared to the estimate from 
the NA51~\protect\cite{NA51} measurement of the Drell-Yan asymmetry.}
\label{fig:ubarodbar}                                              
\end{figure}                       
\newpage    
       
\begin{figure}[H]                                                               
\vspace{2cm}                                                                
\begin{center}     
\epsfig{figure=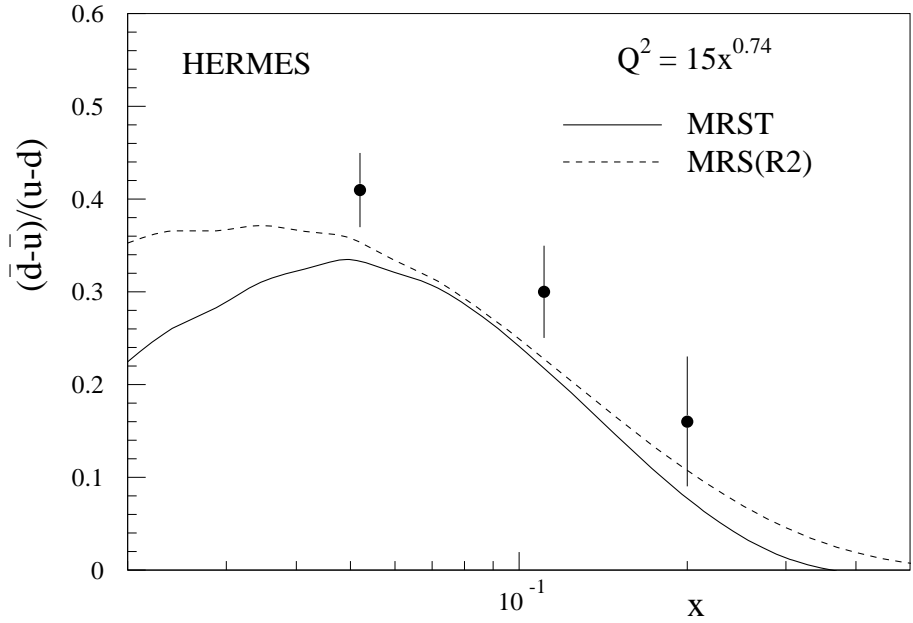,height=20cm}    
\end{center}    
\vspace{-9cm}                                                                
\caption{Predictions for the ratio $(\bar d - \bar u)/(u - d)$ from    
%the parton sets MRST and MRS(R2).}      
the parton sets MRST and MRS(R2) compared with the 
preliminary estimates obtained     
by the     
HERMES collaboration~\protect\cite{HERMES} from their measurements of 
semi-inclusive     
pion production in DIS. Note that there is an additional overall 
systematic error on the data of $\pm 0.07$.
 The $Q^2$ values of the data vary with $x$ approximately    
according to the formula $Q^2\; (\GeV^2) = 15 x^{0.74}$.}
\label{fig:HERMES}                                       
\end{figure}                                             
\newpage    
                                                                 
\begin{figure}[H]                                                               
\vspace{2cm}                                                                
\begin{center}     
\epsfig{figure=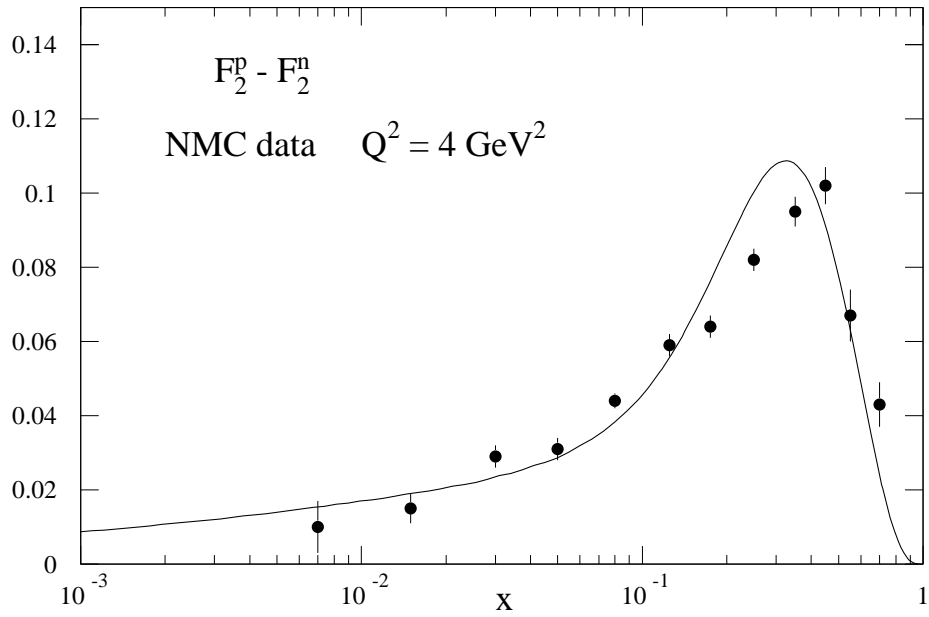,height=20cm}    
\end{center}    
\vspace{-9cm}                                                                
\caption{The MRST description of the difference $F_2^p - F_2^n$ at $Q^2$ = 4    
GeV$^2$    
compared with the measurements from NMC~\protect\cite{NMCGS}.}  
\label{fig:GS}                                                                 
\end{figure}                                                                 
\newpage    
                                                                 
\begin{figure}[H]                                                               
%\vspace{1cm}                                                                
\begin{center}     
\epsfig{figure=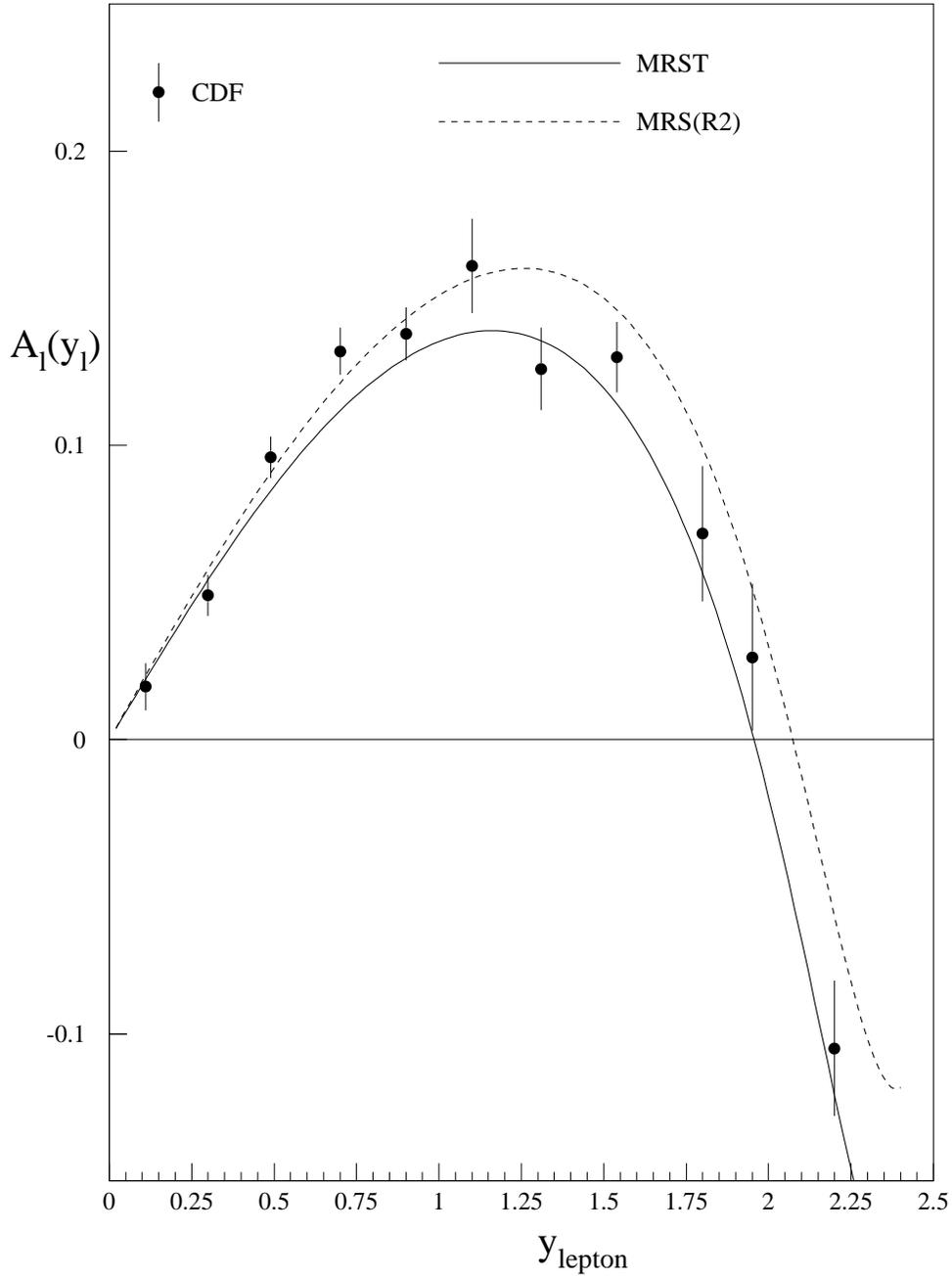,height=20cm}    
\end{center}    
\caption[]{The description of the lepton asymmetry for $W^{\pm}$ production    
in $\bar p p$ collisions at $\sqrt s$ = 1.8 $\TeV$. The data from     
CDF~\protect\cite{CDF} are compared with the new MRST parton set and    
with the previous set MRS(R2).} 
\label{fig:WASY}                
\end{figure}                                                                 
\newpage    
                            
\begin{figure}[H]               
%\vspace{1cm}                                                                 
\begin{center}     
\epsfig{figure=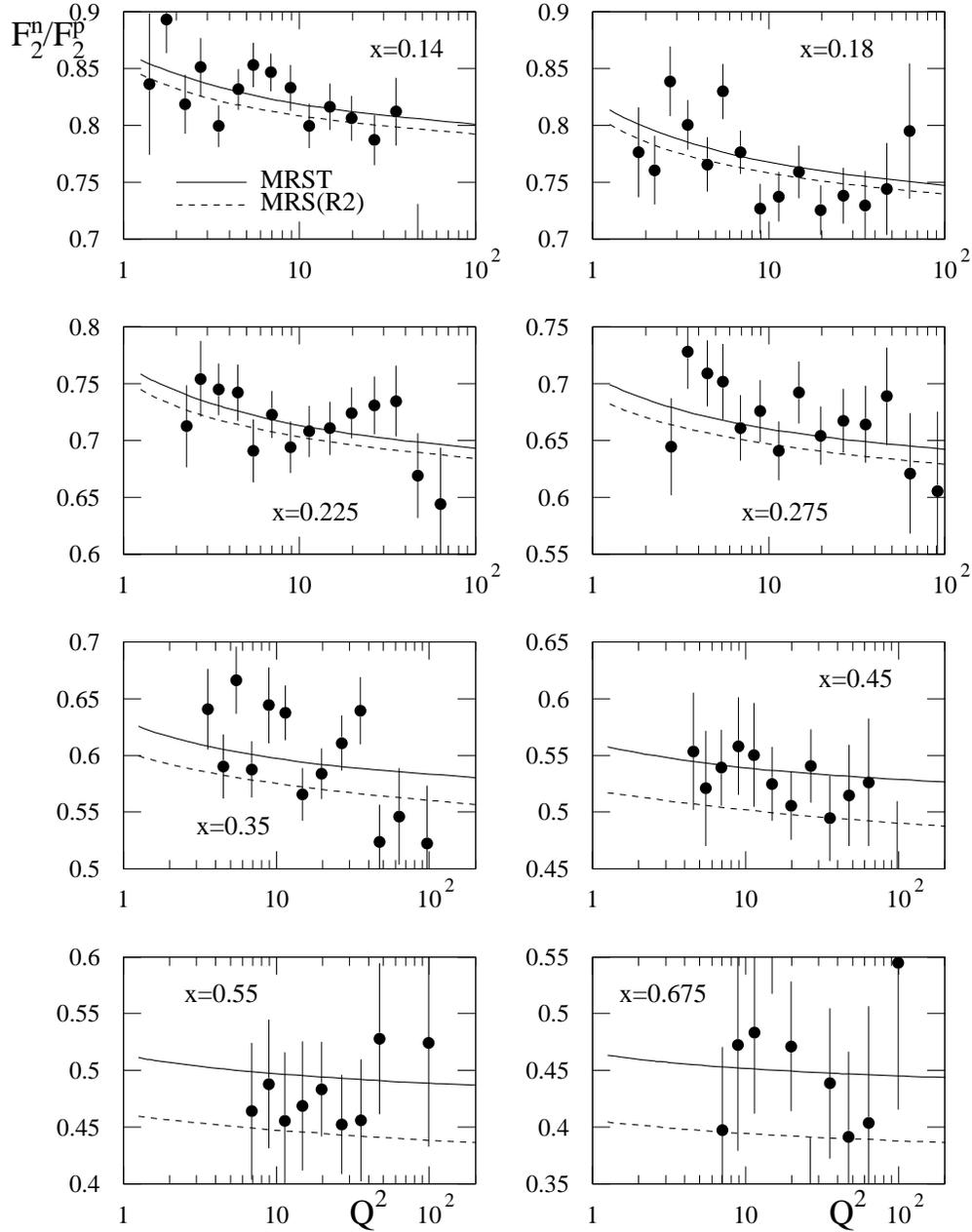,height=20cm}    
\end{center}    
\caption{The large $x$ data on the ratio $F_2^n/F_2^p$ extracted from    
the measurements of $F_2^d/F_2^p$ by NMC~\protect\cite{NMC} compared    
with the MRST and MRS(R2) descriptions.}
\label{fig:NMCratio}                    
\end{figure}                            
\newpage    
                            
\begin{figure}[H]                       
%\vspace{-0.5cm}                                                                
\begin{center}     
\epsfig{figure=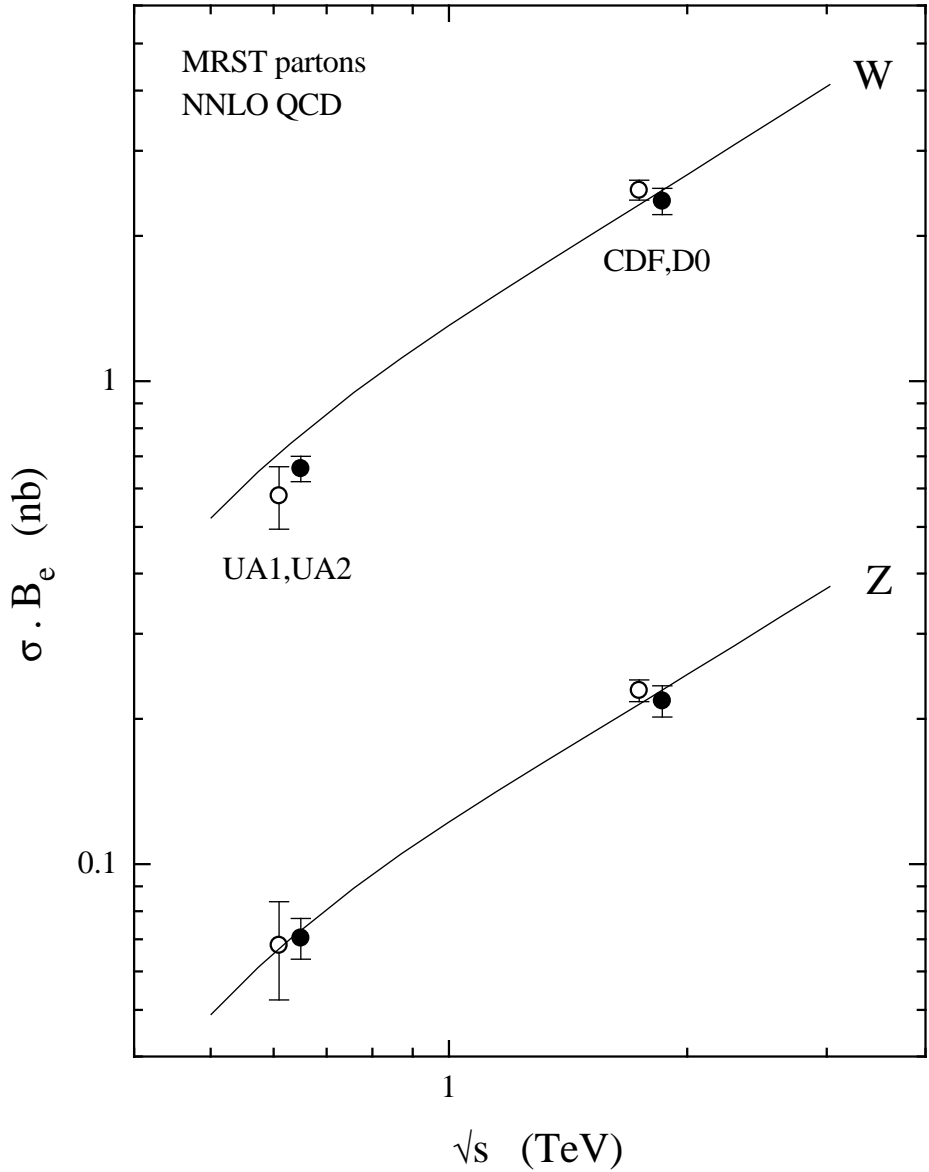,height=20cm}    
\end{center}    
\vspace{-1cm}
\caption[]{Total $W, Z$ production cross sections times leptonic branching 
ratios    
as a function of the ($p \bar p$) collider energy $\sqrt s$, calculated    
using the default MRST partons. Experimental measurements from     
UA1~\protect\cite{UA1SIGW}, UA2~\protect\cite{UA2SIGW}, 
CDF~\protect\cite{CDFSIGW}     
and D0~\protect\cite{D0SIGW} are also shown.}
\label{fig:WZsig}                            
\end{figure}                            
\newpage    
                            
\begin{figure}[H]                            
%\vspace{-0.5cm}                                                                
\begin{center}     
\epsfig{figure=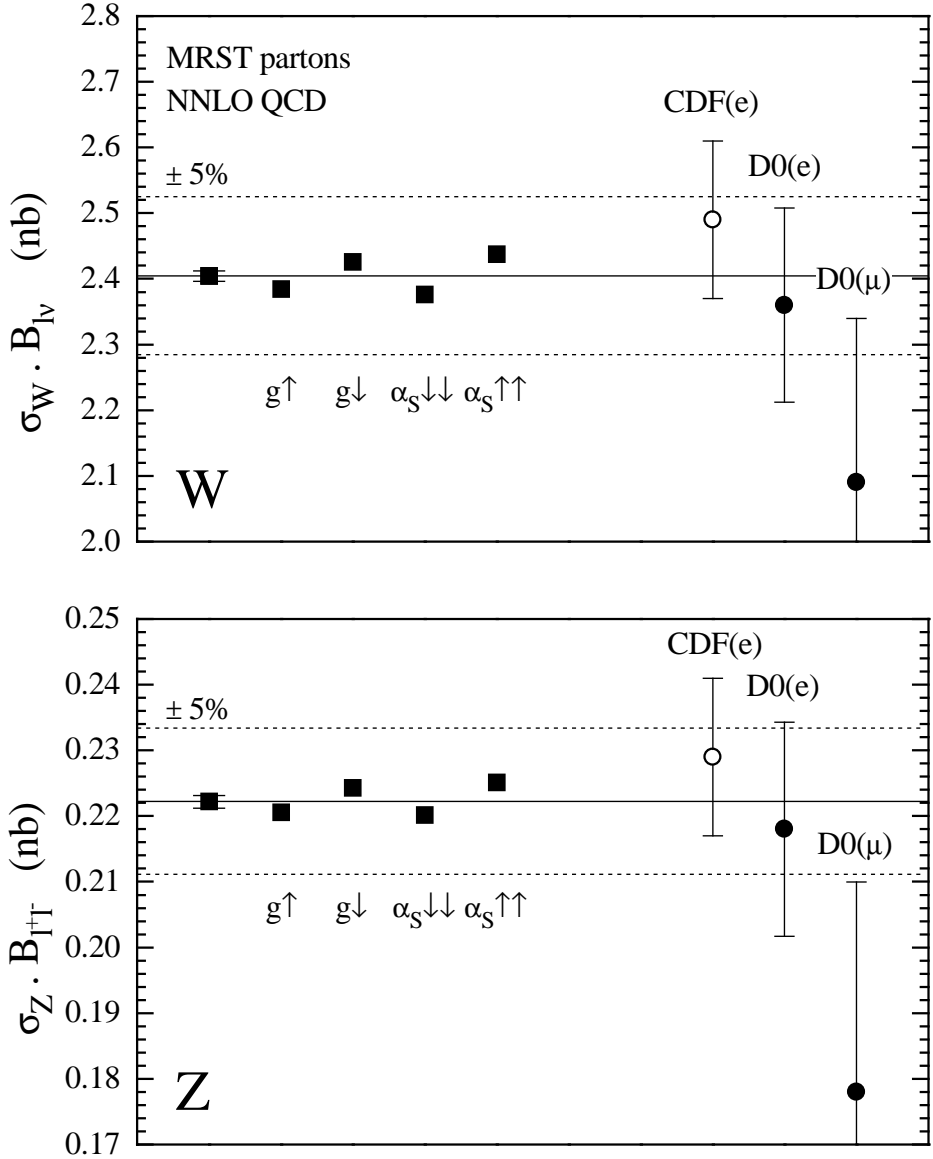,height=20cm}    
\end{center}    
\vspace{-2cm}                                                                
\caption{Predictions for the $W$ and $Z$ production cross sections    
times leptonic branching ratios in     
$p \bar p$ collisions at  1.8 $\TeV$     
using the five MRST parton sets. The error bars on the default MRST prediction 
correspond to a scale variation of $\mu = M_V/2 \to 2 M_V$, $V=W,Z$.    
Experimental measurements from     
 CDF~\protect\cite{CDFSIGW} and D0~\protect\cite{D0SIGW} are shown.}
\label{fig:WZ5}                                                     
\end{figure}                            
\newpage    
                            
\begin{figure}[H]                             
%\vspace{-0.5cm}                                                                
\begin{center}     
\epsfig{figure=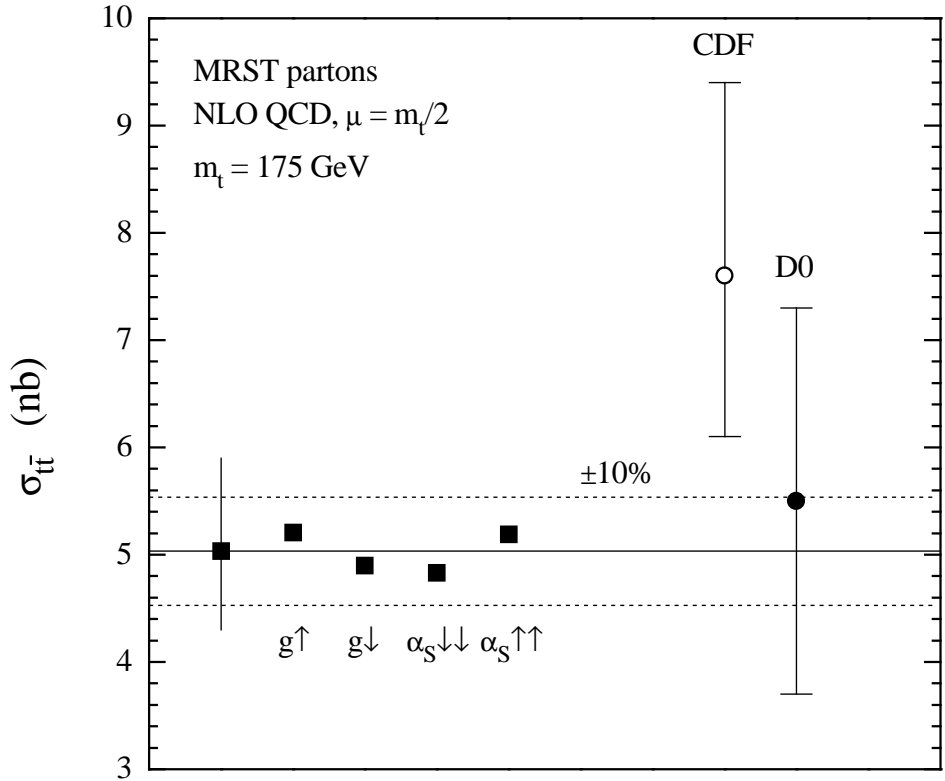,height=20cm}    
\end{center}    
\vspace{-4cm}                                                                
\caption{Predictions for the $t \bar t$ total cross section    
in $p \bar p$ collisions at 1.8 $\TeV$     
using the five MRST parton sets. The error bars on the default MRST prediction 
correspond to a  variation in the top mass of $\pm 5\; \GeV$.    
Experimental measurements from     
 CDF~\protect\cite{topcdf} and D0~\protect\cite{topdzero} are shown.}
\label{fig:top}                                                      
\end{figure}                   
\newpage    
       
\begin{figure}[H]                                                               
%\vspace{1cm}                                                                
\begin{center}     
\epsfig{figure=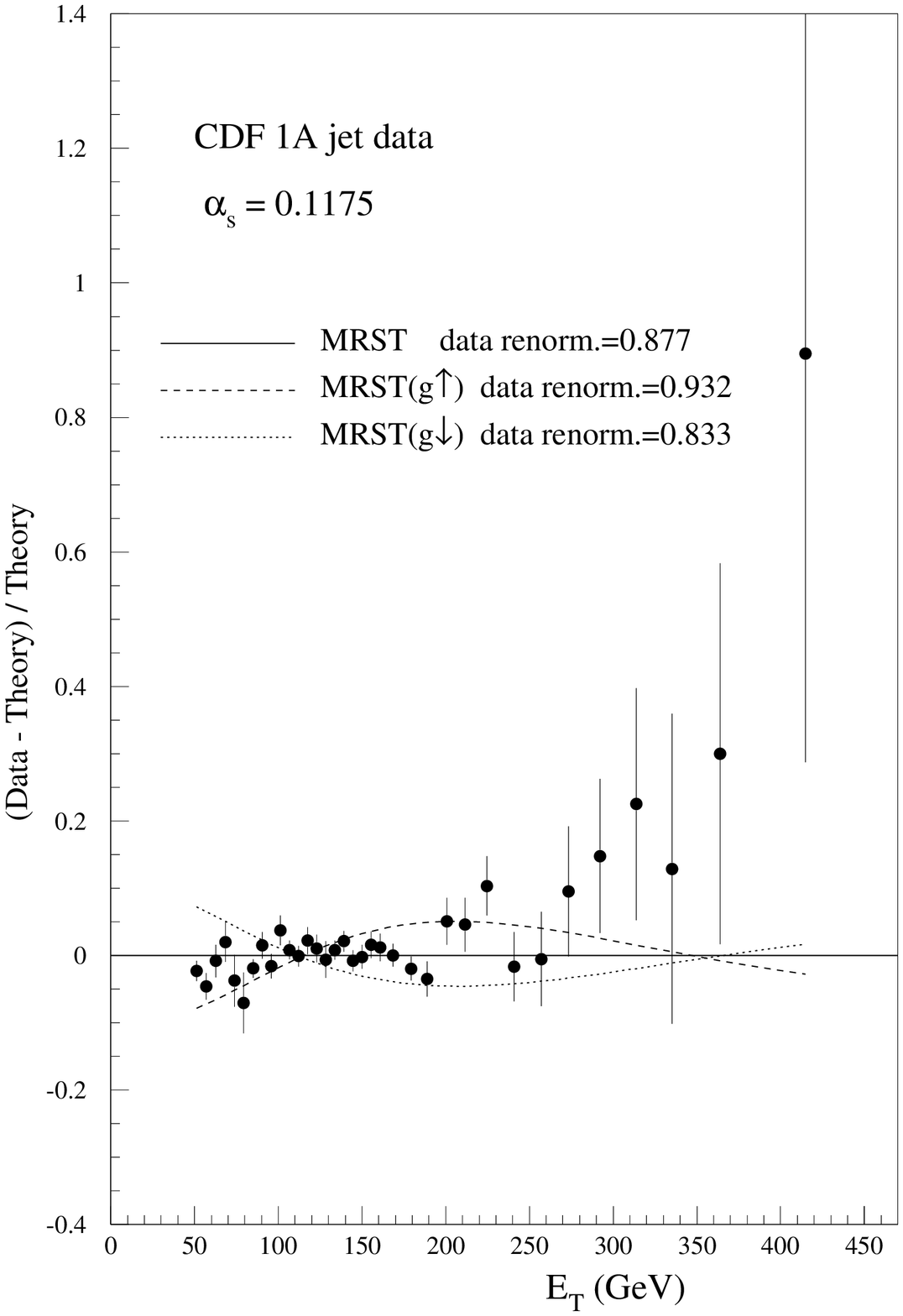,height=20cm}    
\end{center}    
\caption{The next-to-leading order QCD description of the     
CDF~\protect\cite{CDFJ} single jet inclusive $E_T$ distribution    
by the MRST set of partons. The overall normalization of the QCD     
prediction is fitted to the data. The comparisons with the     
MRST($\gup$) and MRST($\gdown$) parton sets are also shown.}                                                                 
\label{fig:J1}                                                                 
\end{figure}                                       
\newpage    
       
\begin{figure}[H]                                                               
%\vspace{1cm}                                                                
\begin{center}     
\epsfig{figure=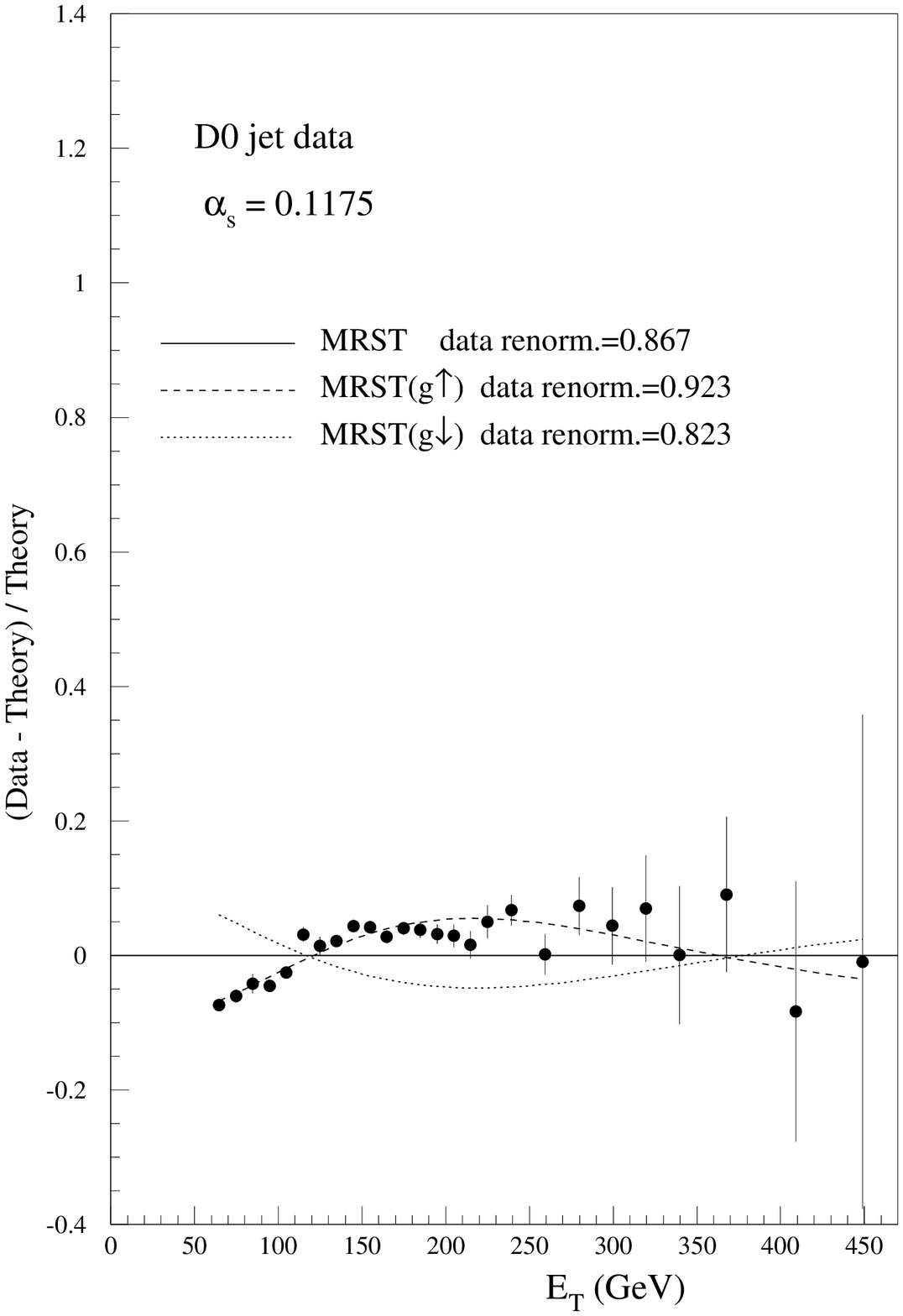,height=20cm}    
\end{center}    
\caption{The next-to-leading order QCD description of the     
D0~\protect\cite{D0J} single jet inclusive $E_T$ distribution    
by the MRST set of partons. The overall normalization of the QCD     
prediction is fitted to the data. The comparisons with the     
MRST($\gup$) and MRST($\gdown$) parton sets are also shown.}         
\label{fig:J2}                                                                 
\end{figure}       
\newpage    
       
\begin{figure}[H]                                                               
%\vspace{1cm}                                                                
\begin{center}     
\epsfig{figure=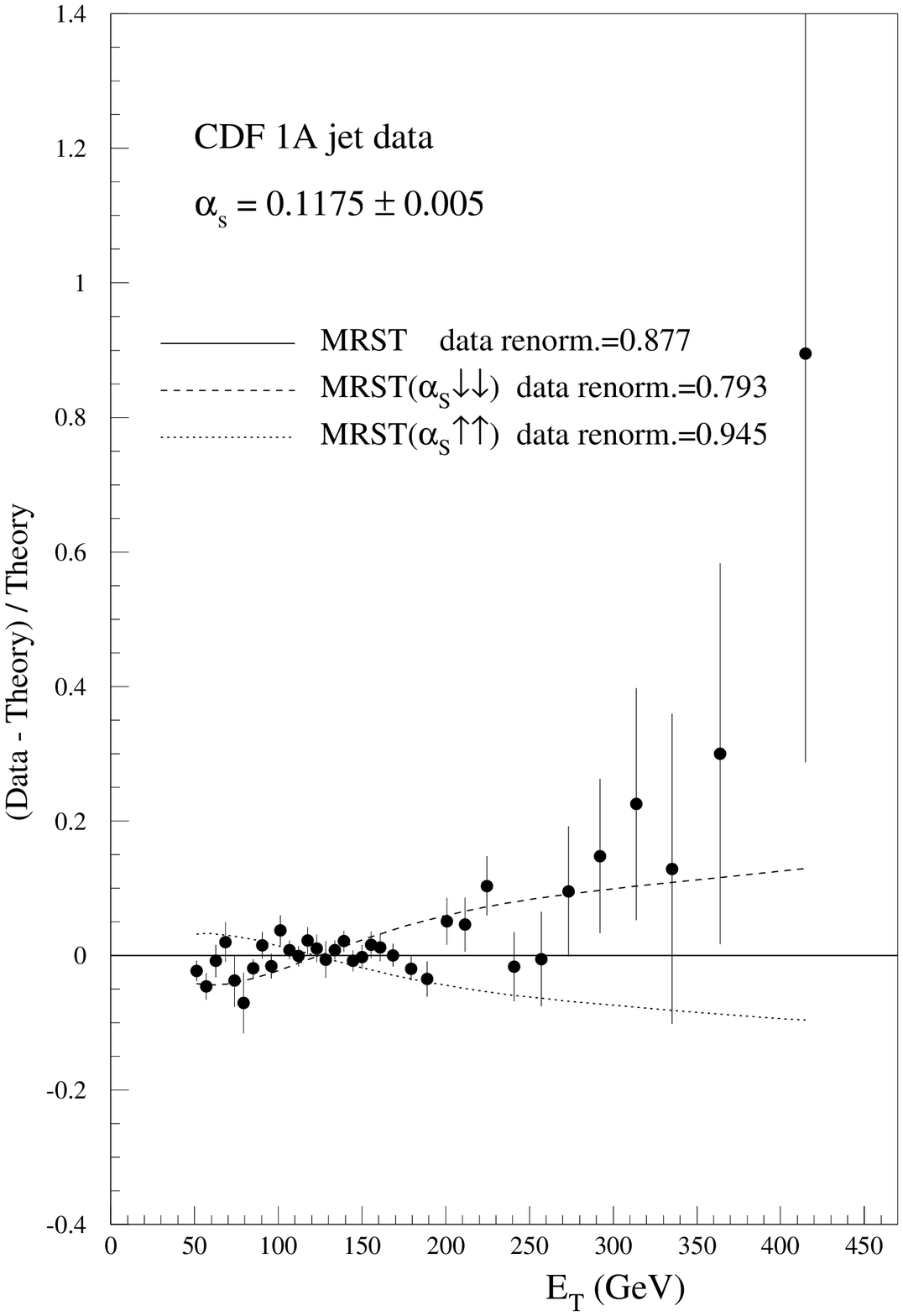,height=20cm}    
\end{center}    
\caption{As for Fig.~\ref{fig:J1} but  including the     
comparisons with the MRST($\protect\asup$)    
and MRST($\protect\asdown$) parton sets.}                            
\label{fig:J3}                                                                 
\end{figure}       
\newpage    
       
\begin{figure}[H]                                                               
%\vspace{1cm}                                                                
\begin{center}     
\epsfig{figure=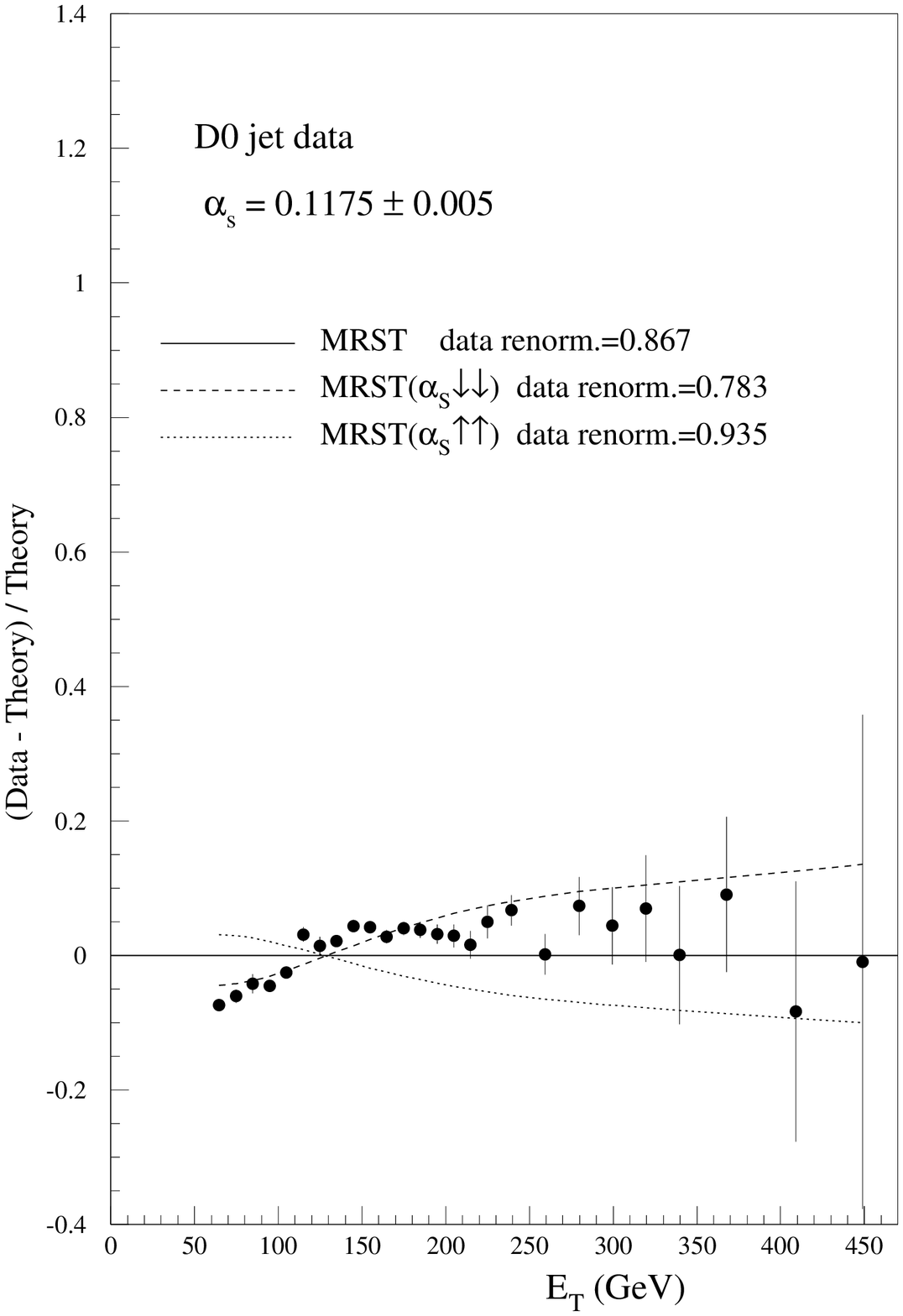,height=20cm}    
\end{center}    
\caption{As for Fig.~\ref{fig:J2} but including the     
comparisons with the MRST($\protect\asup$)    
and MRST($\protect\asdown$) parton sets.}                            
\label{fig:J4}                                                                 
\end{figure}                                                         
\newpage    
                                                                 
\begin{figure}[H]                                                               
%\vspace{1cm}                                                                
\begin{center}     
\epsfig{figure=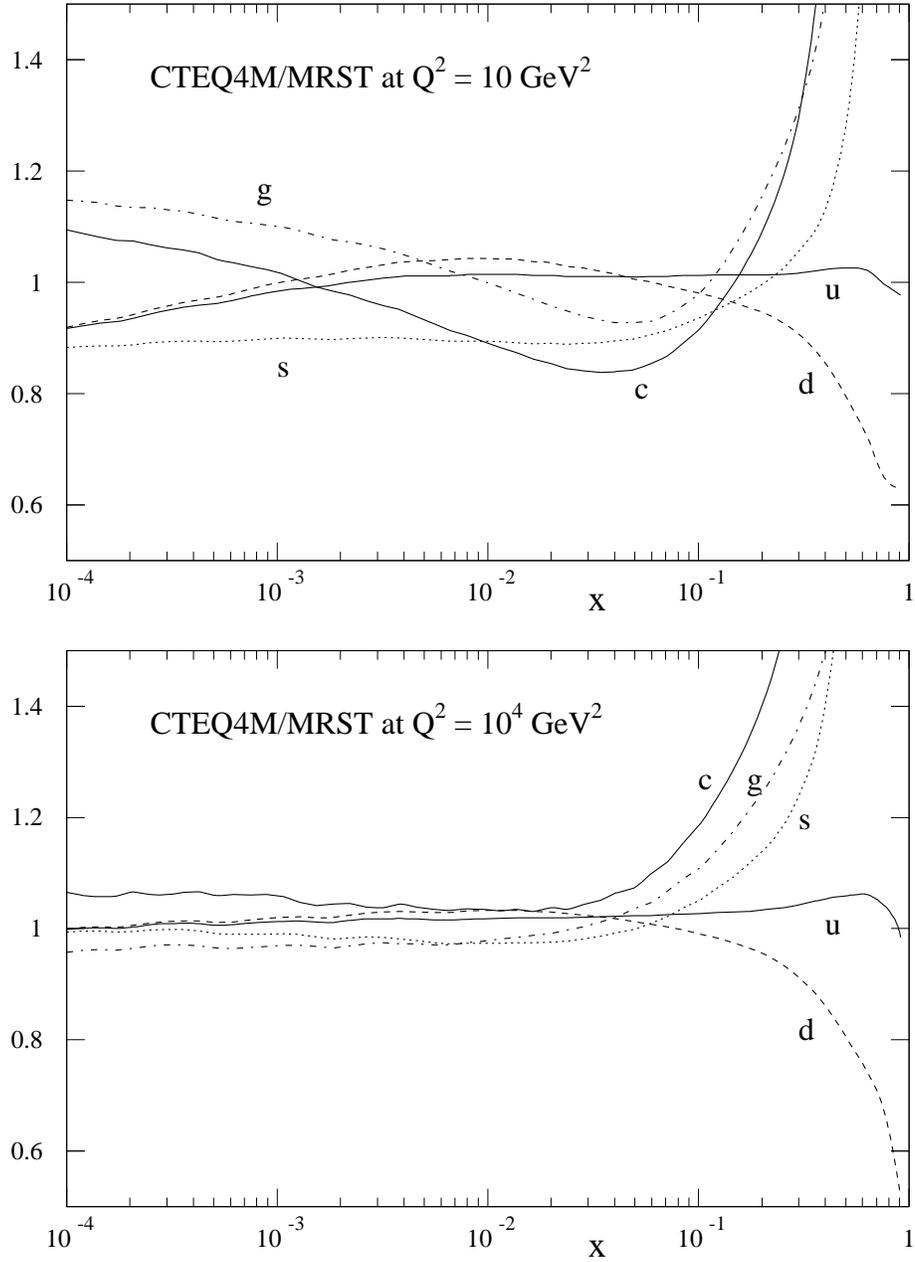,height=20cm}    
\end{center}    
\caption{Ratio of the partons of the CTEQ4M~\protect\cite{CTEQ4M}    
set to those of the MRST set at $Q^2 = 10$ and $10^4\; \GeV^2$.}     
\label{fig:CTEQpart}                                                 
\end{figure}    
                                                                     
}
                                                                             
\end{document}